\documentclass[twocolumn,tighten]{aastex631}
\usepackage{tabto}
\usepackage{capt-of}
\usepackage{upgreek}
\usepackage{makecell}
\usepackage{comment}
\usepackage{amsmath}
\usepackage{mathrsfs}
\usepackage{colortbl}

\shorttitle{VaDAR with the VLBA}
\shortauthors{Schwartzman et al.}

\graphicspath{{./}{Figures/}}

\begin{document}

\title{Varstrometry for Dual AGN using Radio interferometry: VaDAR with the VLBA}

\author[0000-0002-6454-861X]{Emma Schwartzman}
\affiliation{U.S. Naval Research Laboratory, 4555 Overlook Ave SW, Washington, DC 20375, USA}
\affiliation{Department of Physics and Astronomy, George Mason University, 4400 University Drive, MSN 3F3, Fairfax, VA 22030, USA}

\author[0009-0005-4780-991X]{Paula Fudolig}
\affiliation{Department of Physics and Astronomy, George Mason University, 4400 University Drive, MSN 3F3, Fairfax, VA 22030, USA}

\author[0000-0001-6812-7938]{Tracy E. Clarke}
\affiliation{U.S. Naval Research Laboratory, 4555 Overlook Ave SW, Washington, DC 20375, USA}

\author[0000-0003-1991-370X]{Kristina Nyland}
\affiliation{U.S. Naval Research Laboratory, 4555 Overlook Ave SW, Washington, DC 20375, USA}

\author[0000-0002-4902-8077]{Nathan J. Secrest}
\affiliation{U.S. Naval Observatory, 3450 Massachusetts Avenue NW, Washington, DC 20392, USA}

\author[0000-0001-8640-8522]{Ryan W. Pfeifle}
\altaffiliation{NASA Postdoctoral Program Fellow}
\affiliation{X-ray Astrophysics Laboratory, NASA Goddard Space Flight Center, Code 662, Greenbelt, MD 20771, USA}
\affiliation{Oak Ridge Associated Universities, NASA NPP Program, Oak Ridge, TN 37831, USA}

\author[0000-0003-2450-3246]{Henrique Schmitt}
\affiliation{U.S. Naval Research Laboratory, 4555 Overlook Ave SW, Washington, DC 20375, USA}

\author[0000-0003-2277-2354]{Shobita Satyapal}
\affiliation{Department of Physics and Astronomy, George Mason University, 4400 University Drive, MSN 3F3, Fairfax, VA 22030, USA}

\author[0000-0003-2283-2185]{Barry Rothberg}
\affiliation{Department of Physics and Astronomy, George Mason University, 4400 University Drive, MSN 3F3, Fairfax, VA 22030, USA}
\affiliation{U.S. Naval Observatory, 3450 Massachusetts Avenue NW, Washington, DC 20392, USA}

\begin{abstract} 
Multiple active galactic nuclei (multi-AGN) are a known result of galaxy mergers. Therefore, they are an important tool for studying the formation and dynamical evolution of galaxies and supermassive black holes (SMBHs). A novel method for the selection of multi-AGN leverages the exquisite positional accuracy of \textit{Gaia} to detect astrometrically-variable quasars. Previous work has paired this method with radio interferometry on sub-arcsecond scales. In this paper, we present a follow-up study of seven astrometrically-variable quasars from the pilot sample using the Very Long Baseline Array (VLBA). We targeted these seven quasars with the VLBA at 2.0-2.4 GHz (S-band) and 8.0-8.4 GHz (X-band), reaching milliarcsecond resolutions, in order to study the radio properties at smaller scales and to constrain the origin of the astrometric variability. The new observations are also used to identify significant radio-optical offsets in all seven objects, suggesting that many astrometrically-variable quasars also exhibit significant radio-optical offsets. We find that four of the seven sources are possible candidate multi-AGN based on their radio properties and radio-optical offsets. Overall, we use this follow-up study to constrain the smaller-scale radio properties of this sample of astrometrically-variable quasars, and continue to explore the use of this method in the field of multi-AGN.

\end{abstract}


\section{Introduction} \label{sec:intro_vlba}
Within the hierarchical model of galaxy evolution, more massive galaxies are formed via the mergers of their smaller counterparts, in addition to gas and dark matter accretion \cite[e.g.,][]{schweizer1996, toomre1972, rothberg2004}. Most massive galaxies also host a central, supermassive black hole \cite[SMBH;][]{kormendy1995}, with a mass on the order of $10^6 - 10^{10} M_{\odot}$. Galaxy mergers thus result in pairs of gravitationally-bound, synchronously feeding SMBHs that fall to the centers of their host galaxy mergers via a process of dynamical friction. During this process, their separation decreases and they are expected to eventually coalesce \cite[][]{barnes1992, hopkins2008, volonteri2016}. A potential co-evolution exists between a SMBH and its host galaxy, as illustrated in the observed scaling relationship between SMBHs and the properties of their host galaxies \cite[][]{ferrassee2000, gebhardt2000, heckman2014}. As the merger evolves, gravitational and/or hydrodynamical torques drive gas towards the nuclei \cite[e.g.,][]{barnes1996, mihos1996, blumenthal2018, capelo2017}. This can cause gas accretion onto the SMBHs, which then ignite as active galactic nuclei \cite[AGN;][]{hopkins2008, hopkins2010, blecha2018}.

The timescale of a galaxy merger is on the order of hundreds of millions to a few billion years \cite[][]{tremmel2018, Callegari_2009}. Dual AGN represent an earlier stage of evolution, defined as having two AGN with a separation beyond their mutual gravitational spheres of influence \cite[i.e., separations of $\leq$ 110 kiloparsecs,][]{ellison2011, liu2011, pfeifle2023inprep}. Binary AGN systems represent a more advanced evolutionary stage of the merger, having separations within their mutual spheres of influence \cite[i.e., separations $\lesssim$ 30 parsecs,][]{rodriguez2006, liu2018, pfeifle2023inprep}. The process via which a binary AGN decays from separations of $\leq$ 10 pc to $\leq$ 0.1 pc is an area of active research \citep[][]{burkespolaor2018}. Once a binary system reaches separations on scales $<<$1 pc, collapse proceeds efficiently due to gravitational waves \cite[GWs;][]{burkespolaor2018}.

Finding and building larger samples of AGN pairs is necessary to fully understand AGN pair populations and growth across the merger sequence. It is thought that merger-driven SMBH growth represents a crucial stage of galaxy evolution \citep[][]{blecha2018,satyapal2014,ellison2013}. Thus, characterization of AGN pairs at all merger stages is critical. This is made difficult by the relatively few confirmed AGN pairs, particularly at smaller separations. 

A recent review of all confirmed and candidate AGN pairs up to 2020 presents a consolidated list of 156 dual AGN and one binary AGN, amongst thousands of candidates \citep[][]{bigmac}. The vast majority of dual/binary AGN and candidates have been selected and/or confirmed via optical/infrared spectroscopic diagnostics \cite[including multiple velocity peaks;][]{comerford2009, comerford2013, wang2009, liu10b, lyu2016, barrows2013, u2013}, spatially-resolved imaging in the X-ray, optical, infrared, and radio regimes \cite[][]{liu2013, liu2010, komossa2003, koss2011, bianchi2008, piconcelli2010, fu2015}, and mid-infrared colors \cite[][]{pfeifle2019, pfeifle2019b, satyapal2017, ellison2019, barrows2023}. 

The dual AGN population is biased towards local (z $<$ 0.1) redshifts and physical separations $1 < r_p < 100$ kpc \citep[where $r_p$ is the projected separation][]{bigmac}. For binary AGN, the search is significantly more challenging. At very small (pc and sub-pc) scales, the strict spatial requirement necessitates extraordinarily high angular resolution instruments. The only known binary AGN system has a separation of 7.3 pc \cite[][]{rodriguez2006}, and was confirmed via Very Long Baseline Interferometry (VLBI).

Of particular interest are AGN pairs with redshifts at which both the number density of luminous quasars and the global star formation rate density peak, often referred to as ``cosmic noon'' \cite[1 $\leq$ z $\leq$ 3][]{richards2006, madau2014}. However, with respect to the other merger evolutionary stages, the dynamical evolution timescales of dual and binary AGN systems are quite short \cite[$\sim$ 1 Gyr][]{tremmel2018}. Thus, it is predicted that kpc-scale dual AGN represent a short-lived phase, as they quickly progress to the next stage of evolution \cite[][]{tremmel2018, merritt2013, chen2020, yu2002}. This issue is further compounded by sample pre-selection criteria and the limited resolutions and sensitivities of current instruments, leaving an observational gap at high redshifts and small separations \citep[see Fig. 1 in][]{chen_hst}. At redshifts $<$ 1, the vast majority of systems exhibit projected physical separations $>$ 1 kpc \citep[][]{bigmac}. To probe smaller separations (i.e. $\leq$ sub-arcsec), we must move to the higher resolutions of radio interferometry, current space-based imaging, or ground-based imaging with large apertures, adaptive optics, or optical/infrared interferometry. For the realm of binary AGN, only VLBI observations will reach the exceedingly high resolutions required for a direct detection (sub-milliarcsecond scales and smaller). 

The focus of this paper is to further characterize a method that uses astrometry and radio interferometry to select dual and binary AGN candidates. In Section \ref{sec:vars_vlba}, we introduce the varstrometry method. The National Science Foundation's Karl G. Jansky Very Large Array (VLA) was used to study a pilot sample of dual AGN candidates, reaching high enough angular resolution to detect dual AGN systems with separations on sub-arcsecond scales \citep[VaDAR;][]{schwartzman2024}. The results of the pilot VLA study are summarized in Section \ref{sec:vars_vla_vlba}. Follow-up observations at smaller, sub-milliarcsecond scales are made possible with the resolutions of the Very Long Baseline Array (VLBA). In Section \ref{sec:targets_vlba}, we define the follow-up target selection process for the VLBA. In Section \ref{sec:obs_data_vlba}, we describe the new VLBA observations, data reduction, detection criteria, phase positional uncertainty calculations, and analysis. Properties of the VLBA sample are described in Section \ref{sec:sample_vlba}, and are discussed in Section \ref{sec:disc_vlba}. Spectral analysis is presented in Section \ref{sec:specanalysis}. Radio-optical offsets are reported in Section \ref{sec:roo}. Methodology comparisons individual targets are discussed in Section \ref{sec:disc_vlba}. Throughout this paper, all physical separations are the projected separation ($r_p$). A flat $\Lambda$CDM cosmology is adopted, with a $\Omega_{\Lambda}$ = 0.69, $\Omega_{m}$ = 0.31, and $H_{0}$ = 67.7 km s$^{-1}$ Mpc$^{-1}$ \cite[][]{planck2020}.

\section{Varstrometry} \label{sec:vars_vlba}

\textit{Gaia} \cite[][]{gaia2016} is a space-based astrometric mission that has mapped the positions, parallaxes, and proper motions of billions of stars in the Milky Way. The astrometric precision of \textit{Gaia} has provided unprecedented sensitivity to the positions of hundreds of thousands of distant quasars \cite[][]{gaia2021}, and has revealed a population of astrometrically-variable AGN \cite[][]{shen_nature, chen_hst}. A novel astrometric technique leverages \textit{Gaia}'s astrometric precision in the search for AGN pairs. The technique, which was previously used to detect unresolved stellar binaries \citep[e.g.,][]{2016ApJS..224...19M} through photometric variability-induced photocenter pseudo-motion, was applied to the search for unresolved dual AGN by \cite{hwang_initial}. As \textit{Gaia} is progressively source-confused for separations less than $\sim$ 2$\arcsec{}$ (\cite{fabricius2021}; 2$\arcsec{}$ corresponds to $>$ 12.3 kpc for z $>$ 0.5), it is not capable of discerning a close secondary AGN or other extended phenomena commonly associated with AGN, such as jet production. However, the sub-milliarcsecond astrometric precision of \textit{Gaia} is such that pseudo-motion due to flux variability of the individual AGN components in a pair can be measured. This makes astrometric variability a new discovery space for dual and binary AGN. In the context of the search for dual and binary AGN, the technique has been dubbed ``varstrometry'' \citep[variability + astrometry][]{hwang_initial}.

The orbital periods of binary and dual AGN are hundreds to millions of years, respectively \cite[e.g.,][]{dorland2020}, and thus their positions are essentially fixed on the sky, precluding a direct motion measurement via an astrometric mission. However, AGN exhibit intrinsic, stochastic brightness variability on many timescales, some as short as hours \cite[][]{sesar2011variability, MacLeod_2012_variability}. In a dual or binary AGN system with a separation less than the effective angular resolution of \textit{Gaia} ($\sim$0.4$\arcsec{}$), the two component AGN with their varying brightnesses will appear to \textit{Gaia} to have a shifting photocenter \cite[][]{hwang_initial}.

The resolution limits of \textit{Gaia} are such that, for an AGN pair system, individual light curves for each component cannot be observed. A joint variability light curve instead illustrates the stochastic variability of the entire system. In certain cases, an AGN pair system could be identifiable from the joint variability light curve, though in cases where the apparent photocenter of the AGN is offset from that of its host galaxy \citep[e.g., an interacting or merging system or a system in which the AGN obscuration level changes rapidly, confusing the \textit{Gaia} centroid identification;][]{popovic2013}, the stochastic variability will be driven by that offset. Additionally, the varstrometry technique is not only sensitive to dual and binary AGN systems, but can also select for star+quasar superposition systems \cite[][]{pfeifle2023}, lensed quasar systems \cite[e.g.,][]{Mannucci_2022, Ciurlo_2023, Inada_2012, Inada_2014, odowd2015}, and any other morphology that might also drive the excess astrometric noise seen in AGN pair systems (e.g., a single AGN in a host galaxy with bright stellar features such as star formation hotspots that might similarly confuse the \textit{Gaia} centroid identification).

In the case of dual and binary AGN, as the mutual photocenter of the component AGN wanders between the individual AGN, \textit{Gaia} measures a positional ``jitter'', which is representative of the astrometric variations \citep[see Figure 1 in][]{schwartzman2024}. This measurement can also provide a lower limit on the possible physical separation for the component AGN. Assuming a typical fractional rms of $\sim$10\% \cite[e.g.][]{MacLeod_2012_variability}, one can calculate the expected astrometric ``jitter" to be $\sim10$ milliarcseconds for every 0.2$\arcsec{}$ of angular separation. Note that this measurement is a lower limit on the angular separation.

High spatial resolution follow-up has been successful for varstrometry-selected targets. The Varstrometry for Off-nucleus and Dual Sub-kpc AGN \citep[VODKA;][]{shen_nature,chen_hst} program followed up a sample of 84 \textit{Gaia}-identified dual and binary AGN candidates with the Hubble Space Telescope (HST) and Gemini GMOS optical spectroscopy. Their search revealed that $\sim40\%$ of their HST-resolved pairs are likely to be physical quasar pairs or gravitationally-lensed quasars \cite[][]{chen_hst}, including two dual AGN candidates and one triple AGN candidate \cite[][]{chen_hst,shen_nature,chen2022}. \cite{chen2023} presented VLBA observations for 23 radio-bright candidate dual and off-nucleus quasars selected as part of the VODKA program, and used significant offsets detected between the VLBA positions and the \textit{Gaia} positions to further identify candidate dual/binary AGN systems. The VODKA and VaDAR methodology results are compared in Section \ref{subsec:methcomp}.

\cite{schwartzman2024} presented an initial VLA pilot study of 18 \textit{Gaia}-unresolved quasars identified as astrometrically-variable. In combination with significant existing radio and multiwavelength data, the VLA observations were used to constrain the driver of the excess astrometric noise. The pilot study found that $\sim44\%$ of the target sample was likely to be either candidate dual AGN or gravitationally lensed quasars.

\subsection{VLA Results} \label{sec:vars_vla_vlba}

The VLA pilot sample was derived from a cross-match of the SDSS DR16 quasar catalog \cite[DRQ16;][]{sdss} with \textit{Gaia}'s Data Release 3 \citep[DR3;][]{2022arXiv220706849C}. While there are several \textit{Gaia} parameters that can act as indicators of positional noise in a target, the one used in the development of the pilot sample was \texttt{astrometric\_excess\_noise\_significance} (AENS), which is the statistical significance of the \texttt{astrometric\_excess\_noise} (AEN; the previously mentioned astrometric ``jitter'' in units of miilliarcseconds). The sample was limited to sources with AENS $> 5$, ensuring a highly statistically significant measurement of AEN for each target.

The resultant sample was further matched to the VLA Sky Survey \cite[VLASS;][]{vlass} at 3 GHz. As only 1-10\% of AGN are typically radio bright \citep[][]{osterbrock1993,osterbrockbook}, the sample was limited to any targets exhibiting a VLASS detection, thus ensuring the targets would be detected with the VLA. SDSS spectroscopic redshifts were available, and placed the target sample in an interesting observational gap (0.8 $\leq$ z $\leq$ 2.9) in the current population of confirmed and candidate AGN pairs.

In terms of radio morphology at sub-arcsecond scales, the VLA observations identified nine of the eighteen targets as unresolved. Six were identified as multi-component, while the remaining three exhibited jet activity or other extended emission.

In terms of multi-AGN, the multiwavelength data identified four of the eighteen as star+quasar superposition sources, and another four as gravitational lenses. Two were identified as exhibiting jet activity. Overall, eight of the eighteen targets ($\sim$44\% of the overall sample) were identified as either candidate dual AGN or gravitationally lensed quasars.

Given the significant radio survey/archival data available, a thorough review of the radio spectral shapes of each target was possible. This analysis revealed that the majority of the targets, no matter the classification of their spectral shape, exhibited a spectral index consistent with that of optically thin synchrotron emission. The overall sample was also compared to a matched control sample of targets that did not exhibit statistically significant AEN; the target sample was not particularly radio loud in comparison to the matched controls. This likely eliminates the possibility of blazar jet activity as the main driver of the high AEN.

This paper presents milliarcsecond-scale VLBA observations of seven of the original eighteen quasars. The significantly higher resolution of the VLBA observations has been used to probe source structure and properties on smaller scales, and to further constrain the radio morphological and spectral properties the varstrometry-selected sample.

\section{VLBA Target Selection} \label{sec:targets_vlba}

VLBA follow-up observations were made of a portion of the pilot sample. A subset of the brightest targets (with a VLA/A-configuration 3 GHz flux density $>$ 1 mJy) were chosen in order to provide a reasonable flux limit for the VLBA observations. Sources displaying the narrow stellar absorption lines indicative of a star+quasar superposition were excluded, leaving a VLBA sample of seven targets. We note that one of the chosen seven targets observed in the VLBA cycle was later confirmed through analysis as a star+quasar superposition. Though it is likely in this case that the excess astrometric noise is driven by the foreground star, the radio emission from the quasar remains worthy of consideration.

The VLBA sample is similar to the original VLA sample in radio properties (see Table \ref{tab:targetdetails_vlba}), with a spectral index range (taken between the VLA/A-configuration 3 GHz and 10 GHz observations) of $-0.803 < \alpha < 0.816$\footnote{where $S_{\nu} \propto \nu^{\alpha}$}. The radio morphologies as seen in the VLA/A-configuration show a similar diversity, with two multi-component targets, two unresolved targets, two jetted targets, and one target identified as star+quasar superposition.

\begin{deluxetable*}{ccccccccc}
\tablenum{1}
\tablecaption{Target Details\label{tab:targetdetails_vlba}}
\tablewidth{0pt}
\tablehead{
\colhead{Source} & \colhead{Redshift} & \colhead{Scale} & \colhead{G} &
\colhead{AEN} & \colhead{AENS} & \colhead{$VLA/S_{3 GHz}$} & \colhead{$VLA/S_{10 GHz}$} & \colhead{VLA Class}\\
\colhead{[SDSS]} & \colhead{} & \colhead{[kpc/arcsec]} & \colhead{[mag]} &
\colhead{[mas]} & \colhead{} & \colhead{mJy} & \colhead{mJy} & \colhead{}\\
\colhead{(1)} & \colhead{(2)} & \colhead{(3)} &
\colhead{(4)} & \colhead{(5)} & \colhead{(6)} & \colhead{(7)} & \colhead{(8)} & \colhead{(9)}
}
\startdata
J011114.41+171328.5 & 2.198 & 8.5 & 19.29 & 4.46 & 42.3 & 78.7 $\pm$ 0.24 & 70.8 $\pm$ 0.22 & Multi Component\\
J080009.98+165509.4 & 0.708 & 7.4 & 18.30 & 4.51 & 175 & 1.89 $\pm$ 0.04 & 2.23 $\pm$ 0.06 & Unresolved\\
J121544.36+452912.7 & 1.132 & 8.4 & 19.09 & 1.22 & 5.41 & 23.4 $\pm$ 0.05 & 16.4 $\pm$ 0.06 & Jet\\
J143333.02+484227.7 & 1.357 & 8.6 & 19.08 & 1.94 & 13.2 & 34.9 $\pm$ 0.22 & 14.1 $\pm$ 0.44 & Jet\\
J162501.98+430931.6 & 1.653 & 8.7 & 19.23 & 8.20 & 157 & 1.31 $\pm$ 0.08 & 0.63 $\pm$ 0.05 & Multi Component\\
J172308.14+524455.5$\dagger$ & 2.568 & 8.2 & 17.72 & 0.64 & 6.50 & 1.11 $\pm$ 0.03 & 3.01 $\pm$ 0.03 & Star+quasar\\
J173330.80+552030.9 & 1.201 & 8.5 & 18.58 & 1.73 & 17.3 & 7.01 $\pm$ 0.03 & 4.71 $\pm$ 0.03 & Unresolved\\
\enddata
\caption{Column 1: Coordinate names in the form of ``hhmmss.ss$\pm$ddmmss.s" based on SDSS DR16Q; $\dagger$ indicates star+quasar superposition source. Column 2: Spectroscopic redshift from SDSSDR16Q \cite[][]{sdss}, described in detail in \cite{bolton2012}. Column 3: Cosmological scale in kpc per arcsecond. Column 4: \textit{Gaia} \textit{G}-band mean magnitude. Column 5: \textit{Gaia} \texttt{astrometric\_excess\_noise}. Column 6: \textit{Gaia} \texttt{astrometric\_excess\_noise\_significance}. Column 7: VLA 3 GHz (S-band) total flux density $\pm$ error. Column 8: VLA 10 GHz (X-band) uv-tapered total flux density $\pm$ 1$\sigma$ error. Column 9: Radio morphological classification at sub-arcsecond scales, defined using the VLA observations at 10 GHz.}
\end{deluxetable*}

Table \ref{tab:targetdetails_vlba} presents the target properties, including the VLA properties as presented in \cite{schwartzman2024}. The VLA/10 GHz radio morphological classifications fall into one of three categories at sub-arcsecond scales; morphological classifications are defined in \cite{schwartzman2024}. The VLA flux densities are those of the entire source, measured at 3 and 10 GHz. All VLA imaging was done with the Common Astronomy Software Application \citep[CASA;][]{casa}, with Briggs weighting \citep[][]{briggs1995} and a robust factor of 0.5.

\section{Observations and Data Analysis} \label{sec:obs_data_vlba}

In the following sections, we describe the new VLBA observations, their calibration and imaging, how detections were confirmed, and the flux density and positional measurements that were made. 

\subsection{VLBA Observations}

VLBA radio observations of all seven targets were made at S-band (2.2-2.4 GHz, central frequency 2.3 GHz, central wavelength 13 cm) and X-band (8.0-8.8 GHz, central frequency 8.3 GHz, central wavelength 4 cm) under project code BS320. All VLBA data were correlated using the DiFX software correlator \citep[][]{deller2011}. Phase referencing with a switching angle of 2$^\circ$ was used to account for the expected faintness of all sources. Two minute scans on the phase, rate, and delay calibrator preceded and followed three minute scans of the associated target. For each target, a coherence-check calibrator was also observed. Table \ref{tab:obsdetails_vlba} lists the observational details for each target, including all calibrators (absolute flux, complex gain, and coherence check calibrators), in addition to the stations used in each observation (note that an asterisk denotes stations experiencing partial downtime). 

\begin{deluxetable*}{cccccccc}
\tablenum{2}
\tablecaption{Observation Details\label{tab:obsdetails_vlba}}
\tablewidth{0pt}
\tablehead{
\colhead{Source} & \colhead{Band} & \colhead{Obs. Date} & \colhead{Amp cal} &
\colhead{Gain cal} & \colhead{Coherence} & \colhead{Stations}\\
\colhead{[SDSS]} & \colhead{} & \colhead{[UT]} & \colhead{} &
\colhead{} & \colhead{} & \colhead{}\\
\colhead{(1)} & \colhead{(2)} & \colhead{(3)} &
\colhead{(4)} & \colhead{(5)} & \colhead{(6)} & \colhead{(7)}
}
\startdata
J011114.41+171328.5 & S & 02-07-2023 & J0121+1149 & J0101+1639 & J0106+1951 & SC, HN*, NL, FD, LA, PT*, KP, OV, BR, MK* \\
J011114.41+171328.5 & X & 02-09-2023 & J0121+1149 & J0101+1639 & J0106+1951 & SC, HN*, NL, FD, LA, PT*, KP, OV, BR, MK \\
J080009.98+165509.4 & S & 02-06-2023 & J0805+2106 & J0802+1809 & J0750+1823 & SC, HN, NL, FD, LA, PT, KP, OV, BR, MK  \\
J080009.98+165509.4 & X & 02-08-2023 & J0805+2106 & J0802+1809 & J0750+1823 & SC*, NL, FD, LA, PT*, KP*, OV*, BR, MK*  \\
J121544.36+452912.7 & S & 02-03-2023 & J1203+4803 & J1223+4611 & J1224+4335 & HN, NL, FD, LA, PT, OV, BR, MK \\
J121544.36+452912.7 & X & 02-03-2023 & J1203+4803 & J1223+4611 & J1224+4335 & SC, FD, LA, PT, KP, OV, BR, MK \\
J143333.02+484227.7 & S & 02-13-2023 & J1500+4751 & J1439+4958 & J1424+4705 & SC, NL, FD, LA, PT, KP, OV, BR \\
J143333.02+484227.7 & X & 02-18-2023 & J1500+4751 & J1439+4958 & J1424+4705 & SC, HN, NL, FD, LA, PT, KP, OV, BR* \\
J162501.98+430931.6 & S & 03-21-2023 & J1640+3946 & J1625+4347 & J1608+4012 & SC, HN, NL, FD, LA, PT, KP*, BR* \\
J162501.98+430931.6 & X & 03-23-2023 & J1640+3946 & J1625+4347 & J1608+4012 & SC, HN, NL*, FD, LA, PT, KP, OV, BR \\
J172308.14+524455.5 & S & 02-21-2023 & J1740+5211 & J1723+5236 & J1727+5510 & SC, HN, NL, FD, LA, PT, KP, OV, BR \\
J172308.14+524455.5 & X & 03-11-2023 & J1740+5211 & J1723+5236 & J1727+5510 & SC, HN, NL, FD, LA, PT, KP, OV, BR \\
J173330.80+552030.9 & S & 03-16-2023 & J1740+5211 & J1727+5510 & J1722+5856 & SC, NL, FD, LA, PT, KP*, OV, BR \\
J173330.80+552030.9 & X & 03-20-2023 & J1740+5211 & J1727+5510 & J1722+5856 & SC*, HN, NL, FD, LA, PT, KP, BR \\
\enddata
\caption{Note - Column 1: Coordinate names in the form of ``hhmmss.ss$\pm$ddmmss.s" based on SDSS DR16Q. Column 2: VLBA observation band. Column 3: VLBA observation date. Column 4: VLBA amplitude check calibrator. Column 5: VLBA complex gain calibrator. Column 6: VLBA coherence calibrator. Column 7: Participating VLBA stations. Note that the asterisk marks an station experiencing partial downtime due to weather or technical problems. VLBA stations available are Mauna Kea, Hawaii (MK), Owens Valley, California (OV), Brewster, Washington (BR), North Liberty, Iowa (NL), Hancock, New Hampshire (HN), Kitt Peak, Arizona (KP), Pie Town, New Mexico (PT), Fort Davis, Texas (FD), Los Alamos, New Mexico (LA), and St. Croix, Virgin Islands (SC).}
\end{deluxetable*}

\begin{deluxetable*}{ccccccc}
\tablenum{3}
\tablecaption{Observation Details\label{tab:imgdetails_vlba}}
\tablewidth{0pt}
\tablehead{
\colhead{Source} & \colhead{S: $B_{maj}, B_{min}, PA$} & \colhead{X: $B_{maj}, B_{min}, PA$} & \colhead{$\sigma_{2.3 GHz}$} & \colhead{$\sigma_{8.3 GHz}$} & \colhead{Flag$_{2.3 GHz}$} & \colhead{Flag$_{8.3 GHz}$} \\
\colhead{[SDSS]} & \colhead{[mas, mas, $^\circ$]} & \colhead{[mas, mas, $^\circ$]} & \colhead{[$\upmu$Jy/bm]} & \colhead{[$\upmu$Jy/bm]} & \colhead{\%} & \colhead{\%}\\
\colhead{(1)} & \colhead{(2)} & \colhead{(3)} &
\colhead{(4)} & \colhead{(5)} & \colhead{(6)} & \colhead{(7)}
}
\startdata
J011114.41+171328.5 & 9, 3, -13 & 8, 4, 20 & 248 & 443 & 70.0 & 74.1\\
J080009.98+165509.4 & 12, 4, -15 & 2, 1, -9 & 87 & 107 & 73.1 & 10.7\\
J121544.36+452912.7 & 8, 4, -14 & 2, 1, 3 & 447 & 321 & 60.0 & 15.0\\
J143333.02+484227.7 & 10, 8, 28 & 2, 1, 11 & 99 & 47 & 68.4 & 31.4\\
J162501.98+430931.6 & 7, 4, 19 & 2, 1, 32 & 87 & 22 & 53.1 & 40.6\\
J172308.14+524455.5 & 11, 4, -9 & 2, 1, 24 & 106 & 95 & 91.4 & 18.7\\
J173330.80+552030.9 & 8, 3, 57 & 2, 1, 52 & 174 & 186 & 91.5 & 16.8\\
\enddata
\caption{Note - Column 1: Coordinate names in the form of ``hhmmss.ss$\pm$ddmmss.s" based on SDSS DR16Q. Column 2: Restoring beam at 4 cm: major axis, minor axis, position angle. Column 3: Restoring beam at 13 cm: major axis, minor axis, position angle. Column 4: 1-$\sigma$ sensitivity at 4 cm. Column 5: 1-$\sigma$ sensitivity at 13 cm. Column 6: Percentage of target data flagged during calibration of 2.3 GHz observations. Column 7: Percentage of target data flagged during calibration of 8.3 GHz observations.
}
\end{deluxetable*}

\subsection{Data Calibration} \label{sec:calib_vlba}

The observations were calibrated manually using the Common Astronomy Software Application \citep[CASA;][]{casanew2022} following standard VLBA procedures for phase-referenced observations \citep[][]{VLBA_SciMemo38,vanbemmel2022}. An in-depth description of the calibration has been included in Section \ref{sec:vlba_calib_appendix}.

\subsection{Imaging} \label{sec:imaging}

We performed standard VLBA phase referencing \citep[][]{reid2014,wrobel2000,nyland2013}. We verified that all phase referencing calibrators were compact in nature, and performed a round of phase-only self calibration on any phase calibrator with extended structure. All target imaging was completed with CASA \citep[][]{casa}, and was performed with a Clark deconvolver \citep[][]{clark1980}. For the majority of imaging, Briggs weighting was used, with a robust parameter of 0.5 \citep[][]{briggs1995}. In the case of some of the more heavily-flagged visibilities, natural weighting was used in order to improve sensitivity and lessen the impacts of poor uv-coverage and PSF issues. For one visibility, J011114.41+171328.5 at 8.3 GHz, uv-tapering was used to mitigate the elongated beam. The imaging information is indicated in the caption of each target image (see Figures \ref{fig:VLBA_011114} - \ref{fig:VLBA_173330}). Masks for deconvolution were drawn manually, and did not rely on automated masking. Table \ref{tab:imgdetails_vlba} lists the resolutions and sensitivities for each image at each band.

\subsection{Detections and Flux Measurements} \label{sec:fluxmeasurements}

\begin{figure*}
    \centering
     \includegraphics[width=18cm]{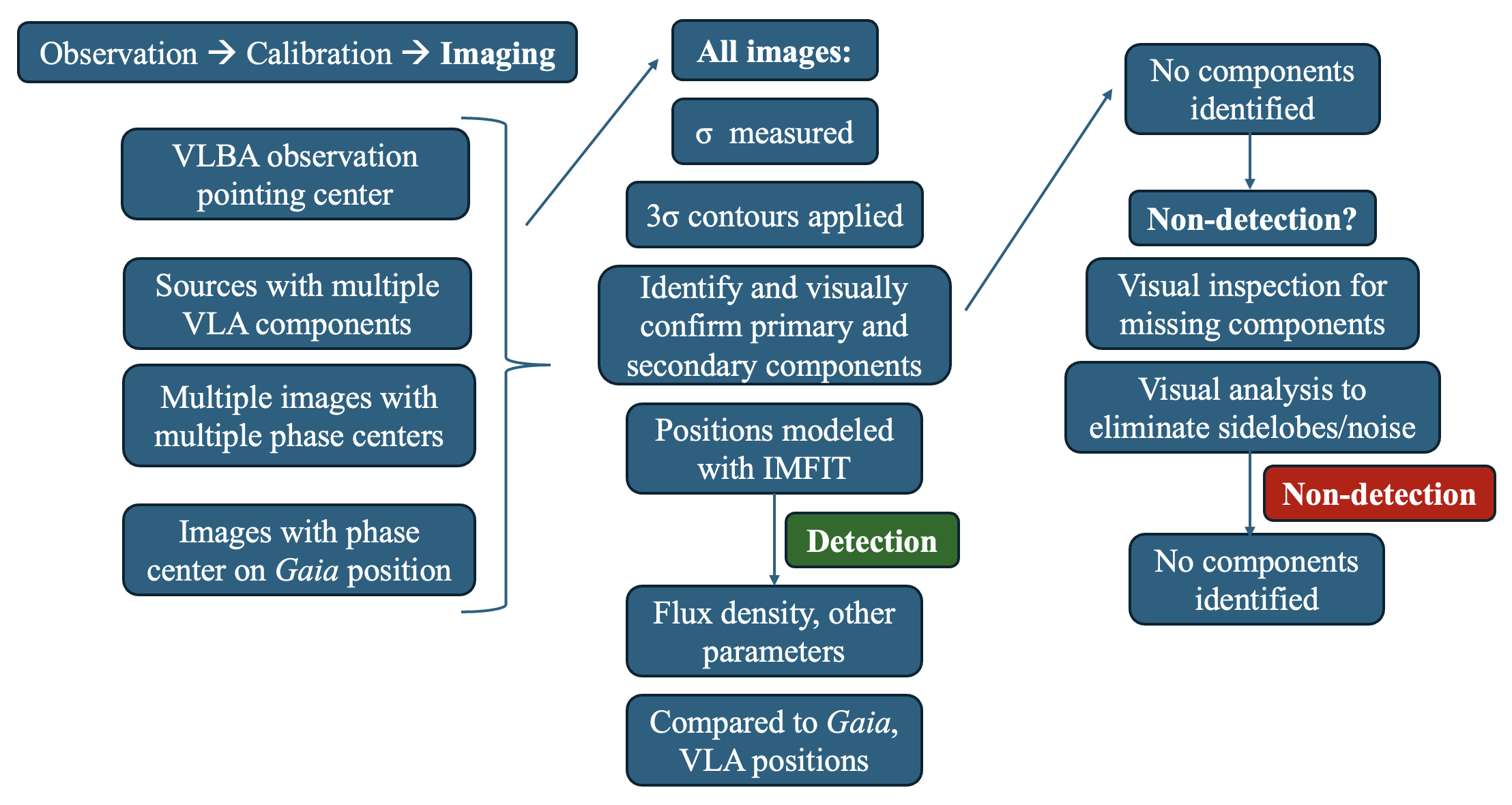}
        \vspace*{1mm}
          \caption{Note: Flowchart illustrating the method and confirmation criteria for detections and non-detections in the new VLBA observations.}
    \label{fig:flowchart}
\end{figure*}

Every observation was carefully inspected to visually confirm a detection. Once imaging was complete, an overall image sensitivity was measured away from the pointing center. The sensitivity measurements are reported in Table \ref{tab:imgdetails_vlba}. Once the sensitivity was determined, 3$\sigma$ contours were applied to each image. Though in many cases the point-like nature of the detected emission made this step unnecessary, this did ensure the identification of any secondary components away from the pointing center, as well as any extended emission. Where necessary, multiple images were created in order to fully image the targets with multiple components. The positions of any significant detections were then compared to the positions of known VLA components. In the case of a non-detection, the new VLBA image was visually inspected for any missing components. Finally, a similar visual analysis of each significant detection was performed for all new VLBA observations in order to eliminate any sidelobes or other noise that was initially identified as significant. Figure \ref{fig:flowchart} illustrates the detection identification process.

Flux densities for every detection were measured with CASA's \textsc{IMFIT} function, which fits one or more elliptical Gaussian components on a defined image region. The image region is defined by the user, and boxes were chosen carefully so as to encompass the detection, in addition to enough surrounding noise to achieve a useful error estimate. Error estimates are based on the work of \cite{condon1997}. Flux density and positional properties were recorded for each detection. All results are reported in Section \ref{sec:fluxdensities}. Flux density scale errors were also taken into account. Following the VLBA Observing Guide\footnote{science.nrao.edu/facilities/vlba/docs/manuals/propvlba/\\
calibration-considerations}, a flux density scale calibration accuracy of 10\% was assumed for both bands \citep[][]{midderlberg2011}. Thus, the flux errors presented in Table \ref{tab:fluxes_VLBA} reflect both the errors in the Gaussian models and the flux density scaling errors added in quadrature. 

\subsection{Positional Uncertainties} \label{sec:phasepositionaluncertainties}

The positional accuracies of each source are dominated by the positional uncertainties of their phase calibrators. We quantified the positional uncertainty by combining the error in the absolute position of the phase calibrator, the uncertainty in the source position, and the phase referencing error. The error in the official position of the phase calibrator, $\sigma_{ref}$\footnote{In some cases, phase-only self calibration was performed on the calibrator in order to account for residual errors caused by source structure. In principle, this could change the absolute position of the calibrator. We account for this with $\sigma_{ref}$ in Equation \ref{equation:pos}}, switching angle error $\sigma_{switch}$, and the Gaussian fit error $\sigma_{imfit}$ were added in quadrature, following standard propagation of errors:

\begin{equation} 
    \sigma_{pos} = \sqrt{(\sigma_{ref})^2 + (\sigma_{switch})^2 + (\sigma_{imfit})^2}.
    \label{equation:pos}
\end{equation}

The official positions for the phase calibrators were taken from the latest 2024d release of the Radio Fundamental Catalog\citep[][]{petrov2024}\footnote{https://astrogeo.org/rfc/}. The switching angle error is the error introduced when switching between the target and the phase calibrator. Section \ref{sec:positionsanduncertainty} presents the final phase calibrator calculations. Section \ref{sec:phasecalimages} presents the final phase calibrator images.

\section{Sample Properties} \label{sec:sample_vlba}

In the following sections, we present new VLBA observations of seven quasars with significantly high astrometric excess noise at 3\,cm and 12\,cm, including the new images and morphology identifications in Section \ref{sec:morphology}, measurements of flux density and errors in Section \ref{sec:fluxdensities}, calculations of compactness and brightness temperature in Section \ref{sec:compactbt}, and determination of positions and uncertainties in Section \ref{sec:positionsanduncertainty}. Spectral properties are presented in Section \ref{sec:specanalysis} and final spectra in Figure \ref{fig:VLBA_spectra}. Positional offsets are discussed in Section \ref{sec:roo} and presented in Figure \ref{fig:roo}.

\subsection{Morphology} \label{sec:morphology}

Figures \ref{fig:VLBA_011114}-\ref{fig:VLBA_173330} in Appendix A present the new VLBA observations of all sources, with images of specific components, where necessary. A basic radio morphology classification reveals a variety of structure, and is presented in Table \ref{tab:fluxes_VLBA}. All images are shown with contours beginning at 3$\sigma$ and proceeding in integer multiples of $\sqrt2$. Each image is labeled with the appropriate frequency, a scale bar labeled with the appropriate number of milliarcseconds and kpc or pc (for scales, see Table \ref{tab:targetdetails_vlba}), and a beam in the lower left-hand corner. Images at both frequencies include the \textit{Gaia} position, marked with a grey ellipse to reflect the error in the position. The size of the ellipse marks the error in the \textit{Gaia} position. In the 2.3 GHz images, the blue crosses mark the position of the VLBA detection at 8.3 GHz, and vice versa. The size of the cross reflects the error in the VLBA position. In the 2.3 GHz images, a dark pink square outlines the size of the 8.3 GHz images. In the cases of J162501.98+430931.6 and J143333.02+4834227.7, multiple components were observed in the VLA images. Only images in which a source is detected above the 3$\sigma$ level are included.

Five of the seven targets display unresolved, point-like radio emission at sub-milliarcsecond scales. The new VLBA imaging has provided a significantly smaller-scale view of the compact emission, compared to the VLA observations, but it is still possible for there to be structure on still smaller scales that remain inaccessible with the new observations. One of the unresolved targets, J172308.14+524455.5, was identified in \citep{schwartzman2024} as a star+quasar superposition via an analysis of the SDSS spectra, which showed the presence of stellar absorption lines consistent with being at a redshift $z = 0$. 

For this target, it is likely that the high AENS is driven by the superimposed star. It is possible for star+quasar superposition sources to show similar morphologies to multi-AGN at optical, IR, and radio wavelengths. Thus, the radio observations of the quasar are still worthy of consideration as understanding the radio properties of these types of contaminants will be useful in constraining varstrometry as a multi-AGN selection methodology. 

The angular and projected physical sizes of each source are included in Table \ref{tab:fluxes_VLBA}, measured using CASA's IMFIT function. For unresolved VLBA sources, the measured angular extent of the source acts as an upper limit on the possible separation of any components. While constraints can be placed on unresolved targets with the new VLBA observations, further multi-wavelength and smaller scale radio follow-up at different frequencies will be required to continue to characterize the astrometric driver of each system, and to confirm or reject the existence of any smaller scale radio components. 

Two of the seven targets display smaller-scale jets or other extended emission more commonly associated with a single AGN, though multi-AGN systems are similarly capable of exhibiting jet activity \cite[e.g. 3C75 or CSO 0402+379;][]{owen1985, rodriguez2006}. This morphology is generally an excellent indicator of the driver of the astrometric variability \citep[][]{hwang_initial}. Of the two targets displaying milliarcsecond-scale jet activity, SDSSJ143333.02+484227.7 (see Figure \ref{fig:VLBA_143333}) displays jet activity at both arcsecond and milliarcsecond-scales. Component A in the VLA observations is detected at both 2.3 and 8.3 GHz, though it is an unresolved, point-like source at both frequencies. Component B, which was identified in the VLA observations as the ``central core'', is similarly detected at both frequencies, though it displays extended emission at both, including extensive, clear jet activity at 2.3 GHz. Components C and D are not detected in the VLBA observations. Given that in the original VLA observations, Components A and C exhibit similar flux densities, it is likely that the emission from Component C exists on spatial scales too large to be detected with the VLBA.

The other target exhibiting jet activity in the new VLBA sample is SDSSJ011114.41+171328.5 (see Figure \ref{fig:VLBA_011114}). This target exhibits clear extension to the northeast at 2.3 GHz, though at 8.3 GHz it appears as a point source. At 2.3 GHz, the extension appears more as a secondary peak to the northeast, perhaps indicating the existence of a companion AGN. It could also be indicative of a smaller-scale jet exhibiting hotspot activity. 

\subsection{Flux Densities} \label{sec:fluxdensities}

Table \ref{tab:fluxes_VLBA} presents the flux density information for each source. This includes the peak flux measurements, as well as their errors, extracted from the \textsc{IMFIT} results, in addition to a morphology label and luminosity calculations. Finally, angular and projected physical source sizes are included; for the unresolved, point-like sources, these act as upper limits to the projected separation, assuming the AENS is driven by an AGN pair. Though several sources exhibit multiple components or extension in their VLA observations, only components detected in the VLBA observations are listed in Table \ref{tab:fluxes_VLBA}. For the cases of a detection in only one of two bands, an upper limit of the RMS value is listed for the non-detection.

Fluxes of all targets are measured as peak flux densities, in units of Janskys per beam, and all flux errors take into account the errors in the Gaussian modeling and the flux density scaling errors. All fluxes are presented in Section \ref{sec:fluxmeasurements}. The seven targets in the new VLBA observations range in angular size from 12.1 - 84.5 milliarcseconds at 2.3 GHz and 3.22 - 7.21 milliarcseconds at 8.3 GHz, and in projected physical size from 102 - 727 pc at 2.3 GHz and 25.2 - 61.2 pc at 8.3 GHz, in the redshift range $0.708 < z < 2.568$. This redshift and pair separation parameter space has gone without significant study or detections in the known population of multi-AGN systems \citep[][]{pfeifle2023inprep,voggel2022,koss2023}, and is thus of great interest. We note that the new VLBA observations are also sensitivity-limited, and thus it is possible that there exist other components in the observations that have gone undetected.

\begin{deluxetable*}{ccccccccc}
\tablenum{4}
\tablecaption{Sample Properties for VLBA Sources\label{tab:fluxes_VLBA}}
\tablewidth{0pt}
\tablehead{
\colhead{Source} & \colhead{Morphology} &
\colhead{$S_{13 cm}$} & \colhead{log($L_{13 cm}$)} & \colhead{$S_{4 cm}$} &
\colhead{log($L_{4 cm}$)} & \colhead{$\Delta \theta r_{p,13 cm}$} & \colhead{$\Delta \theta r_{p,4 cm}$} & \colhead{$\alpha_{8.3GHz}^{2.3 GHz}$}\\
\colhead{[SDSS]} & \colhead{} &
\colhead{[mJy/bm]} & \colhead{[W Hz$^{-1}$]} & \colhead{[mJy/bm]} & \colhead{[W Hz$^{-1}$]} & \colhead{[mas(pcs)]} & \colhead{[mas(pcs)]} & \colhead{}\\
\colhead{(1)} & \colhead{(2)} & \colhead{(3)} &
\colhead{(4)} & \colhead{(5)} & \colhead{(6)} & \colhead{(7)} & \colhead{(8)} & \colhead{(9)}
}
\startdata
J011114.41+171328.5 & Extended & 2.06$\pm$0.17 & 26.30 & 34.1$\pm$0.16 & 26.81 & 19.4(164) & 11.7(99.5) & 2.17\\
J080009.98+165509.4 & Unresolved & 1.01$\pm$0.08 & 24.34 & 1.63$\pm$0.05 & 24.50 & 16.1(119) & 3.41(25.2) & 0.37\\
J121544.36+452912.7 & Unresolved & 8.50$\pm$0.20 & 25.62 & 9.36$\pm$0.10 & 25.64 & 12.1(102) & 4.32(36.1) & 0.07\\
J143333.02+484227.7 (A) & Extended & 0.75$\pm$0.06 & 24.76 & 0.78$\pm$0.02 & 24.74 & 15.5(133) & 3.31(28.4) & 0.03\\
J143333.02+484227.7 (B) & Extended & 0.33$\pm$0.04 & 25.49 & 0.08$\pm$0.02 & 24.39 & 84.5(727) & 3.22(27.5) & -1.09\\
J162501.98+430931.6 (S) & Unresolved & $<$0.05 & - & 0.14$\pm$0.02 & 24.26 & - & 3.43(29.6) & 0.79\\
J172308.14+524455.5$\dagger$ & Unresolved & 0.72$\pm$0.09 & 25.37 & 1.38$\pm$0.05 & 25.59 & 19.1(157) & 3.34(27.1) & 0.50\\
J173330.80+552030.9 & Unresolved & 1.72$\pm$0.26 & 24.96 & 3.46$\pm$0.61 & 25.31 & 14.7(124) & 4.01 (34.0) & 0.54\\
\enddata
\caption{Column 1: Coordinate names in the form of ``hhmmss.ss$\pm$ddmmss.s" based on SDSS DR16Q; $\dagger$ indicates target identified as star+quasar superposition via SDSS spectrum; if necessary, component is listed in parentheses. Column 2: Morphology designation based on new VLBA observations. Column 3: VLBA 13 cm (S-band) peak flux density $\pm$ 1$\sigma$ error; upper limit reported in the case of a non-detection. Column 4: VLBA 13 cm (S-band) peak luminosity. Column 5: VLBA 4 cm (X-band) peak flux density $\pm$ 1$\sigma$ error. Column 6: VLBA 4 cm (X-band) peak luminosity. Column 7: Angular and projected physical size, in milliarcseconds and pc, at 13cm. Column 8: Angular and projected physical size, in milliarcseconds and pc, at 4cm. Column 9: Quasi-instantaneous spectral index as measured between 2.3 GHz and 8.3 GHz.}
\end{deluxetable*}

\subsection{Compactness and Brightness Temperature} \label{sec:compactbt}

Both compactness and brightness temperature in radio sources are useful indicators of emission from quasars. Though the sample of seven targets presented here are all SDSS-identified quasars, it is nevertheless useful to characterize these parameters for the overall sample. Below, the calculation of both parameters is described. Table \ref{tab:ctb_VLBA} lists the brightness temperature and compactness for each target at each frequency.

Compactness is a measure of how compact a source is \citep[][]{kellerman1981}. It indicates what fraction of the total emission from a source is effectively within the peak of the source. The compactness parameter \textit{C} can be described as:

\begin{center}
\begin{equation} \label{eq:1}
    C = \frac{S_{integrated}}{I_{peak}},
\end{equation}
\end{center}

or the ratio of the integrated flux density and peak intensity of each source. If \textit{C} = 1, then the peak and integrated intensities are very similar, and thus the source is very compact. If \textit{C} $>$ 1, then the source becomes more extended for increasing values of \textit{C}. For quasars, compactness values close to 1 are expected, as the observed collimated jets should be quite compact. However, for objects exhibiting extended emission, larger values of C are expected.

In the current sample, the majority of the targets and their components exhibit compactness parameters very near to one. The full range of compactness for the entire sample is 1.01 to 14.89. There are three components with compactness parameters greater than 5, and as expected these are J011114.41+171328.5 at 2.3 GHz, and Component B in J143333.02+484227.7 at both 2.3 GHz and 8.3 GHz. These are the components that have been identified as extended, and thus significant emission is located outside of the peak of the source. 
 
Brightness temperature is particularly useful as the new VLBA observations of this sub-sample provide milliarcsecond-scale spatial resolution radio continuum images. Sufficiently high resolution radio continuum images can provide a reliable means of identifying AGN based on their brightness temperature. Brightness temperature is defined as:

\begin{equation}
    T_b = \frac{S}{\Omega_{beam}}\frac{c^2}{2k\nu^2},
\end{equation}

where $\nu$ is the observing frequency, $S$ is the integrated flux density, and $\Omega_{beam}$ is the beam solid angle. The brightness temperature limit that separates AGN emission from star formation or otherwise weaker emission is generally around $T_b  = 10^5 K$ \citep[][]{condon1991}. A brightness temperature calculated below that value would indicate that the emission is driven by star formation or a compact starburst, amongst other drivers, while a value above that would indicate emission driven by accretion onto a central SMBH. In this sample of seven, the brightness temperatures range from 0.16 - 18.1 $\times 10^7 K$, significantly above the accretion limit. This is strong evidence that all of the sources in the sample are exhibiting accretion onto an SMBH, and generally excludes other origins, such as SNR.

The brightness temperature and compactness values for the VLBA sample can be found in Table \ref{tab:ctb_VLBA}. Note that in the case of a non-detection (where an upper limit on the integrated flux density has been taken as the RMS value of the observation), no compactness nor brightness temperature has been calculated, and thus those targets have been omitted from the table. 

\begin{deluxetable}{cccc}
\tablenum{5}
\tablecaption{VLBA Properties: Compactness and $T_{B}$\label{tab:ctb_VLBA}}
\tablewidth{0pt}
\tablehead{
\colhead{Source} &
\colhead{Band} & \colhead{C} & \colhead{$T_{b}$}\\
\colhead{[SDSS]} &
\colhead{} & \colhead{} & \colhead{[$10^7$ K]}\\
\colhead{(1)} & \colhead{(2)} & \colhead{(3)} &
\colhead{(4)}
}
\startdata
J011114.41+171328.5 & S & 5.19 & 9.03\\
J011114.41+171328.5 & X & 1.01 & 18.1\\
J080009.98+165509.4 & S & 1.44 & 0.68\\
J080009.98+165509.4 & X & 1.29 & 1.83\\
J121544.36+452912.7 & S & 1.15 & 6.94\\
J121544.36+452912.7 & X & 1.01 & 8.94\\
J143333.02+484227.7 (A) & S & 1.21 & 0.25\\
J143333.02+484227.7 (A) & X & 1.09 & 0.75\\
J143333.02+484227.7 (B) & S & 14.89 & 2.49\\
J143333.02+484227.7 (B) & X & 4.99 & 0.34\\
J162501.98+430931.6 (S) & X & 1.38 & 0.16\\
J172308.14+524455.5$\dagger$ & S & 1.24 & 0.46\\
J172308.14+524455.5$\dagger$ & X & 1.07 & 1.28\\
J173330.80+552030.9 & S & 1.09 & 1.79\\
J173330.80+552030.9 & X & 1.22 & 3.67\\
\enddata
\caption{Note - Column 1: Coordinate names in the form of ``hhmmss.ss$\pm$ddmmss.s" based on SDSS DR16Q; $\dagger$ indicates target identified as star+quasar superposition via SDSS spectrum; if necessary, component is listed in parentheses. Column 2: Observation band. Column 3: Source compactness as calculated from the total flux to peak flux ratio. Column 4: Source brightness temperature as calculated from the total flux.}
\end{deluxetable}

\subsection{Positions and Uncertainty} \label{sec:positionsanduncertainty}

Table \ref{tab:phase_pos_unc} presents the Radio Fundamental Catalog information for each calibrator, the \textsc{IMFIT} model position results for each source and component at both frequencies (non-detections omitted), and the final calculated positional uncertainties for the VLBA sources. It includes all information necessary to calculate the phase positional uncertainties, following the method presented in Section \ref{sec:phasepositionaluncertainties}. Specifically, column [8] in Table \ref{tab:phase_pos_unc} presents the final phase positional uncertainties, in milliarcseconds. They are all below one milliarcsecond, as is expected for VLBA observations with standard phase calibration, and range between 0.21 and 0.96 milliarcseconds.

Images of the phase calibrators themselves, for each target at both frequencies, are presented in Figures \ref{fig:011114_phase}-\ref{fig:173330_phase}. They are shown with the official Radio Fundamental Catalog position marked with a green cross. All of the phase calibrators are unresolved, point-like sources, as expected, with the exception of J1723+5236, the phase calibrator for J172308.14+524455.5, which exhibits faint but clear extension to the southeast at 8.3 GHz. 

\begin{deluxetable*}{cccccccc}
\tablenum{6}
\tablecaption{Positional Uncertainty} \label{tab:phase_pos_unc}
\tablewidth{0pt}
\tablehead{
\colhead{Phase} & \colhead{Position} & \colhead{$\sigma_{dec}$} & \colhead{$B_{maj},B_{min}$} & \colhead{$S_{peak}$} & \colhead{$\sigma_s$} & \colhead{Position} & \colhead{$\sigma_{pos}$}\\
\colhead{Calibrator} & \colhead{Official} & \colhead{mas} & \colhead{[mas,mas]} & \colhead{mJy/bm} & \colhead{mJy} & \colhead{VLBA} & \colhead{mas}\\
\colhead{[1]} & \colhead{[2]} & \colhead{[3]} & \colhead{[4]} & \colhead{[5]} & \colhead{[6]} & \colhead{[7]} & \colhead{[8]}\\
}
\startdata
J0101+1639 (S) & \makecell{01:01:57.719552\\ +16:39:40.95356} & 0.28 & 19, 6 & 82.6 & 8.70 & \makecell{01:01:57.7195550\\ +16:39:40.9531157} & 0.96\\
J0101+1639 (X) & \makecell{01:01:57.719552\\ +16:39:40.95356} & 0.28 & 5, 1 & 66.3 & 5.67 & \makecell{01:01:57.71955121\\ +16:39:40.95324031} & 0.67\\
J0802+1809 (S) & \makecell{08:02:48.031972\\ +18:09:49.24934} & 0.15 & 13, 4 & 317 & 27.1 & \makecell{08:02:48.03197397\\ +18:09:49.24922732} & 0.43\\
J0802+1809 (X) & \makecell{08:02:48.031972\\ +18:09:49.24934} & 0.15 & 3, 1 & 385 & 32.7 & \makecell{08:02:48.03196765\\ +18:09:49.24934747} & 0.23\\
J1223+4611 (S) & \makecell{12:23:39.336669\\ +46:11:18.60262} & 0.15 & 8, 5 & 157 & 8.67 & \makecell{12:23:39.33664902\\ +46:11:18.60276660} & 0.38\\
J1223+4611 (X) & \makecell{12:23:39.336669\\ +46:11:18.60262} & 0.15 & 3, 2 & 132 & 8.38 & \makecell{12:23:39.33665829\\ +46:11:18.60271371} & 0.30\\
J1439+4958 (S) & \makecell{14:39:46.976235\\ +49:58:05.45577} & 0.13 & 10, 8 & 148 & 10.4 & \makecell{14:39:46.97622359\\ +49:58:05.45581411} & 0.38\\
J1439+4958 (X) & \makecell{14:39:46.976235\\ +49:58:05.45577} & 0.13 & 3, 1 & 317 & 22.7 & \makecell{14:39:46.976225719\\ +49:58:05.455792016} & 0.21\\
J1625+4347 (S) & \makecell{16:25:53.307166\\ +43:47:13.84170} & 0.28 & 8, 4 & 84.1 & 5.29 & \makecell{16:25:53.30714219\\ +43:47:13.84123938} & 0.81\\
J1625+4347 (X) & \makecell{16:25:53.307166\\ +43:47:13.84170} & 0.28 & 3, 1 & 81.9 & 5.79 & \makecell{16:25:53.307142162\\ +43:47:13.841248502} & 0.79\\
J1723+5236 (S) & \makecell{17:23:39.746459\\ +52:36:48.39565} & 0.15 & 13, 6 & 147 & 9.04 & \makecell{17:23:39.7466226\\ +52:36:48.3974905} & 0.46\\
J1723+5236 (X) & \makecell{17:23:39.746459\\ +52:36:48.39565} & 0.15 & 3, 2 & 134 & 10.5 & \makecell{17:23:39.74648855\\ +52:36:48.39549679} & 0.35\\
J1727+5510 (S) & \makecell{17:27:23.469314\\ +55:10:53.53543} & 0.13 & 8, 3 & 98.1 & 5.92 & \makecell{17:27:23.4693144\\ +55:10:53.5354385} & 0.24\\
J1727+5510 (X) & \makecell{17:27:23.469314\\ +55:10:53.53543} & 0.13 & 2, 1 & 236 & 20.2 & \makecell{17:27:23.46931612\\ +55:10:53.53545857} & 0.21\\
\enddata
\caption{Note - Column 1: Coordinate names of the VLBA phase calibrators from the Radio Fundamental Catalog 2024c. Band denoted in parentheses. Column 2: Official phase calibrator position from the RFC. Column 3: Error in official declination from the RFC. Column 4: Restoring beam of the VLBA phase calibrator imaging. Column 5: VLBA phase calibrator peak flux density in milliJanskys per beam. Column 6: 1-$\sigma$ RMS measurement of the VLBA phase calibrator image. Column 7: VLBA phase calibrator position as modeled with CASA's \textsc{IMFIT}. Column 8: Positional uncertainty in the VLBA sources; the error in the official position of the phase calibrator, the switching angle error, and the Gaussian fit error, added in quadrature.}
\end{deluxetable*}

\section{Spectral Analysis} \label{sec:specanalysis}

\begin{figure*}
\centering
\begin{minipage}{0.32\textwidth}
    \includegraphics[width=\linewidth]{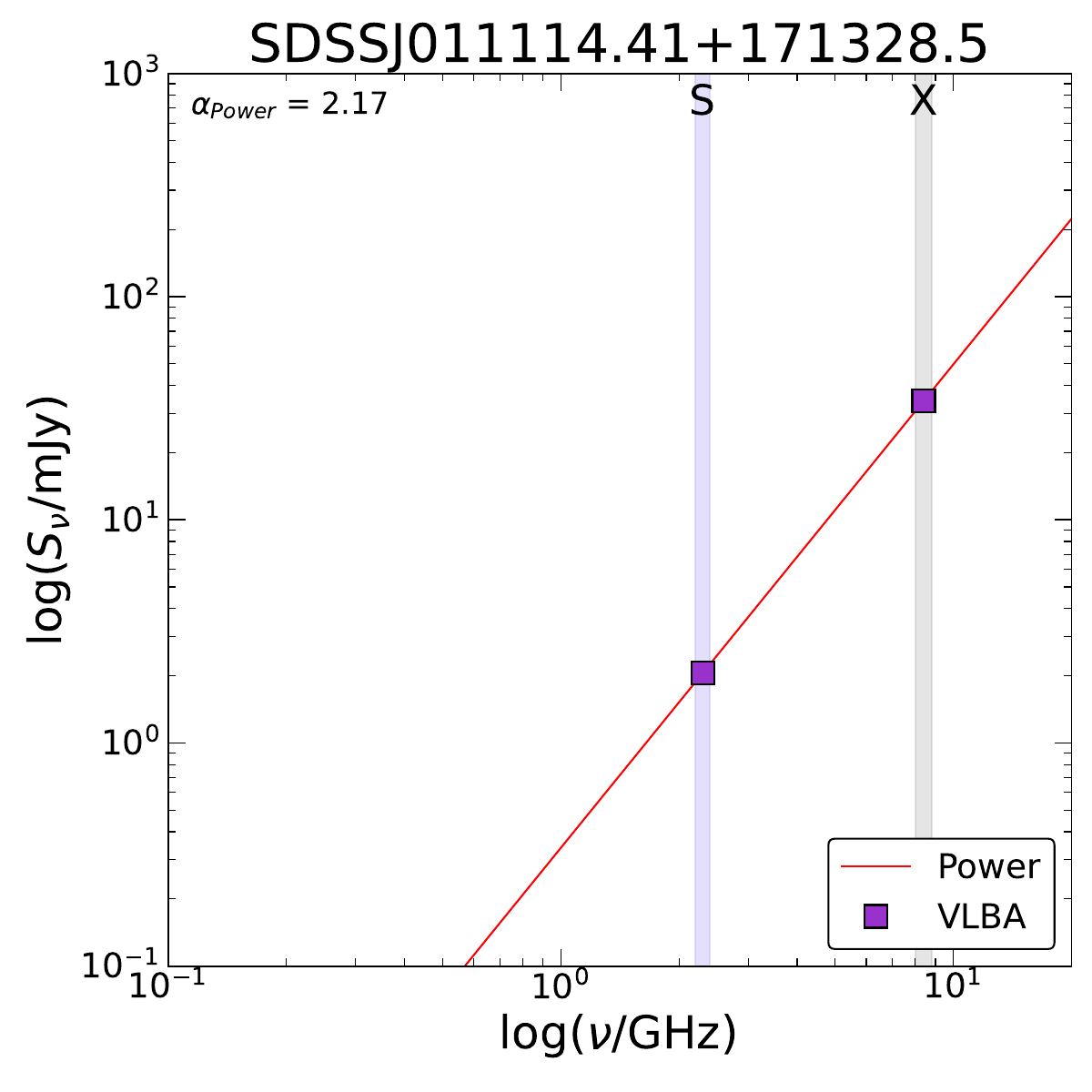}
\end{minipage}
\begin{minipage}{0.32\textwidth}
    \includegraphics[width=\linewidth]{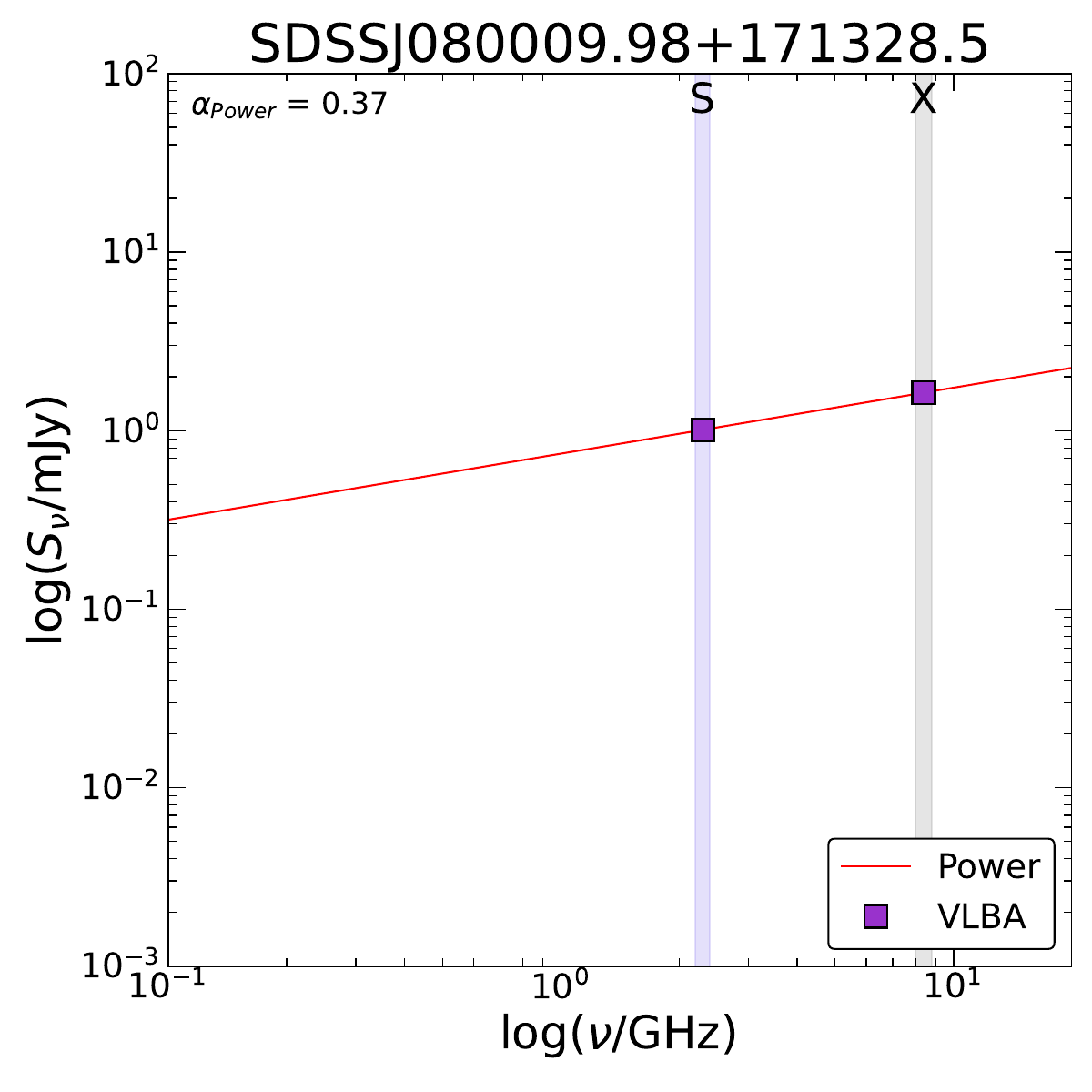}
\end{minipage}

\begin{minipage}{0.32\textwidth}
    \includegraphics[width=\linewidth]{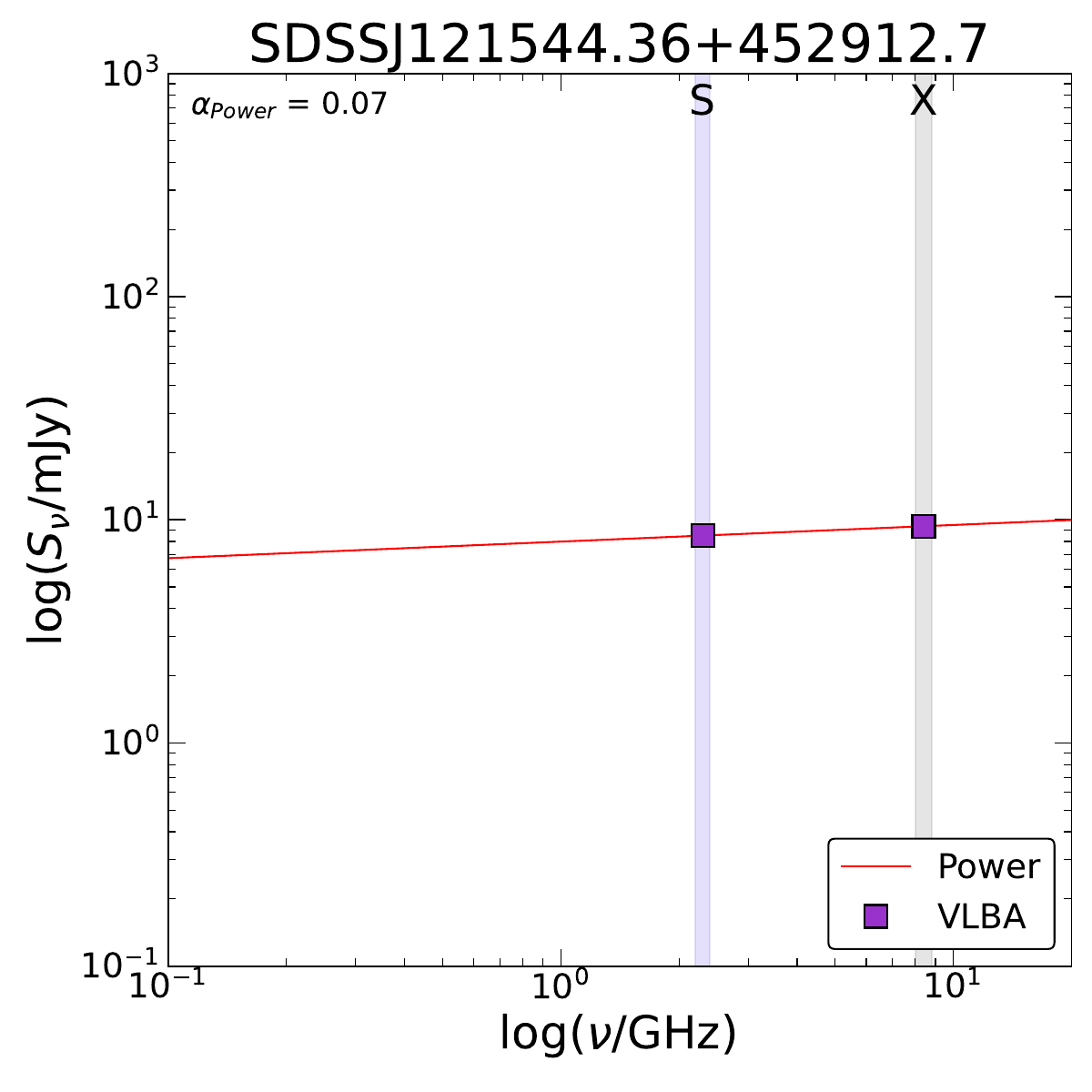}
\end{minipage}
\begin{minipage}{0.32\textwidth}
    \includegraphics[width=\linewidth]{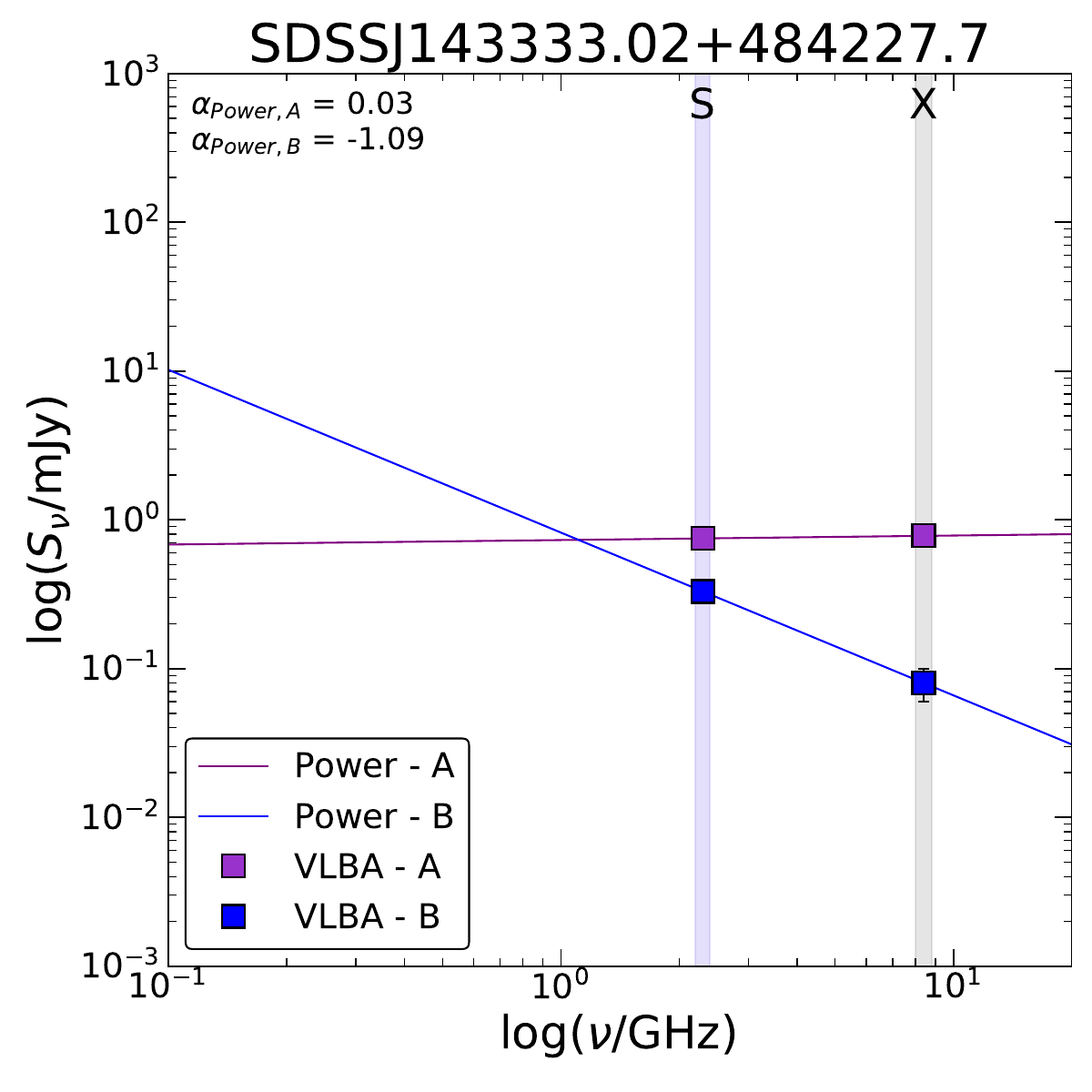}
\end{minipage}
\begin{minipage}{0.32\textwidth}
    \includegraphics[width=\linewidth]{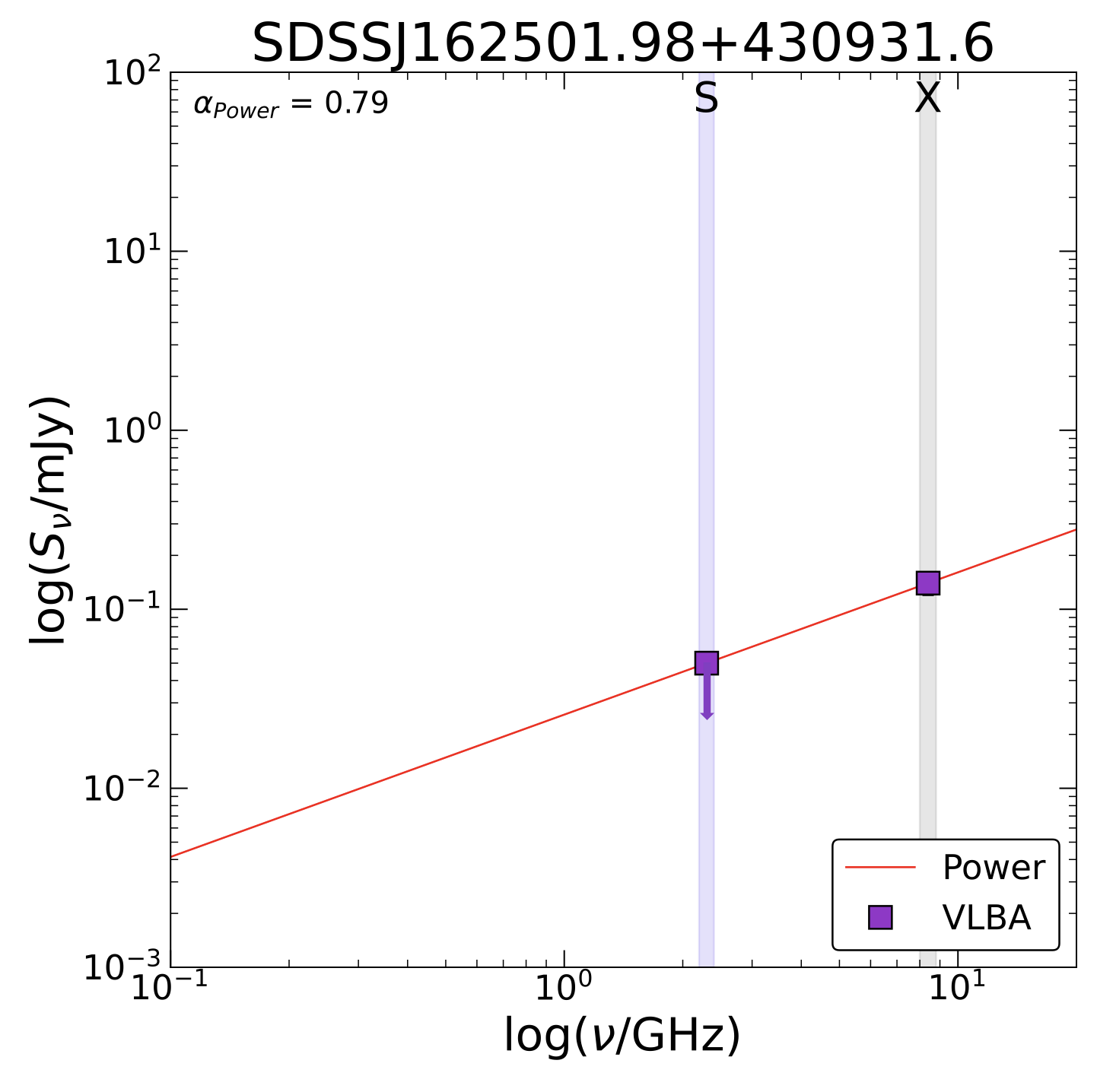}
\end{minipage}

\begin{minipage}{0.32\textwidth}
    \includegraphics[width=\linewidth]{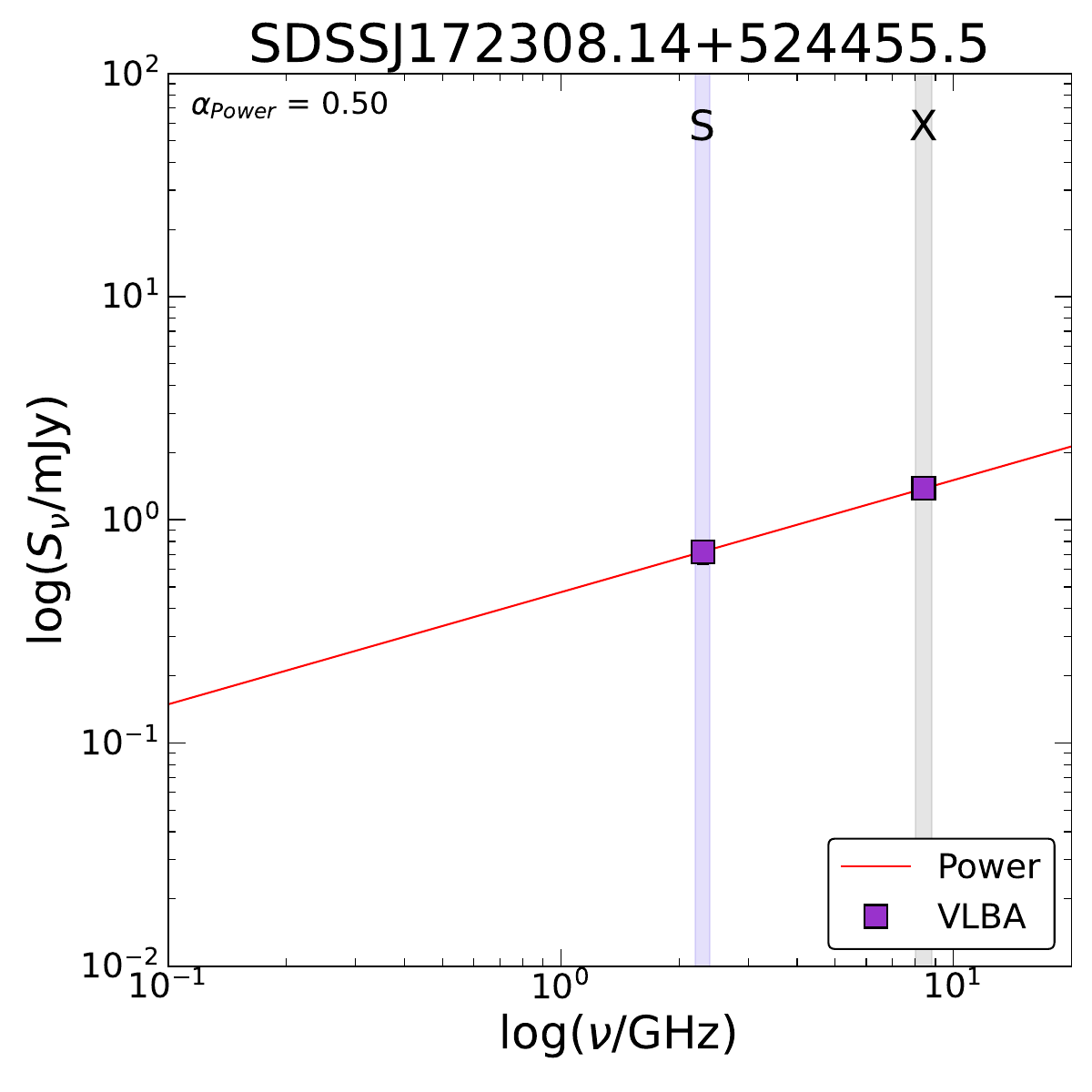}
\end{minipage}
\begin{minipage}{0.32\textwidth}
    \includegraphics[width=\linewidth]{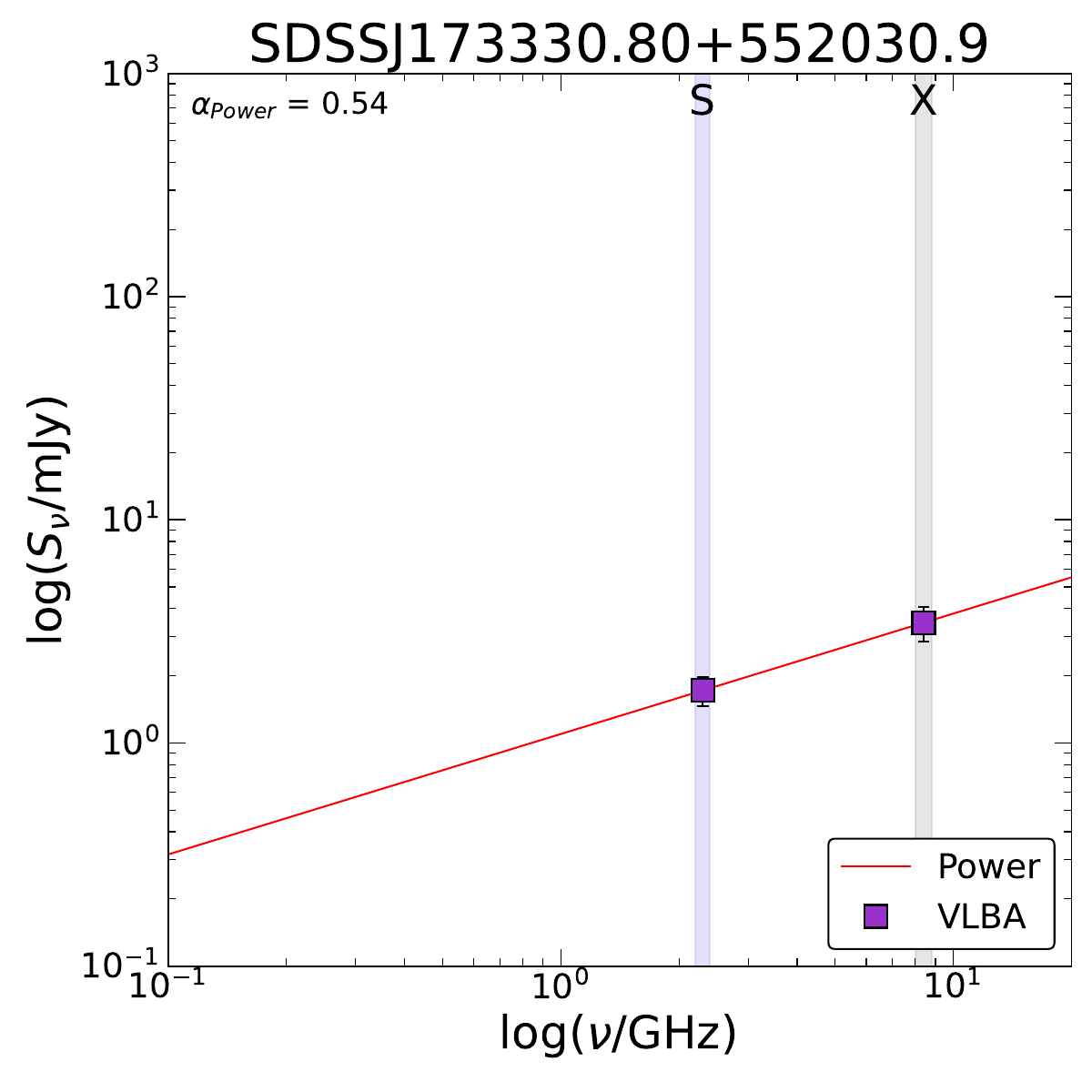}
\end{minipage}

\caption{Radio spectra for each of the seven new VLBA observations showing the multi-frequency quasi-simultaneous data. In most panels, the new VLBA observations are in purple squares, and are modeled with a standard power law (red line), and the resultant spectral indices are included in the upper left hand corner. SDSSJ143333.02+484227.7 includes the spectra for both components A and B in purple and blue, respectively. The power law models are the same color as their respective points, and both spectral indices are labeled and listed in the upper left hand corner. For SDSSJ1625081.98+430931.6, the 2.3 GHz observations were a non-detection, and thus the flux value is an upper limit measured from the noise of the image. However, the true flux could be significantly less than the upper limit, which would impact the spectral slope. Error bars representing peak flux density errors have been included but in most cases are too small to be visible.}
\label{fig:VLBA_spectra}
\end{figure*}

Radio spectra of each target showing the new, quasi-simultaneous VLBA observations  are presented in Figure \ref{fig:VLBA_spectra}. In most panels, the new VLBA observations are illustrated by purple squares, representing the peak flux measurements as modeled with CASA's \textsc{IMFIT}. All have been modeled with a standard power law, drawn in red, following the relation $S \propto \nu^{\alpha}$, and the resultant spectral indices are included in the upper left hand corner. 

In the case of SDSSJ143333.02+484227.7, spectra for both components A and B are included, denoted by purple and blue squares, respectively. The power law models are drawn in the same color as their respective points, and both spectral indices are labeled and listed in the upper left hand corner. In the case of SDSSJ1625081.98+430931.6, the 2.3 GHz observations did not reveal a detection, and thus the flux value is an upper limit, measured as the noise of the image. However, the true flux could be significantly less than the upper limit, which would likely increase the spectral index, making the current measurement a lower limit on the spectral index. 

The spectral indices display a similar diversity to the morphologies of the targets on sub-milliarcsecond scales. All have been fitted with a standard power law, where a spectral index of $|\alpha| <$ 0.5 is considered to be a flat spectrum source, while optically-thin synchrotron emission is expected to produce a spectral index of $\alpha < -0.7$. The spectral indices of the VLBA components range from -1.09 to 2.17. One component exhibits a steep spectral index of -1.09, likely due to jet activity. Four of the components exhibit flat spectral indices between $-0.5 < \alpha < 0.5$, which is expected for an AGN. Finally, three components exhibit inverted (more steeply positive) spectral indices, including J0111+1713, which has a spectral index of $\alpha = 2.17$. The inverted spectral index is likely driven by absorption, whether extrinsic (free-free absorption, FFA) or intrinsic (synchrotron self-absorption, SSA). In idealized sources, both SSA and FFA can lead to spectral indices as high as 2.5, in the optically-thick regime. 

\section{Radio-Optical Offsets} \label{sec:roo}

Given the resolution of the VLBA, and the importance of milliarcsecond-scale radio VLBI observations of quasars to the International Celestial Reference Frame \citep[][]{charlot2020}, it is interesting to compare the optical positions identified by \textit{Gaia} with the positions of the new VLBA detections, as modeled using \textsc{IMFIT}. This comparison is done to determine the significance of any radio-optical position offsets. The final VLBA positional uncertainties are between 0.21 and 0.96 milliarcseconds.

\begin{figure}
    \centering
     \includegraphics[width=8cm]{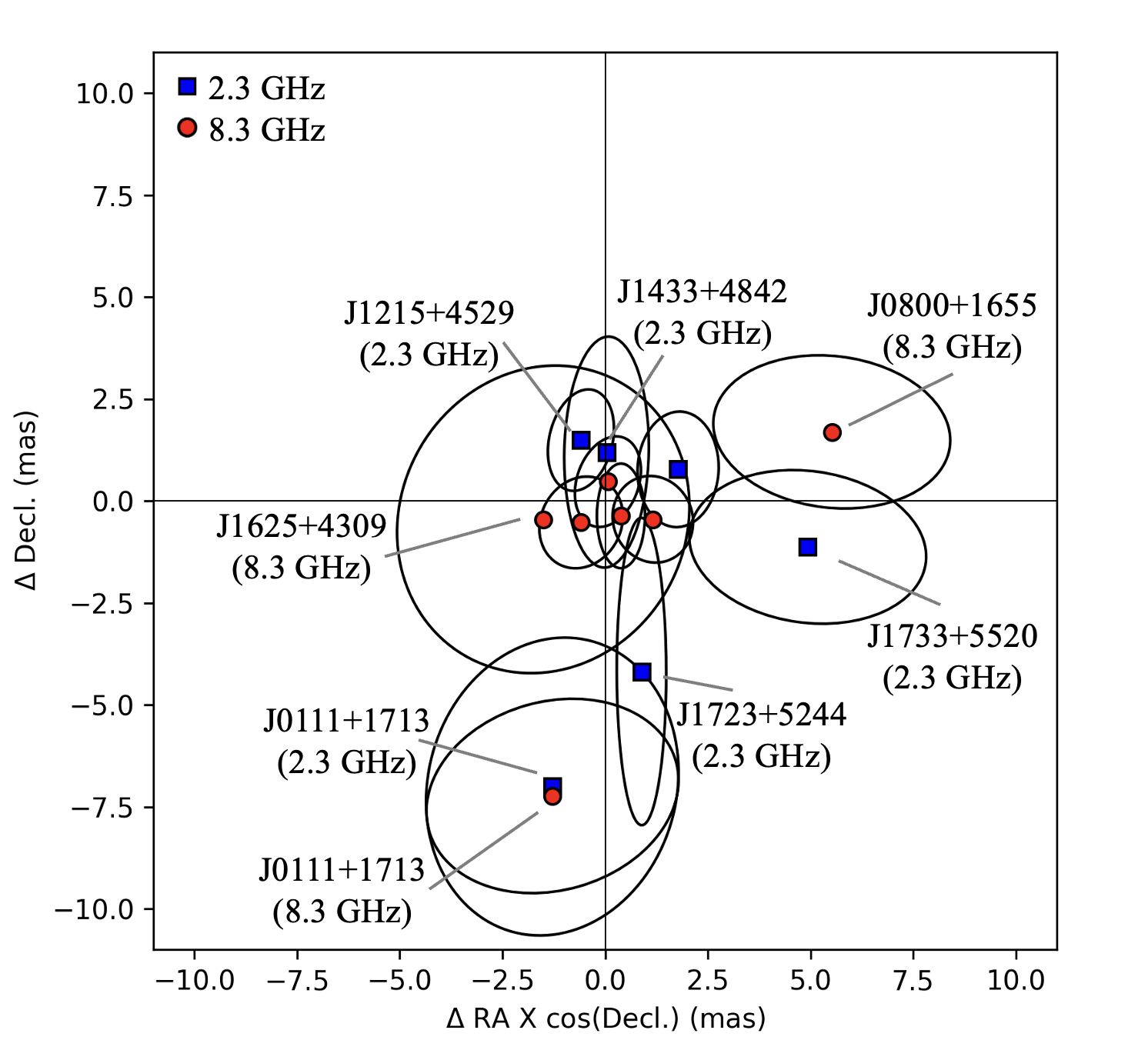}
        \vspace*{1mm}
          \caption{Distribution of positional offsets between the \textit{Gaia} DR3 optical position and the VLBA radio source positions for the seven VaDAR-selected VLBA sources. Blue squares represent 2.3 GHz observations, while red circles represent 8.3 GHz observations. The error ellipses enclose a 95\% probability, taking into account the VLBA position (statistical) error, the error in the phase calibrator position, and the \textit{Gaia} position error, as well as the covariance matrices for each.}
    \label{fig:roo}
\end{figure}

Figure \ref{fig:roo} displays the distribution of the \textit{Gaia}-VLBA positional offsets for the seven targets in this sub-sample, shown as $\Delta$\,decl.\ versus $\Delta$\,R.A.\,$\cdot$\,cos(decl.), in milliarcseconds. Blue squares denote offsets at 2.3 GHz, while red circles denote offsets at 8.3 GHz. 

The errors in the offset are calculated as the error in the \textsc{IMFIT} statistical error, the systematic error arising from the position uncertainty of the phase calibrator, and the \textit{Gaia} position error. Covariance matrices including the correlation information for each source of error were added to form a total covariance matrix, from which the properties of each ellipse were calculated. The ellipses enclose a 95\% probability.

The plot shows 13 points corresponding to all seven targets. Only the ``core'' component of J143333.02+484227.7, and since J162501.98+430931.6 was not detected at 2.3 GHz only its 8.3~GHz position, is shown.

\begin{deluxetable*}{cc|c}
\tablenum{7}
\tablecaption{Radio-Optical Offsets} \label{tab:roo}
\tablewidth{0pt}
\tablehead{
\colhead{Target} & \colhead{Position} & \colhead{X}\\
\colhead{[SDSS]} & \colhead{\makecell{\textit{(Gaia)}\\ (VLBA - S)\\ (VLBA - X)}} & \colhead{}\\
\colhead{[1]} & \colhead{[2]} & \colhead{[3]}\\
}
\startdata
J011114.41+171328.5 & 01:01:14.415241$\pm$0.627445, +17:13:28.589276$\pm$0.466315 & - \\
 & 01:11:14.415331$\pm$0.021, +17:13:28.596282$\pm$0.582 & 89.7\\
 & 01:11:14.4153308$\pm$0.002, +17:13:28.596513$\pm$0.135 & 222\\
\hline
J080009.98+165509.4 & 08:00:09.969485$\pm$0.589728, +16:55:09.630438$\pm$0.381477 & - \\
  & 08:00:09.969141$\pm$0.015, +16:55:09.631569$\pm$0.731 & 77.6\\
 & 08:00:09.96910150$\pm$0.001, +16:55:09.62874948$\pm$0.040 & 126\\
\hline
J121544.36+452912.7 & 12:15:44.365233$\pm$0.163693, +45:29:12.799523$\pm$0.225414 & - \\
 & 12:15:44.3652894$\pm$0.004, +45:29:12.7980366$\pm$0.117 & 78.6\\
 & 12:15:44.36522678$\pm$0.001, +45:29:12.79905461$\pm$0.022 & 4.3\\
\hline
J143333.02+484227.7 & 14:33:33.031602$\pm$207057, +48:42:27.775180$\pm$227361 & - \\
 & 14:33:33.031599$\pm$0.067, +48:42:27.773987$\pm$0.550 & 4.3\\
 & 14:33:33.03166187$\pm$0.013, +48:42:27.77570867$\pm$0.032 & 21.1\\
\hline
J162501.98+430931.6 & 16:25:01.990430$\pm$0.726625, +43:09:31.394014$\pm$0.742520 & - \\
 & -, - & - \\
 & 16:25:01.9905661$\pm$0.011, +43:09:31.3944695$\pm$0.203 & 8.2\\
\hline
J172308.14+524455.5$\dagger$ & 17:23:08.089233$\pm$0.091141, +52:44:56.064262$\pm$0.089132 & - \\
 & 17:23:08.089136$\pm$0.028, +52:44:56.068448$\pm$0.755 & 263\\
 & 17:23:08.08919149$\pm$0.001, +52:44:56.06463083$\pm$0.216 & 51.0\\
\hline
J173330.80+552030.9 & 17:33:30.843820$\pm$0.198159, +55:20:30.852561$\pm$0.218067 & - \\
 & 17:33:30.843611$\pm$0.037, +55:20:30.851790$\pm$0.190 & 245\\
 & 17:33:30.8436851$\pm$0.002, +55:20:30.8530128$\pm$0.016 & 113\\
\enddata
\caption{Column 1: Coordinate names in the form of ``hhmmss.ss$\pm$ddmmss.s" based on SDSS DR16Q; $\dagger$ indicates source with star+quasar superposition. Column 2: \textit{Gaia} EDR3 position $\pm$ errors in milliarcseconds, VLBA position $\pm$ error in milliarcseconds at S-band, and VLBA position $\pm$ error in milliarcseconds at X-band. Both VLBA positions are modeled by CASA's \textsc{IMFIT}. Column 3: The normalized separation X between the \textit{Gaia} and VLBA positions, calculated using Equation 4 in \cite{mignard2016}.}
\end{deluxetable*}

Table \ref{tab:roo} presents both the \textit{Gaia} and VLBA positions and their errors, and the normalized separation $X$, calculated following the method presented in \cite{mignard2016}:

\begin{equation}
     X^{2} = \begin{bmatrix} X_{\alpha} & X_{\delta} \\ \end{bmatrix}\begin{bmatrix} 1 & C \\ C & 1 \\ \end{bmatrix}^{-1} \begin{bmatrix} X_{\alpha} \\ X_{\delta}\end{bmatrix},
\end{equation}

\noindent where $X_{\alpha}$ and $X_{\delta}$ are the uncertainty-normalized coordinate differences and $C$ is the correlation coefficient between R.A.\ ($\alpha$) and decl.\ ($\delta$). Correlation values were drawn from the Radio Fundamental Catalog \citep{2025ApJS..276...38P} and the \textit{Gaia} DR3. The normalized separation quantity takes into account the errors in all optical and radio positions, as well as the correlations between the errors.

The significance of the normalized separation measured in each source is understood as follows. Assuming no intrinsic offset and accurate positional uncertainties, \textit{X} should follow the Rayleigh distribution $\mathcal{R}(1)$. The survival function, $1 - \mathrm{CDF}_{\mathcal{R}(1)}(X)$, gives the probability of a value of $X$ drawn from $\mathcal{R}(1)$ being larger than the observed $X$. 

With a sample size of seven, we calculate that an \textit{X} limit of $X > 3.6$ ensures that the chance of having a false positive in at least one of these seven targets is $< 1\%$. Thus, any offset in this sample with $X > 3.5$ is considered to be significant. 

\subsubsection{Origins of VLBA-\textit{Gaia} Offsets} 

One interpretation of a significant radio-optical offset, along with the detection of a single compact radio core at milliarcsecond scales, is that the target is a multi-AGN candidate \citep[][]{chen2023,cigan2024,pfeifle2023inprep}. In this case, the radio-optical offset can be considered in the context of the radio spectral index: if a target displays a flat-spectrum core with significant radio-optical offset, it is considered to be a multi-AGN candidate. 

The physical explanation is that one of the cores is radio-loud, but without existing optical emission, perhaps due to obscuration. The secondary core is radio-quiet \citep[not unexpected, as only $\sim 10$\% of all single AGN exhibit associated radio-loud emission;][]{osterbrockbook}, but is more dominant in the optical regime by \textit{Gaia}. In this case, the observed excess astrometric variability might be driven by single AGN variability (in the optically-dominant core), or by the centroid shifting between the AGN and the host galaxy, if the AGN is offset due to a merger. 

However, these systems must remain candidates, as there are other physical explanations. It may be possible to observe a single core driving both optical and radio jets exhibiting hotspots with significant separation. These systems could be identified via a steep ($\alpha \leq -1$) radio spectral index. Another explanation is the frequency dependence of the core itself. The position of the core has been shown to be a function of the observational frequency \citep[][]{sokolovsky2011}, and a frequency-dependent ``core shift'' has been observed, induced by flaring \citep[][]{plavin2019}.

Of the seven VaDAR targets, four display a flat spectrum generally associated with an AGN, which includes the core of J143333.02+484227.7. One component, the jetted emission in J143333.02+484227.7, exhibits a steep spectral index indicative of jet activity. The remaining three show positive slopes, likely driven by absorption mechanisms in the core. Using the normalized separation $X$, all seven targets also show significant radio-optical offset. Thus, based on the sub-mas scale spectral indices and the positional offsets alone, all seven targets could be classified as multi-AGN candidates. 

This classification needs to be considered alongside other results. Target J172308.14+524455 has been identified via the SDSS spectrum as a star+quasar superposition. We note that the radio emission likely traces the quasar, and not the superimposed star. However, though the significant radio-optical offset deserves some consideration, the observed astrometric variability of this target is likely being driven by the star+quasar superposition. Both J121544.36+452912.7 and J143333.02+484227.7 show signs of jet activity that could similarly explain the excess astrometric noise. 

Of the remaining four sources, all four show spectral indices consistent with quasars, and unresolved emission at milliarcsecond scales. Though this supports the candidate multi-AGN interpretation, it is possible to place a lower limit on the predicted angular separation, assuming a typical fractional RMS of 10\% \citep[][]{hwang_initial}. Lower limits on angular separations for these sources were calculated in \cite{schwartzman2024}, and are on the order of tens of milliarcseconds. Thus, if a secondary radio core does exist and emit at the scales accessible to the VLBA, it should be detected in our observations. We note that it remains possible for the secondary core to be radio-silent. 

It is possible to further understand if the radio-optical offsets are driven by multiple cores by searching for evidence of extended structure indicating a recent merger in the optical images from the Dark Energy Camera Legacy Survey \citep[DECaLs;][]{dey2019}. The DECaLs images are presented in Appendix \ref{sec:decals}, Figure \ref{fig:decals}. Each image has a resolution of approximately 1'', and is shown with the VLBA 8.3 GHz positions marked with red squares, and the \textit{Gaia} optical positions marked with blue triangles.

\begin{deluxetable*}{ccccc}
\tablenum{8}
\tablecaption{DECaLS Gaussian Models\label{tab:optical}}
\tablewidth{0pt}
\tablehead{
\colhead{Source} & \colhead{Gaussian Source Model} & \colhead{$e_{source}$} & \colhead{Gaussian Star Model} & \colhead{$e_{star}$}\\
\colhead{[SDSS]} & \colhead{[asec, asec, $^\circ$]} & \colhead{} & \colhead{[asec, asec, $^\circ$]} & \colhead{}\\
\colhead{[1]} & \colhead{[2]} & \colhead{[3]} & \colhead{[4]} & \colhead{[5]}
}
\startdata
J011114.41+171328.5 & 1.37, 1.29, 26 & 0.06 & 1.43, 1.37, 16 & 0.04\\
J080009.98+165509.4 & 1.67, 1.32, 129 & 0.21 & 1.29, 1.24, 29 & 0.03\\
J121544.36+452912.7 & 1.54, 1.47, 128 & 0.05 & 1.53, 1.45, 111 & 0.05\\
J143333.02+484227.7 & 1.31, 1.08, 52 & 0.18 & 1.21, 1.15, 61 & 0.04\\
J162501.98+430931.6 & 2.01, 1.92, 116 & 0.04 & 2.30, 2.26, 92 & 0.02\\
J172308.14+524455.5$\dagger$ & 2.36, 1.82, 156 & 0.23 & 1.30, 1.26, 65 & 0.03\\
J173330.80+552030.9 & 1.95, 1.65, 174 & 0.15 & 1.71, 1.67, 178 & 0.02\\
\enddata
\caption{Note - Column 1: Coordinate names in the form of ``hhmmss.ss$\pm$ddmmss.s" based on SDSS DR16Q; $\dagger$ indicates target identified as star+quasar superposition via SDSS spectrum. Column 2: DECaLS \textit{r}-band Gaussian target model: major axis, minor axis, position angle. Column 3: DECaLS \textit{r}-band target ellipticity. Column 4: DECaLS  \textit{r}-band Gaussian star model: major axis, minor axis, position angle. Column 5: DECaLS \textit{r}-band star ellipticity.}
\end{deluxetable*}

A careful examination of the DeCaLs \textit{ugz} image for each target exhibiting significant radio-optical offset has revealed that four of the seven optical hosts are slightly elliptical. The optical hosts in the \textit{r}-band were modeled with a single Gaussian using CASA's IMFIT. The results, including ellipticity, are presented in Table \ref{tab:optical}. As a control, a star nearby each target was also modeled, and the results are included in Table \ref{tab:optical}. None of the stars show significant ellipticity. 

The four targets with optical ellipticity and significant radio-optical offset are J080009.98+165509.4, J172308.14+524455.5, J173330.80+552030.9, and J143333.02+484227.7. For all three targets, the ellipticity might be evidence for extended structure that indicates a recent merger. Overall, however, further optical observations at significantly higher resolutions are required to confirm the presence of any sub-milliarcsecond scale optical jets, or higher sensitivity observations to identify any faint optical jets at the sub-arcsecond scales accessible in the DeCaLs imaging. Even the best ground-based AO observations are still unable to match the VLBA resolutions.

\subsubsection{Other \textit{Gaia} Parameters}

\textit{Gaia}'s DR3 includes several other parameters that can indicate multiplicity in sources \citep[][]{makarov2022}. These include Renormalized Unit Weight Error (RUWE), $\texttt{astrometric\_n\_bad\_obs\_al}$, $\texttt{ipd\_frac\_multi\_peak}$, and $\texttt{phot\_bp\_rp\_excess\_factor}$. In this section, we explore how these parameters might indicate source multiplicity, and what impacts they have on our analysis. The parameter values for the VLBA sample are included in Table \ref{tab:gaia_params}.

\begin{deluxetable*}{ccccc}
\tablenum{9}
\tablecaption{\textit{Gaia} Source Multiplicity Parameters\label{tab:gaia_params}}
\tablewidth{0pt}
\tablehead{
\colhead{Source} &
\colhead{RUWE} & \colhead{BP/RP Excess} & \colhead{Bad Obs AL} & \colhead{IPD Multi}\\
\colhead{[1]} & \colhead{[2]} & \colhead{[3]} & \colhead{[4]} & \colhead{[5]}
}
\startdata
J011114.41+171328.5 & \multicolumn{1}{>{\columncolor{lightgray}}c}{2.35} & 1.41 & 2 & 0\\
J080009.98+165509.4 & \multicolumn{1}{>{\columncolor{lightgray}}c}{4.18} & \multicolumn{1}{>{\columncolor{lightgray}}c}{1.89} & \multicolumn{1}{>{\columncolor{lightgray}}c}{20} & \multicolumn{1}{>{\columncolor{lightgray}}c}{23}\\
J121544.36+452912.7 & 1.20 & 1.19 & 1 & 0\\
J143333.02+484227.7 & \multicolumn{1}{>{\columncolor{lightgray}}c}{1.48} & \multicolumn{1}{>{\columncolor{lightgray}}c}{1.49} & 3 & 1\\
J162501.98+430931.6 & \multicolumn{1}{>{\columncolor{lightgray}}c}{3.64} & \multicolumn{1}{>{\columncolor{lightgray}}c}{1.83} & 1 & 5\\
J172308.14+524455.5$\dagger$ & 1.20 & \multicolumn{1}{>{\columncolor{lightgray}}c}{1.81} & \multicolumn{1}{>{\columncolor{lightgray}}c}{10} & \multicolumn{1}{>{\columncolor{lightgray}}c}{16}\\
J173330.80+552030.9 & \multicolumn{1}{>{\columncolor{lightgray}}c}{1.62} & \multicolumn{1}{>{\columncolor{lightgray}}c}{1.54} & 1 & \multicolumn{1}{>{\columncolor{lightgray}}c}{20}\\
\hline
Target Mean & 2.239(0.459) & 1.594(0.098) & 5.429(2.716) & 9.286(3.803)\\
Control Mean & 1.014(0.005) & 1.233(0.010) & 2.057(0.262) & 0.038(0.026)\\
\textit{p}-value & $< 0.001$ & $< 0.001$ & 0.171 & $< 0.001$\\
\enddata
\caption{Note - Column 1: Coordinate names in the form of ``hhmmss.ss$\pm$ddmmss.s" based on SDSS DR16Q; $\dagger$ indicates target identified as star+quasar superposition via SDSS spectrum. Column 2: RUWE value. Column 3: BP/RP excess value. Column 4: \texttt{astrometric\_n\_bad\_obs\_al} value. Column 5: \texttt{ipd\_frac\_multi\_peak} value. Cells highlighted in grey indicate a value for that parameter that exceeds the range for a single-peak observation. This could be driven by source multiplicity. To compare the target and control samples, the rows below each source list the mean and standard error in the mean for each sample for each \textit{Gaia} parameter. Additionally, \textit{p}-values calculated using the Anderson-Darling test are listed.}
\end{deluxetable*}

The RUWE parameter (see Column 2 in Table \ref{tab:gaia_params}) is a dimensionless quantity that defines how well the astrometric observations of a \textit{Gaia} source match the typical single-star model. A value near 1.0 is expected for a good fit; values $> 1.4$ indicate that the astrometric behavior of the source deviates from the single-star model, possibly due to source multiplicity \citep[][]{lindegren2021}. 

The $\texttt{phot\_bp\_rp\_excess\_factor}$ (see Column 3 in Table \ref{tab:gaia_params}) is calculated as the ratio of the combined flux from the \textit{b} and \textit{r}-bands to the flux in the \textit{g}-band. For a well-behaved source, this factor should fall within a narrow, color-dependent range with values typically around $\sim 1.0-1.3$. Values significantly higher than this expected range can indicate excess flux in the \textit{b} and \textit{r}-bands, which may be a result of extended emission or source multiplicity, but can also be driven by variability or extended host galaxies \citep[][]{riello2021,gaiaDR3documentation}.

The $\texttt{astrometric\_n\_bad\_obs\_al}$ parameter (see Column 4 in Table \ref{tab:gaia_params}) records the number of along-scan (AL) observations that were flagged as problematic or rejected during the astrometric data reduction. A higher count indicates that many observations did not conform well to the single-star model, and might be affected by multiplicity or blending \citep[][]{gaiaDR3documentation}.

Finally, the $\texttt{ipd\_frac\_multi\_peak}$ parameter (see Column 5 in Table \ref{tab:gaia_params}) reports the percentage of successful Image Parameter Determination (IPD) windows with more than one peak \citep[][]{gaiaDR3documentation}. It reflects how often source multiplicity (visual or real) is seen in a successful IPD window. Higher percentages indicate that a larger fraction of windows show signs of source multiplicity \citep[multiple peaks;][]{gaiaDR3documentation}. 

Taken together with the significant excess astrometric noise in each source, these four parameters can further indicate source multiplicity. In all cases, we note that there remain other physical explanations for the measured \textit{Gaia} parameters. These include crowded fields, extended host galaxies, photometric variability, calibration and modeling issues, low S/N objects, color-dependent effects, and scanning geometry issues.

Table \ref{tab:gaia_params} also gives a visual representation of significant values of the four parameters. Cells highlighted in grey have values exceeding the range expected for a normal single-peak observation. This could be driven by source multiplicity. Both RUWE and BP/RP excess are significant across the majority of the sample. Parameters $\texttt{astrometric\_n\_bad\_obs\_al}$ and $\texttt{ipd\_frac\_multi\_peak}$ are most significant in J080009.98+165509.4. 

J121544.36+452912.7 shows no noteworthy value across all four \textit{Gaia} parameters. Similarly, J011114.41+171328.5 shows only a noteworthy value of RUWE. All four parameters in Table \ref{tab:gaia_params} are included in the analyis of each individual target presented in Section \ref{sec:individualtargets}.

In order to better understand these results, a control sample was generated. Similar to the pilot sample presented in \cite{schwartzman2024}, the control sample was generated from a crossmatch of the SDSS DR16Q and \textit{Gaia} DE3 catalogs, to within 1.5''. The sample \textit{Gaia} \textit{G} magnitude ($G < 20$) and redshift ($z > 20$) limits were also applied. As the target sample was selected to exhibit high $\texttt{astrometric\_excess\_noise\_significance}$ ($AENS > 5$), for the control sample, an $AENS < 4$ limit was applied. This generates a control sample of sources that do not display significant astrometric noise. The control sample was also spatially matched to the VLBA target sample, so as to ensure that the controls covered the same portion of the sky. This minimizes systematics driven by \textit{Gaia}'s scanning law. For the seven VLBA targets, this results in a control sample of about 50 sources. 

Table \ref{tab:gaia_params} also shows the mean and standard error in the mean for the control sample and the target sample, for each of the four \textit{Gaia} parameters of interest. The two samples were compared using an Anderson-Darling test, and the \textit{p}-values are also included in Table \ref{tab:gaia_params}. Three of the parameters have \textit{p}-values $< 0.001$, rejecting the null hypothesis, and indicating that the control and target samples vary by more than the parameter of interest. 

It is also possible that \textit{Gaia}'s astrometric solution for each target has been impacted by optical emission from host galaxies. As described in \cite{schwartzman2024}, the original sample (and thus the sub-sample further explored in this paper) was carefully selected so as to have redshifts $>$ 0.5, drastically lowering the possibility of spurious astrometric excess noise detections due to the presence of extended structures in the host galaxies \citep[][]{hwang2020varstrometry}. A similar pattern has been identified in a systematic study by \cite{makarov2019}, in which the fraction of sources exhibiting significant VLBA-\textit{Gaia} offsets is found to be higher at redshifts $z < 0.5$. This is again attributed to the impact of extended structures of host galaxies \citep[][]{hwang_initial,chen2023}. 

Though the $z > 0.5$ cut placed on the sample should limit the contributions from host galaxies, the DECaLS images were further investigated for significant host galaxy contribution (e.g., supernovae or star formation), and none were found. This indicates that the \textit{Gaia} astrometric noise solutions were not contaminated by host galaxy features and are primarily based on the emission from the quasar(s).

\section{Discussion} \label{sec:disc_vlba}

The full analysis of the radio and multiwavelength parameters of the subset of seven astrometrically-variable quasars chosen for VLBA follow-up has revealed a variety of interesting trends in the radio observations. Here, we present a discussion of each of the seven VLBA targets. Additionally, the VaDAR method can now be compared to similar varstrometry-based selection methods for multi-AGN, including the Varstrometry for Offset and Dual sub-Kpc AGN \citep[VODKA;][]{shen_centered,chen_hst} method, for which significant VLBA work has been done.

\subsection{Individual Targets} \label{sec:individualtargets}

In the following subsections, each target is discussed in the context of all of the new and existing information presented above. All VLA observations referenced were presented in \cite{schwartzman2024}.

\subsubsection{J011114.41+171328.5}

In the VLA observations at sub-arcsecond scales, this target exhibits multiple bright components, and it was tentatively identified as a candidate dual AGN with a relatively flat spectral index. However, it is noteworthy that the flux ratio between the primary and secondary peaks was quite large, and possibly indicative of jet activity. 

In the VLBA observations at milliarcsecond scales, this target exhibits one source with clear extension to the northeast at 2.3 GHz. These observations are associated with the northern component identified at sub-arcsecond scales. The source exhibits a brightness temperature indicative of a quasar at both frequencies, and while it is compact at 8.3 GHz (C = 1.01), it is extended at 2.3 GHz (C = 5.19). It shows a steeply positive spectral index of $\alpha_{2.3 GHz}^{8.3 GHz}$ = 2.17, and displays significant radio-optical offset. We note that this target exhibits a significant RUWE value, though the remaining indicators of possible multiplicity remain outside the noteworthy range. In the context of the radio-optical offset and the spectral index, J011114.41+171328.5 is tentatively identified as a candidate multi-AGN in which the quasar visible in the radio regime is observed in the optically-thick regime.

However, given the extension visible at sub-milliarcsecond scales and the two components with a significant flux difference visible at sub-arcsecond scales, this source could also be identified as exhibiting jet activity. The position angle of the extension on sub-milliarcsecond scales at both frequencies is aligned with the separation angle of the two sub-arcsecond components. Jet activity can also explain the significant radio-optical offset. 

\subsubsection{J080009.98+165509.4}

In the VLA observations at sub-arcsecond scales, this target exhibits a single unresolved, point-like source. It was also identified as having an upturned spectral shape, indicative of some aging or absorption activity in the core, with a spectral index of $\alpha_{3 GHz}^{10 GHz} = -1.46$. Given the morphology, no further conclusions are drawn for this source from the VLA observations.

In the VLBA observations at sub-milliarcsecond scales, this target once again exhibits unresolved, point-like radio emission. It exhibits a brightness temperature indicative of a quasar at both frequencies, and it is also compact at both frequencies (C (2.3 GHz) = 1.44; C (8.3 GHz) = 1.29). It is found to have a relatively flat spectral index of $\alpha_{2.3 GHz}^{8.3 GHz} = 0.37$, and significant radio-optical offset. We note that all four \textit{Gaia} parameters presented in Table \ref{tab:gaia_params} are noteworthy for this target, suggesting that intrinsic source multiplicity is possible.

The difference in spectral indices on sub-arcsecond to milliarcsecond scales is notable. It is likely that the VLA emission includes contributions from larger-scale emission, possibly associated with jet activity. This would explain the significantly steeper spectral index. The VLBA is not sensitive to emission on those scales, and thus the milliarcsecond spectral index is reflective of emission at smaller scales, likely impacted by an absorption mechanism.

The observed radio-optical offset is significant for J080009.98+165509.4. Both the VLA and VLBA observations indicate that, while it is a quasar, there is no secondary radio core observed. While it is possible that the separation is simply too small to be resolved, even on sub-milliarcsecond scales, it is also possible that the second component is radio-silent but emitting in the optical. This is indicated by the significant radio-optical offset, in addition to the flat spectrum. 

We note that the target exhibits no morphological signs of jet activity. However, given the steep spectral index on sub-arcsecond scales, one viable explanation is that the jet emission exists on scales too small for the VLA to resolve, but too large for the VLBA to detect. Thus, the VLA spectral index remains steep, even without morphological evidence for extended emission. 

\subsubsection{J121544.36+452912.7}

In the VLA observations at sub-arcsecond scales, this target features a pair of canonical radio jets, between which a compact point source is observed. The jets themselves are diffuse, large-scale structure, while the core is significantly more compact. Though the jets are observed to be steep, the spectral index of the core is about $\alpha_{3 GHz}^{10 GHz} = -0.5$, indicative of a quasar. The target is thus identified as exhibiting jet activity, potentially the driver of the astrometric variability.

In the VLBA observations at sub-milliarcsecond scales, only the core of the jet is observed. The lobes are too diffuse and too large in angular size to be detected in the VLBA observations. However, the core is detected as an unresolved, point-like source at both frequencies. It exhibits a brightness temperature indicative of a quasar at both frequencies, and it was found to be compact (C (2.3 GHz) = 1.14; C (8.3 GHz) = 1.005). It exhibits a flat spectral index of $\alpha_{2.3 GHz}^{8.3 GHz} = 0.07$, and significant radio-optical offset. We note that none of the \textit{Gaia} parameters presented in Table \ref{tab:gaia_params} are noteworthy for this source, indicating that it is possible there is another explanation for the observed astrometric excess noise.

The likely interpretation for J121544.36+452912.7 remains the same: the canonical jet activity seen at sub-arcsecond scales is likely the driver of the astrometric variability. While there was some possibility of smaller-scale jets at sub-milliarcsecond scales, no such jets are observed. The observed compact emission is not unexpected for the central core of a canonical jet. While the significant radio-optical offset is notable, the existence of large-scale jets implies that the separation is likely to be driven by jet activity, in this case.

\subsubsection{J143333.02+484227.7}

In the VLA observations at sub-arcsecond scales, this target features clear jet activity. A ``core'' is identified as Component B, while an extension to the east is labeled as Component A, and two components to the west are labeled Components C and D. The core exhibits a standard power law spectral index of $\alpha_{3 GHz}^{10 GHz}$ = -0.86, which is expected for a quasar. Due to resolution limitations, it was not possible to characterize the spectral index of the jetted components. However, the target is identified as exhibiting jet activity, potentially the driver of the astrometric variability.

In the VLBA observations at sub-milliarcsecond scales, only Components A and B are observed, though the positions of all four components were used as phase centers in separate imaging cycles at both frequencies. Components C and D are not detected at either 2.3 and 8.3 GHz. Component A appears as an unresolved, point-like source in the VLBA images at both frequencies. Component B is clearly extended at 2.3 GHz, exhibiting obvious jet activity, though at 8.3 GHz, the extension is less clear. At both frequencies, Component A exhibits a brightness temperature indicative of a quasar, as does Component B. Component A is seen to be compact at both frequencies (C (2.3 GHz) = 1.21; C (8.3 GHz = 1.21), but Component B is considerably more extended (C (2.3 GHz) = 14.89; C (8.3 GHz) = 4.99). The spectral index of Component A is flat, at $\alpha_{2.3 GHz}^{8.3 GHz}$ = 0.03, while the spectral index of Component B is significantly steeper, at $\alpha_{2.3 GHz}^{8.3 GHz}$ = -1.09. We note that the RUWE and BP/RP excess parameters are found to be noteworthy for this source. Finally, the source shows significant radio-optical offset. 

In the case of Components C and D, which are not detected on milliarcsecond scales, we note that there are two possibilities to explain the non-detection. First, that the VLBA observations are sensitivity-limited, and the components were too faint to be detected. While this is possibly the case for Component C, Component D exhibits a similar flux density to Component A at sub-arcsecond scales. Thus, the second possibility applies in the case of Component D, and the emission observed with the VLA likely exists only on scales too large for the VLBA to detect.

The likely interpretation for the morphology seen in J143333.02+484227.7 is jet activity. It displays clear extension on both sub-arcsecond and sub-milliarcsecond scales, and the more diffuse jet emission is clearly visible at 2.3 GHz with the VLBA. Jet activity can explain the significant astrometric excess noise and the significant radio-optical offset. We note that the sub-milliarcsecond spectral index of Component B is steeper than what is expected for a quasar, though this is likely attributable to the jet extension visible at 2.3 GHz. Finally, we note that the spectral index at sub-milliarcsecond scales of Component A is flatter than is expected for a jet hotspot. While it is possible for this component to represent a secondary quasar core, it is considered unlikely due to the obvious jet activity.

\subsubsection{J162501.98+430931.6}

In the VLA observations at sub-arcsecond scales, this source exhibits two components at 10 GHz, a northern and a southern core. A spectral index was calculated for the system as a whole (including both components), and the source exhibits a standard power law with a spectral slope of $\alpha_{3 GHz}^{10 GHz}$ = -0.89, which is expected for a quasar. The target is identified as a candidate multi-AGN, with the secondary component the potential driver of the astrometric variability.

In the VLBA observations at sub-milliarcsecond scales, only the southern component was detected at 8.3 GHz, though the positions of both components were used as phase centers in separated imaging cycles at both frequencies. The northern component is a non-detection at both frequencies, while the southern component is not detected at 2.3 GHz. However, at 8.3 GHz, the southern component is an unresolved, point-like source that exhibits a brightness temperature indicative of a quasar, and is considered compact (C = 1.38). The spectral index of the source is $\alpha_{2.3 GHz}^{8.3 GHz}$ = 0.79, though the measurement at 2.3 GHz was taken as the noise of the image, and so likely represents an upper limit to the flux. We note that the RUWE and BP/RP excess parameters are significant for this source. Finally, the source exhibits significant radio-optical offset.

In the context of the sub-arcsecond observations and the sub-milliarcsecond observations of the southern core, the interpretation for J162501.98+430931.6 is that it is a candidate multi-AGN. This is primarily supported by the VLA observations and the significant radio-optical offset. We also consider jet activity, which might explain why some components of the source are undetected with the VLBA (due to brightness temperature limitations, angular size, etc.). 

\subsubsection{J172308.14+524455.5}

In the VLA observations at sub-arcsecond scales, this target exhibits a single, unresolved, point-like source. However, careful modeling of the SDSS spectrum for this source reveals evidence for stellar absorption consistent with a foreground star, and thus the source is identified as a star+quasar superposition, a designation that was made after the VLBA observations were taken. It is likely that the astrometric variability is being driven by this superposition. However, the radio emission from the quasar is still worthy of consideration. This target exhibits an upturned spectrum at sub-arcsecond scales, with a spectral slope of $\alpha_{3 GHz}^{10 GHz}$ = -1.29. 

In the VLBA observations at sub-milliarcsecond scales, this target is observed as a single, unresolved, point-like source at both frequencies. Also at both frequencies, the emission exhibits a brightness temperature indicative of a quasar, and is compact (C (2.3 GHz) = 1.24; C (8.3 GHz) = 1.07). The source displays a relatively flat spectral index of $\alpha_{2.3 GHz}^{8.3 GHz}$ = 0.5, and is identified as having significant radio-optical offset. We note that all \textit{Gaia} parameters presented in Table \ref{tab:gaia_params} are noteworthy for this source, save the RUWE value. 

As a driver of the astrometric variability, the star+quasar superposition seems more likely for J172308.14+524455.5. The quasar component is compact at both scales and both frequencies, with a flat spectral index at sub-milliarcsecond scales. Though the upturned spectral index displayed at sub-arcsecond scales is interesting, it is possibly driven by observational limitations and low spectral sampling \citep[given the lack of flux density measurements available for the spectrum][]{patil2022}. The most interesting component is the observed radio-optical offset. Since the astrometric information from the star was incorporated into \textit{Gaia}'s astrometric solution for the source (thus driving the excess astrometric noise), it follows that the optical position identified by \textit{Gaia} could be similarly contaminated by the star, rather than any optical emission from the quasar. This would explain the observed radio-optical offset. Thus, this target is likely only a star+quasar superposition.

\subsubsection{J173330.80+552030.9}

In the VLA observations at sub-arcsecond scales, this target exhibits a single unresolved, point-like source. It was also identified as having a standard power law spectral slope of $\alpha_{3 GHz}^{10 GHz}$ = -0.33, as expected for a quasar. Given the morphology, no further conclusions are drawn for this source from the VLA observations.

In the VLBA observations at sub-milliarcsecond scales, this target also exhibits unresolved, point-like radio emission at both frequencies. It has a brightness temperature indicative of a quasar at both frequencies, and it also compact (C (2.3 GHz) = 1.09; C (8.3 GHz) = 1.22). It has a relatively flat spectral slope of $\alpha_{2.3 GHz}^{8.3 GHz}$ = 0.5, and is observed to exhibit significant radio-optical offset. We note that all \textit{Gaia} parameters presented in Table \ref{tab:gaia_params} are noteworthy for this source, save the $\texttt{ipd\_frac\_multi\_peak}$ value.  

The observed radio-optical offset is significant for J173330.80+552030.9. Both the VLA and VLBA observations indicate that, while this source is a quasar, there is no secondary radio core observed. While it is possible that the separation is simply too small to be resolved, even on sub-milliarcsecond scales, it is also possible that the second component is radio-silent, but emitting in the optical regime. As previously mentioned, it is also possible that the secondary radio core exists below the sensitivity of the VLBA observations. Both scenarios align with the radio-optical offset, in addition to the flat spectrum. The target exhibits no signs of jet activity.

\subsection{Methodology Comparison} \label{subsec:methcomp}

To assess the VaDAR method in the context of the many pre-selection methods for multi-AGN, we now compare these results to those from the VODKA \citep[][]{shen_nature,chen2023} program. We note that there are important similarities and differences between the two samples. Both samples are composed of SDSS-identified quasars, and are selected with similar AENS, redshift, \textit{Gaia G} magnitude cuts. From there, however, the VaDAR sample was crossmatched with the VLASS catalog \citep[][]{schwartzman2024}, while the VODKA sample was not. Additionally, the VODKA sample is composed of an IR crossmatch, among other important differences \citep[][]{chen_hst}.

In this context, we compare the results of these two programs as the only two programs to thoroughly investigate samples selected using the varstrometry technique. Though the VLBA observations presented in this paper have provided useful insight into the VaDAR pilot sample on sub-milliarcsecond scales, they do not further constrain the relative fractions of multi-AGN and gravitational lenses in the sample. Combined, these objects account for $\sim$44\% of the VaDAR pilot sample \citep[][]{schwartzman2024}. The result is comparable to that of the VODKA program, as presented in \cite{chen_hst}.

\cite{chen2023} presented VLBA observations for 23 radio-bright candidate dual and off-nucleus quasars selected as part of the VODKA program, a combination of new VLBA observations (18 targets) and archival observations (6 targets). Of the 18 targets observed in the new VLBA observations, 16 were detected (5$\sigma$ threshold), and 2 were non-detections. This is a similar detection fraction to that achieved for the VaDAR sample (see Section \ref{sec:sample_vlba}), though both samples were selected to be bright enough in the radio regime to be detectable with the VLBA (the VaDAR sample was limited to those objects with 3 GHz peak flux greater than 1 mJy at VLA scales; the VODKA sample was limited to those with peak fluxes greater than 15 mJy in the Faint Images of the Radio Sky at Twenty-Centimeters \citep[FIRST;][]{becker1995,chen_hst}).

With respect to the radio morphology at sub-milliarcsecond scales, two of the seven targets presented in this paper exhibit jet activity or otherwise extended emission. This is a comparable, though slightly higher, fraction to that identified in \cite{chen2023}. As part of the VODKA VLBA program, they identified three of their 16 detected targets as exhibiting secondary compact emission, jet activity, or otherwise extended structure. For both samples, however, the majority of the targets are unresolved, point-like detections at the sub-milliarcsecond scales probed by the VLBA. 

Finally, \cite{chen2023} compared the optical positions from \textit{Gaia} DR3 and the VLBA radio positions for the VODKA sample, similar to the analysis presented for the VaDAR sample in Section \ref{sec:roo} of this paper. Of the 21 VODKA targets detected with the VLBA (both new and archival observations), six were found to display significant ($> 3\sigma$) radio-optical offsets. In the VaDAR sample, 100\% of the targets (7/7) were identified as exhibiting significant, $> 3\sigma$ radio-optical offsets, as described in Section \ref{sec:roo}. It is perhaps unsurprising that varstrometry-selected samples display significant radio-optical offsets, as they are selected to be astrometrically-variable. Though disparate, both results highlight the need for a dedicated study of the radio-optical offsets of a larger sample of astrometrically-variable sources, and vice versa.

Overall, varstrometry as applied in both the VaDAR and the VODKA samples appears to select for samples with similar parameters, though we note that the small number of targets makes any statistical comparison impossible. Additionally, there are important differences between the two selection methodologies. However, both strategies successfully select for a high fraction of either multi-AGN or gravitationally-lensed quasars, though separating out the gravitationally-lensed quasars from the multi-AGN continues to prove difficult. Both have similar success with VLBA radio observations probing sub-milliarcsecond scales, including detections with similar distributions of radio morphology. Finally, both methods appear to produce samples with a relatively high fraction of sources with significant radio-optical offsets.

\section{Summary and Conclusions}

We present new, targeted VLBA observations of a subset of seven of the 18 \textit{Gaia}-unresolved quasars from the original VaDAR pilot sample \citep[][]{schwartzman2024}. These new observations further constrain the smaller-scale radio properties of this sample, and highlight the necessity of high-resolution observations in confirming multi-AGN. They additionally illustrate the prevalence of significant radio-optical offsets amongst astrometrically-variable quasars. The high redshifts ($0.7 < z < 2.6$) of the targets represent an observational gap in the current population of multi-AGN, while the sub-milliarcsecond resolutions observed with the VLBA probe a separation regime (10's-100's of pc) in which there currently only exists one confirmed binary AGN \citep[][]{rodriguez2006}. This VLBA follow-up study is the next in a series of studies designed to fully characterize the VaDAR sample. Our conclusions are as follows:

\begin{itemize}
    \item All seven targets were detected with the VLBA in at least one of the two observed frequencies, 2.3 GHz and 8.3 GHz. Two of the seven sources exhibit extension indicative of jet activity or extended emission otherwise associated with an AGN. While astrometric variability can be attributed to jet activity, it is more difficult to ascertain in the case of the unresolved sources.
    \item In the context of these follow-up observations, 4/7 of the targets have been identified as potential candidate multi-AGN, 2/7 are found to exhibit signs of jet activity, and 1/7 has been classified as star+quasar superposition.
    \item The compactness of each target was measured as the ratio of the integrated and total flux. All unresolved targets are found to be compact, with values close to 1. The targets exhibiting extension have higher values (as expected), ranging from 4.99 to 14.89.
    \item All targets detected at both 2.3 GHz and 8.3 GHz are found to exhibit brightness temperatures indicative of quasar activity.
    \item The two-band quasi-instantaneous spectral indices $\alpha_{2.3 GHz}^{8.3 GHz}$ range from -1.09 to 2.17. Four of the components (including the second component observed in J143333.02+484227.7) have been classified as flat spectrum, three exhibit steeply positive spectral indices ($>$ 0.5), and one exhibits a steep spectral index of -1.09. The flat spectral slopes are attributable to quasars, the steeply negative spectral slope is indicative of jet activity, and the steeply positive spectral indices are perhaps representative of quasars in the optically-thick regime.
    \item Significant radio-optical offsets were seen in all seven targets. One interpretation is sub-milliarcsecond scale jet activity that is too small or too faint for the existing optical observations of these targets. Another interpretation is a second component AGN.    
    \item Gaussian modeling of the DECaLS optical host images shows ellipticity in four of seven targets. Further optical observations at significantly higher resolutions will be required to confirm the presence of any extended optical structure, including interaction features or optical jets.
    \item A study of four other \textit{Gaia} parameters that can indicate multiplicity in sources shows values exceeding the range expected for a normal single-peak source for most parameters in all targets. Control and target samples vary by more than the parameter of interest for three of the four parameters. 
    \item An in-depth comparison to the work done on the VODKA sample \citep[][]{chen_hst,chen2023} reveals that both varstrometry-based methods select for similar samples. At sub-arcsecond scales, both VODKA and this study find that gravitationally-lensed quasars and multi-AGN make up $\sim$40\% of varstrometry-selected samples. Both the VODKA and the VaDAR samples have similar detection fractions at sub-milliarcsecond scales with the VLBA, and the majority of the targets are found to be unresolved radio cores. Both samples have a significant fraction of sources with observed radio-optical offsets.
\end{itemize}

Overall, we add further constraints to the VaDAR sample using new higher-resolution VLBA observations. These results demonstrate the potential of using varstrometry in tandem with radio VLBI to identify candidate multi-AGN at low separations in an interesting and relatively unexplored redshift regime. We show that these astrometrically-variable systems show significant radio-optical offsets, which are another possible indicator of multi-AGN. This result is significant with respect to the accuracy of the International Celestial Reference Frame, as source with intrinsic structure increase the noise floor of the ICRF. Future work understanding the astrometric parameters of both the ICRF and the entire population of multi-AGN will be beneficial, including understanding the astrometric properties of sources with significant radio-optical offset. Overall, this study will inform future work on the VaDAR sample, which will include further follow-up at high resolutions, and continue to grow the sample of confirmed and candidate multi-AGN.

\begin{acknowledgements}
E. S. gratefully acknowledges support from the National Radio Astronomy Observatory's Student Observing Support fellowship and George Mason University's Doctoral Research Scholars program. R. W. P. gratefully acknowledges support through an appointment to the NASA Postdoctoral Program at Goddard Space Flight Center, administered by ORAU through a contract with NASA. 
This work has made use of data from the European Space Agency (ESA) mission Gaia (https://www.cosmos.esa.int/gaia), processed by the Gaia Data Processing and Analysis Consortium (DPAC, https://www.cosmos.esa.int/web/gaia/dpac/consortium). Funding for the DPAC has been provided by national institutions, in particular the institutions participating in the Gaia Multilateral Agreement.
This research made use of Astropy, a community-developed core Python package for Astronomy (\cite{astropy}), $\mathrm{TOPCAT}$ (\cite{topcat}), the Common Astronomy Software Application (\cite{casanew2022}), and the Python Blob Detector and Source Finder (\cite{pybdsf}). 
Funding for the Sloan Digital Sky Survey IV has been provided by the Alfred P. Sloan Foundation, the U.S.
Department of Energy Office of Science, and the Participating Institutions.
The National Radio Astronomy Observatory is a facility of the National Science Foundation operated under cooperative agreement by Associated Universities, Inc. Basic research in radio astronomy at the U.S. Naval Research Laboratory is supported by 6.1 Base Funding. 
\end{acknowledgements}

\vspace{5mm}
\facilities{\textit{Gaia}, VLA (NRAO), Sloan, HST, VLBA}

\software{Astropy (\cite{astropy}), CASA (\cite{casanew2022}), PyBDSF (\cite{pybdsf}), $\mathrm{TOPCAT}$ (\cite{topcat})}

\clearpage
\bibliography{citation.bib} 

\bibliographystyle{aasjournal}

\section{Appendix A: Target Images}

This section presents the final images for all targets. All images have been formatted to emphasize the observed radio structure, and include a scalebar labeled with the appropriate number of milliarcseconds and corresponding parsecs. Images at both frequencies include the \textit{Gaia} position, marked with a grey ellipse to reflect the error in the position. The size of the ellipse marks the error in the \textit{Gaia} position. On the 2.3 GHz (13cm) images, blue crosses mark the position of the VLBA detection at 8.3 GHz (4cm), and vice versa. The size of the cross reflects the error in the VLBA position. In the 2.3 GHz images, a dark pink square outlines the size of the 8.3 GHz images. In the case of J162501.98+430931.6 and J143333.02+4834227.7, multiple components were observed in the VLA images. Multiple phase centers were used during cleaning to examine all components in the new VLBA observations. Only components detected in the VLBA observations are imaged. All images shown with contours beginning at 3$\sigma$ and proceeding in integer multiples of $\sqrt2$. Dashed contours denote emission at -3$\sigma$.

\begin{figure*}[!htb]
\gridline{\fig{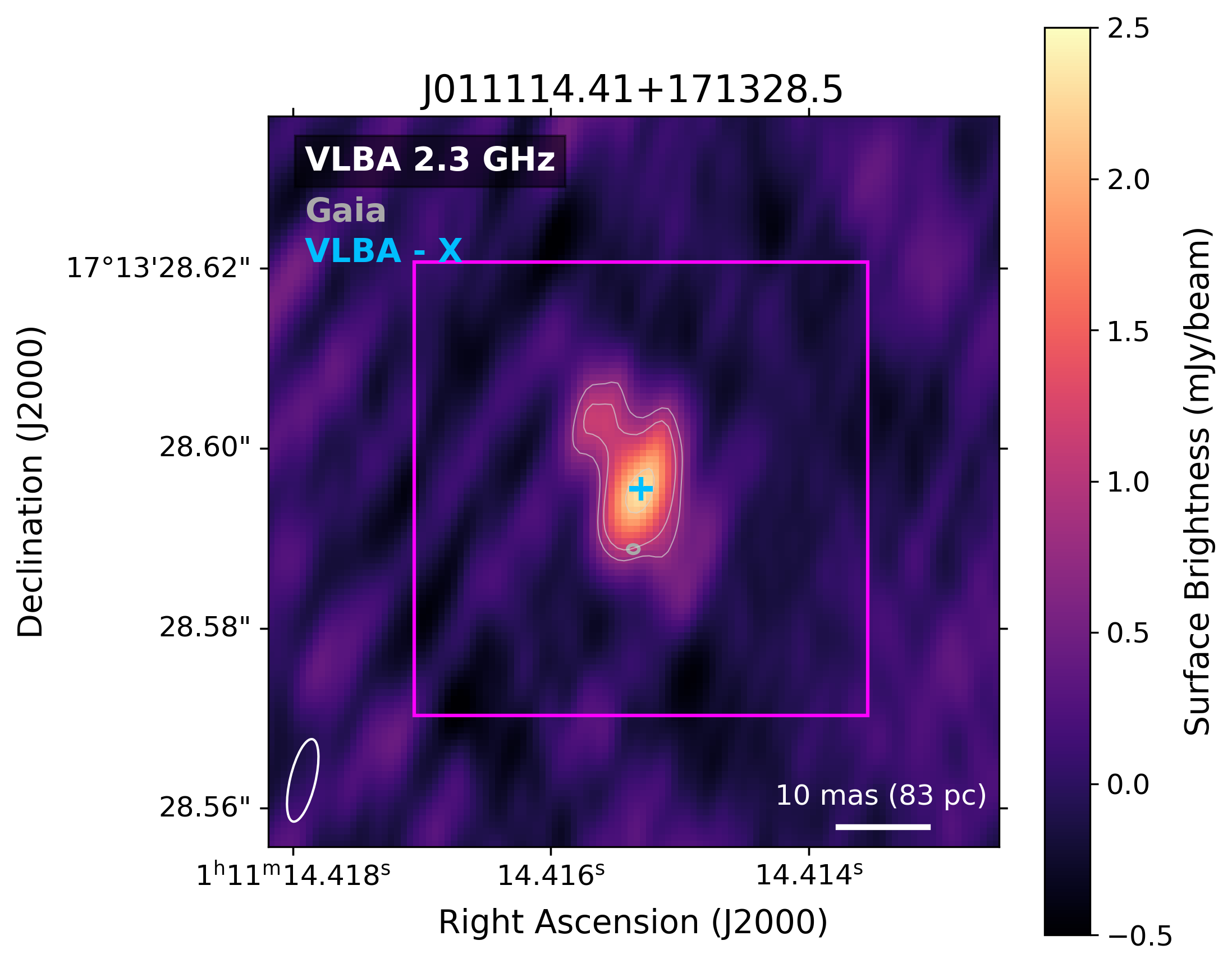}{0.5\textwidth}{(a) Observations at 2.3 GHz with a 0.009$\arcsec{}$ $\times$ 0.003$\arcsec{}$ beam. Image is 0.075'' square. Briggs weighting with a robust factor of 0.5 was used.}
          \fig{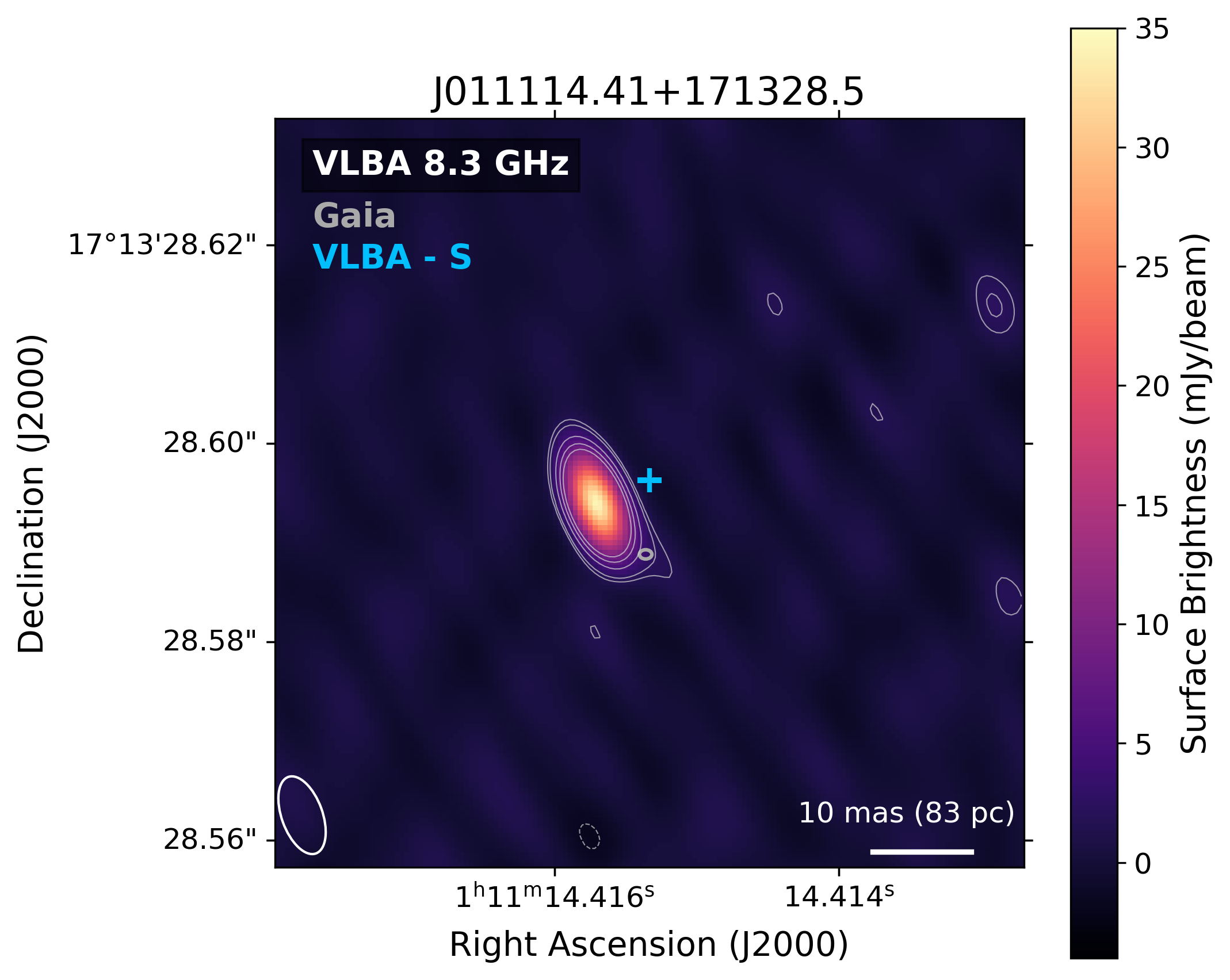}{0.5\textwidth}{(b) Observations at 8.3 GHz with a 0.006$\arcsec{}$ $\times$ 0.001$\arcsec{}$ beam. Image is 0.05'' square. Natural weighting was used, with a uv-taper of 3 milliarcseconds.}}
\caption{VLBA observations of \textbf{J011114.41+171328.5}. At milliarcsecond scales, this target shows some extension at 2.3 GHz. It also displays significant radio-optical offset, and is thus identified as a candidate multi-AGN. We note that jet activity is possible but unlikely given the target's steeply positive spectral index of $\alpha_{2.3 GHz}^{8.3 GHz}$ = 2.17.} 
\label{fig:VLBA_011114}
\end{figure*}

\begin{figure*}[!htb]
\gridline{\fig{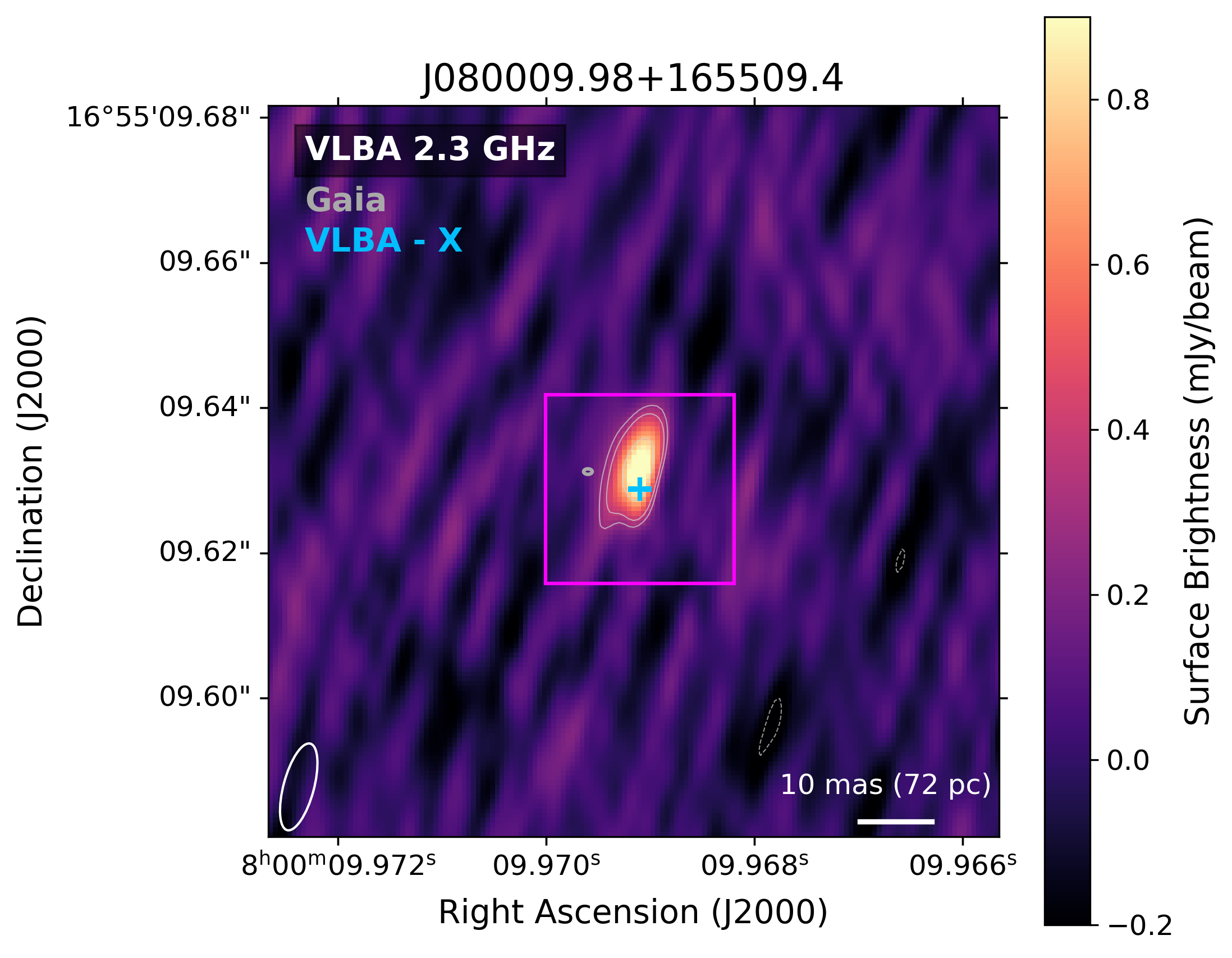}{0.5\textwidth}{(a) Observations at 2.3 GHz with a 0.010$\arcsec{}$ $\times$ 0.003$\arcsec{}$ beam. Image is 0.1'' square. Natural weighting was used.}
          \fig{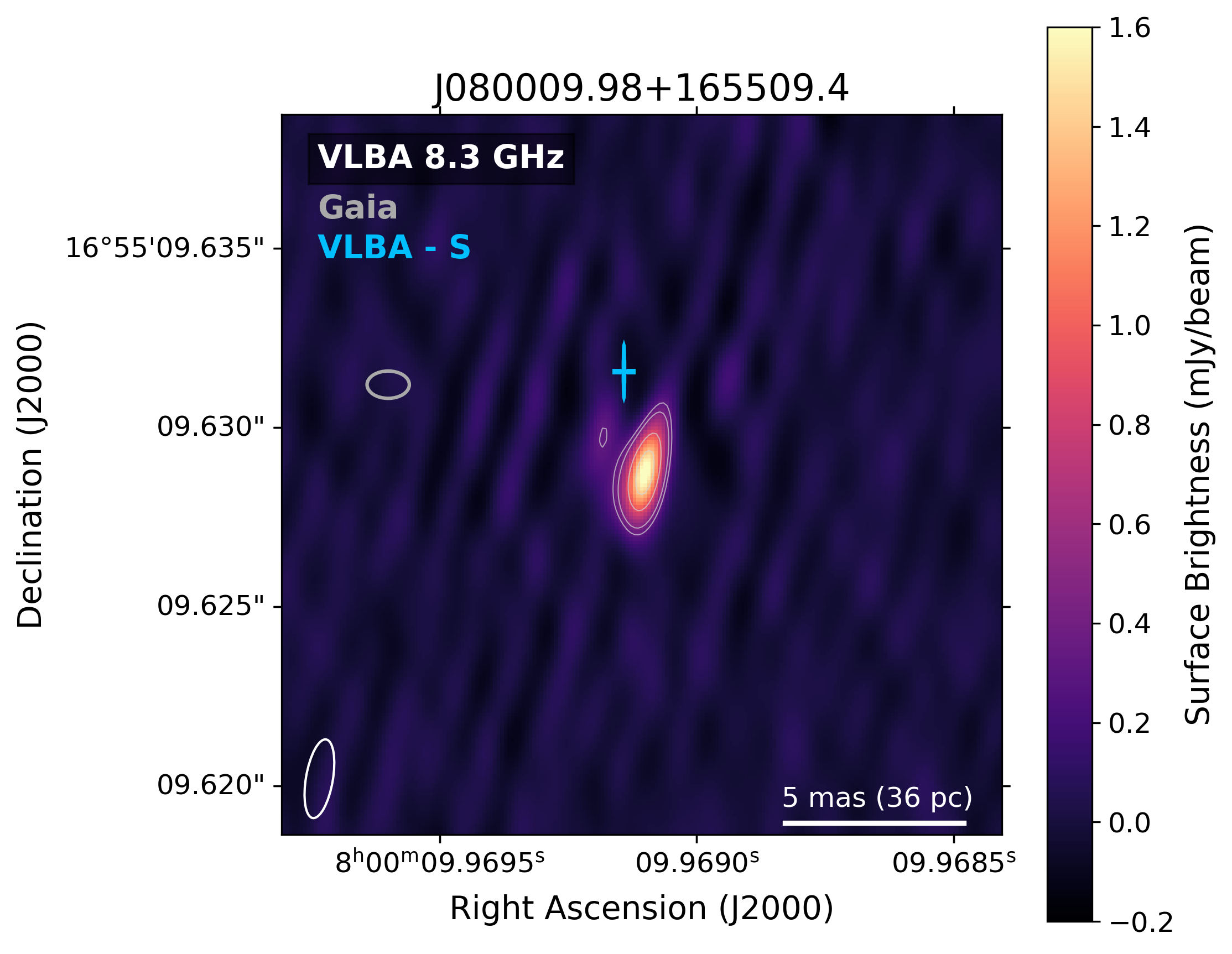}{0.5\textwidth}{(b) Observations at 8.3 GHz with a 0.002$\arcsec{}$ $\times$ 0.001$\arcsec{}$ beam. Image is 0.02'' square. Briggs weighting with a robust factor of 0.5 was used.}}
\caption{VLBA observations of \textbf{J080009.98+165509.4}. At milliarcsecond scales, this target is unresolved. It also displays significant radio-optical offset, and is thus identified as a candidate multi-AGN. This is further supported by the target's flat spectral index of $\alpha_{2.3 GHz}^{8.3 GHz}$ = 0.37.} 
\label{fig:VLBA_080009}
\end{figure*}

\begin{figure*}[!htb]
\gridline{\fig{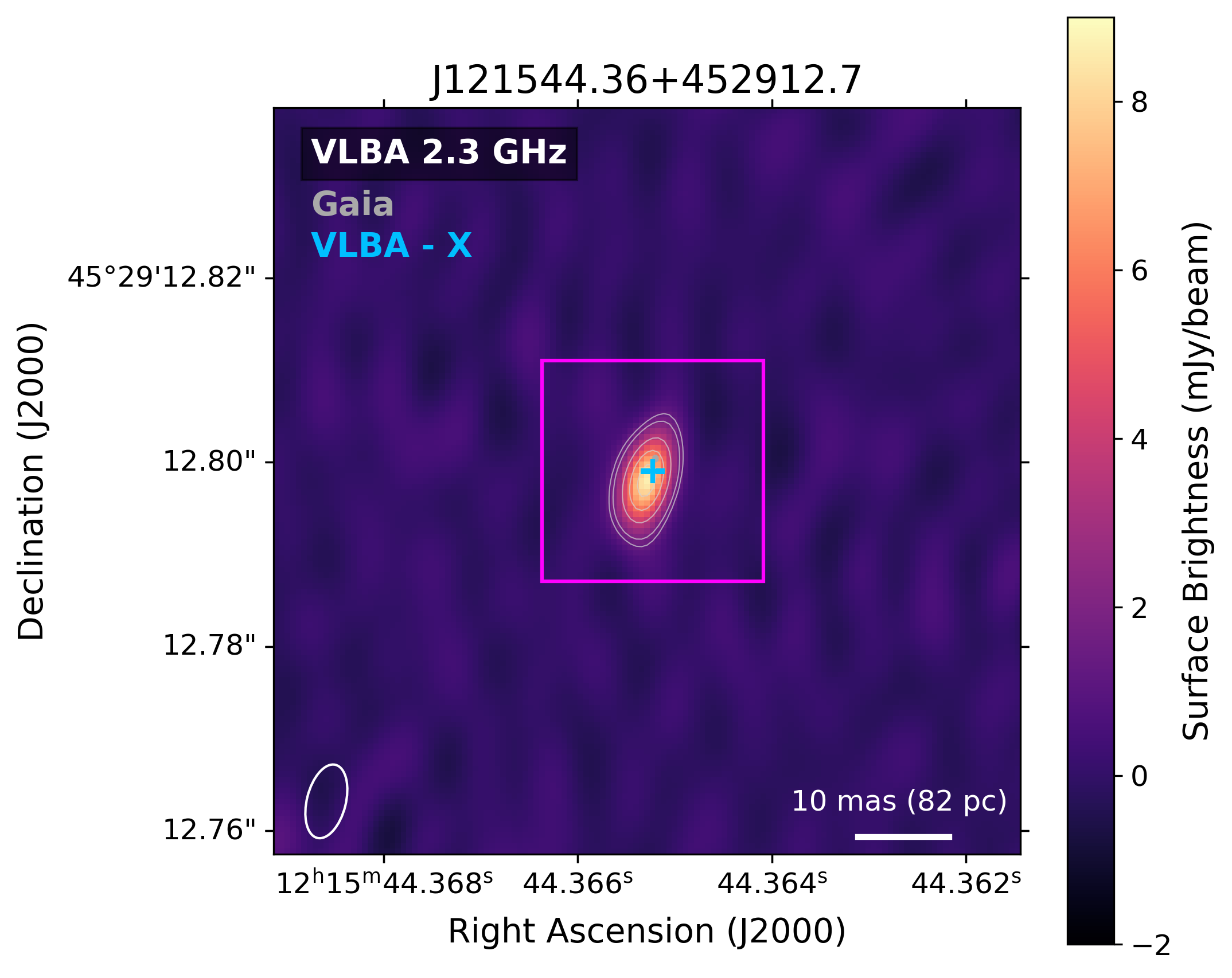}{0.5\textwidth}{(a) Observations at 2.3 GHz with a 0.008$\arcsec{}$ $\times$ 0.004$\arcsec{}$ beam. Image is 0.08'' square. Briggs weighting with a robust factor of 0.5 was used.}
          \fig{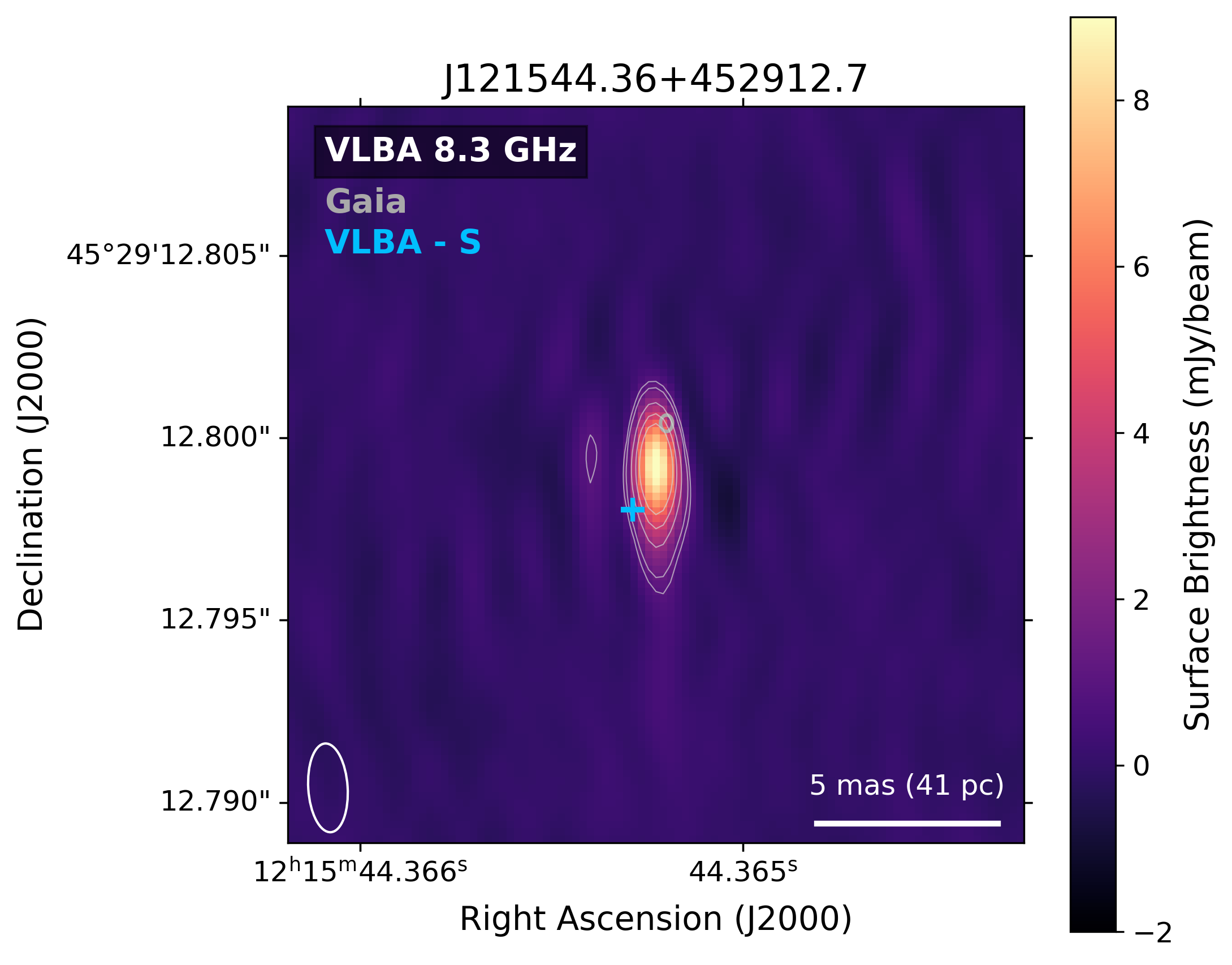}{0.5\textwidth}{(b) Observations at 8.3 GHz with a 0.002$\arcsec{}$ $\times$ 0.001$\arcsec{}$ beam. Image is 0.02'' square. Briggs weighting with a robust factor of 0.5 was used.}}
\caption{VLBA observations of \textbf{J121544.36+452912.7}. At milliarcsecond scales, this target is unresolved. It also displays significant radio-optical offset. However, because the target displays large-scale jet activity at sub-arcsecond scales, it is likely that the jets are responsible for both the excess astrometric variability and the observed radio-optical offset. The VLBA observations focus on the core, and this is further supported by the target's flat spectral index of $\alpha_{2.3 GHz}^{8.3 GHz}$ = 0.07. Note that at 2.3 GHz, the \textit{Gaia} and VLBA 8.3 GHz positions (and thus their markers) overlap.} 
\label{fig:VLBA_121544}
\end{figure*}

\begin{figure*}[!htb]
\gridline{\fig{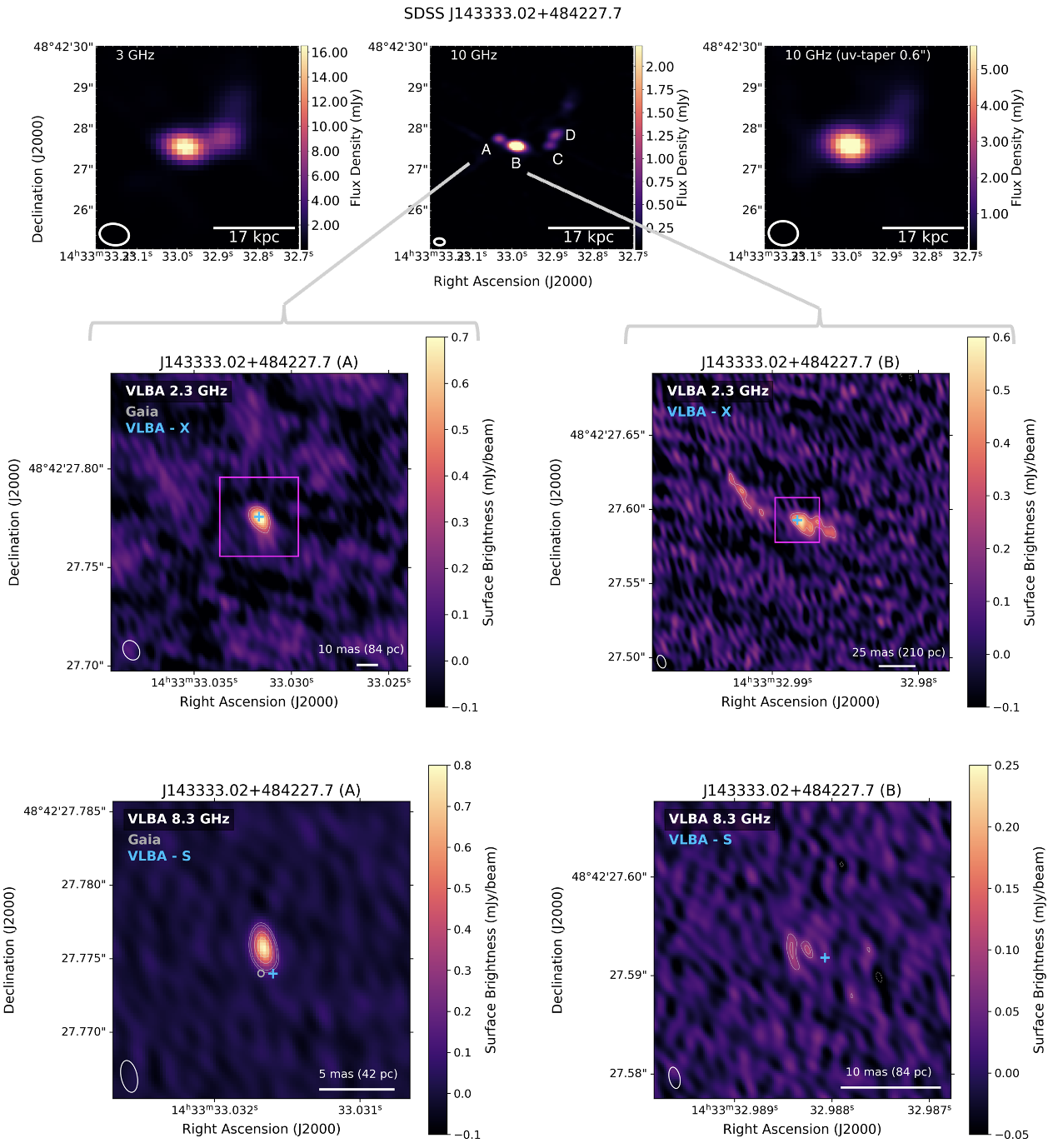}{0.95\textwidth}{}}
\caption{VLBA observations of \textbf{J143333.02+484227.7}. The three-panel set of VLA images is presented in \cite{schwartzman2024}. The VLBA images presented on the left focus on component A; the images presented on the right focus on component B. Components C and D were not detected with the VLBA. Observations at 2.3 GHz have a 0.009$\arcsec{}$ $\times$ 0.005$\arcsec{}$ beam; Observations at 8.3 GHz have a 0.002$\arcsec{}$ $\times$ 0.001$\arcsec{}$ beam. The component A image at 2.3 GHz is 0.1'' square, the component A image at 8.3 GHz is 0.02'' square, the component B image at 2.3 GHz is 0.2'' square, and the component B image at 8.3 GHz is 0.03'' square. For Component A at 2.3 GHz, natural weighting was used. For all other images, Briggs weighting with a robust factor of 0.5 was used. Due to the large offset, the \textit{Gaia} positions have been omitted from the Component B images. Note that in the 2.3 GHz image of Component A, the \textit{Gaia} and VLBA 8.3 GHz positions (and thus their markers) overlap.
This object is extended at milliarcsecond scales, particularly in the case of Component B. Component A remains unresolved. It does not show significant radio-optical offset, and so it is likely that the excess astrometric variability is driven by the observed jet activity. This is supported by the flat core seen in Component A ($\alpha_{2.3 GHz}^{8.3 GHz}$ = 0.03), and the steep Component B ($\alpha_{2.3 GHz}^{8.3 GHz}$ = -1.09).} 
\label{fig:VLBA_143333}
\end{figure*}

\begin{figure*}[!htb]
\gridline{\fig{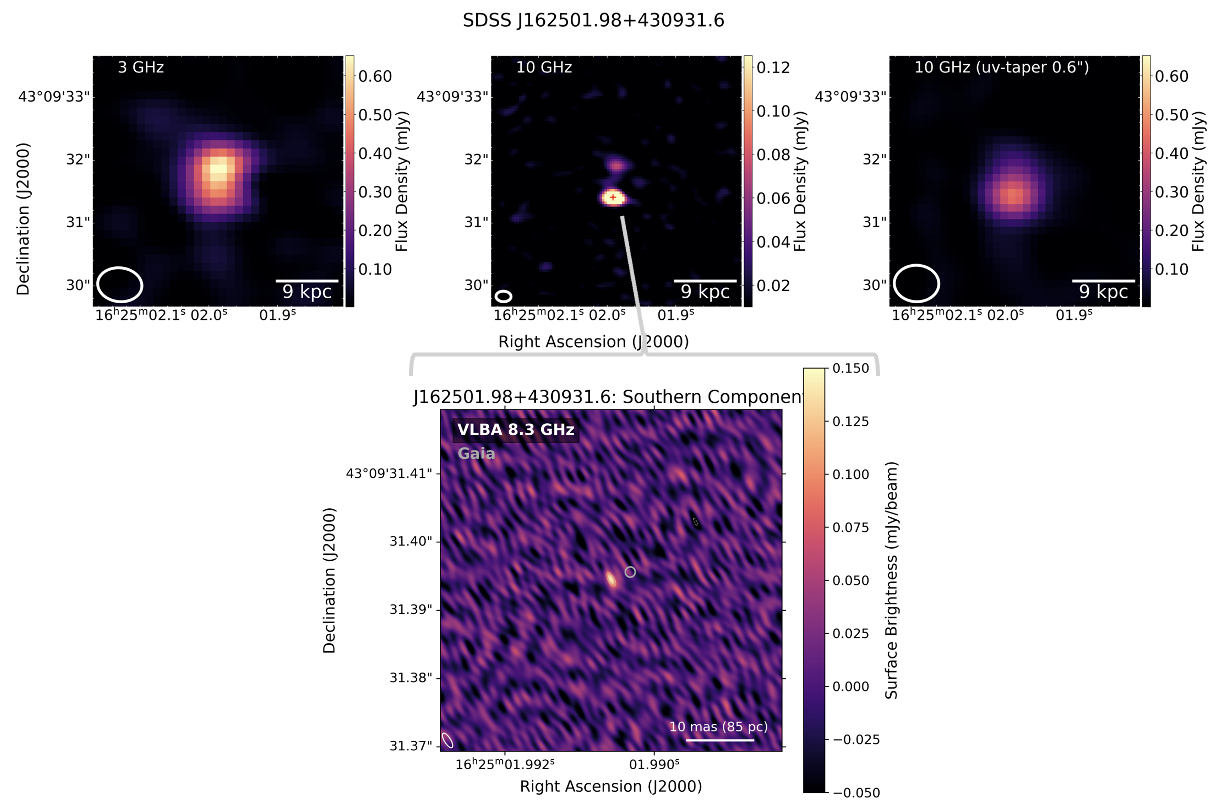}{0.95\textwidth}{}
          }
\caption{VLBA observations of \textbf{J162501.98+430931.6}. The three-panel set of VLA images is presented in \cite{schwartzman2024}. The VLBA image presented focuses on the southern component at 8.3 GHz. The southern component at 2.3 GHz, as well as the northern component at both frequencies, were not detected with the VLBA. Observations at 8.3 GHz have a 0.002$\arcsec{}$ $\times$ 0.001$\arcsec{}$ beam. The southern component image at 8.3 GHz is 0.01'' square, and uses Briggs weighting with a robust factor of 0.5. 
This southern component of this object is unresolved at 8.4 GHz. It does not show significant radio-optical offset. Though the two observed components at sub-arcsecond scales are possibly indicative of a multi-AGN, they could also be driven by jet activity. This target displays a spectral index of $\alpha_{2.3 GHz}^{8.3 GHz}$ = 0.79.} 
\label{fig:VLBA_162501_N}
\end{figure*}

\begin{figure*}[!htb]
\gridline{\fig{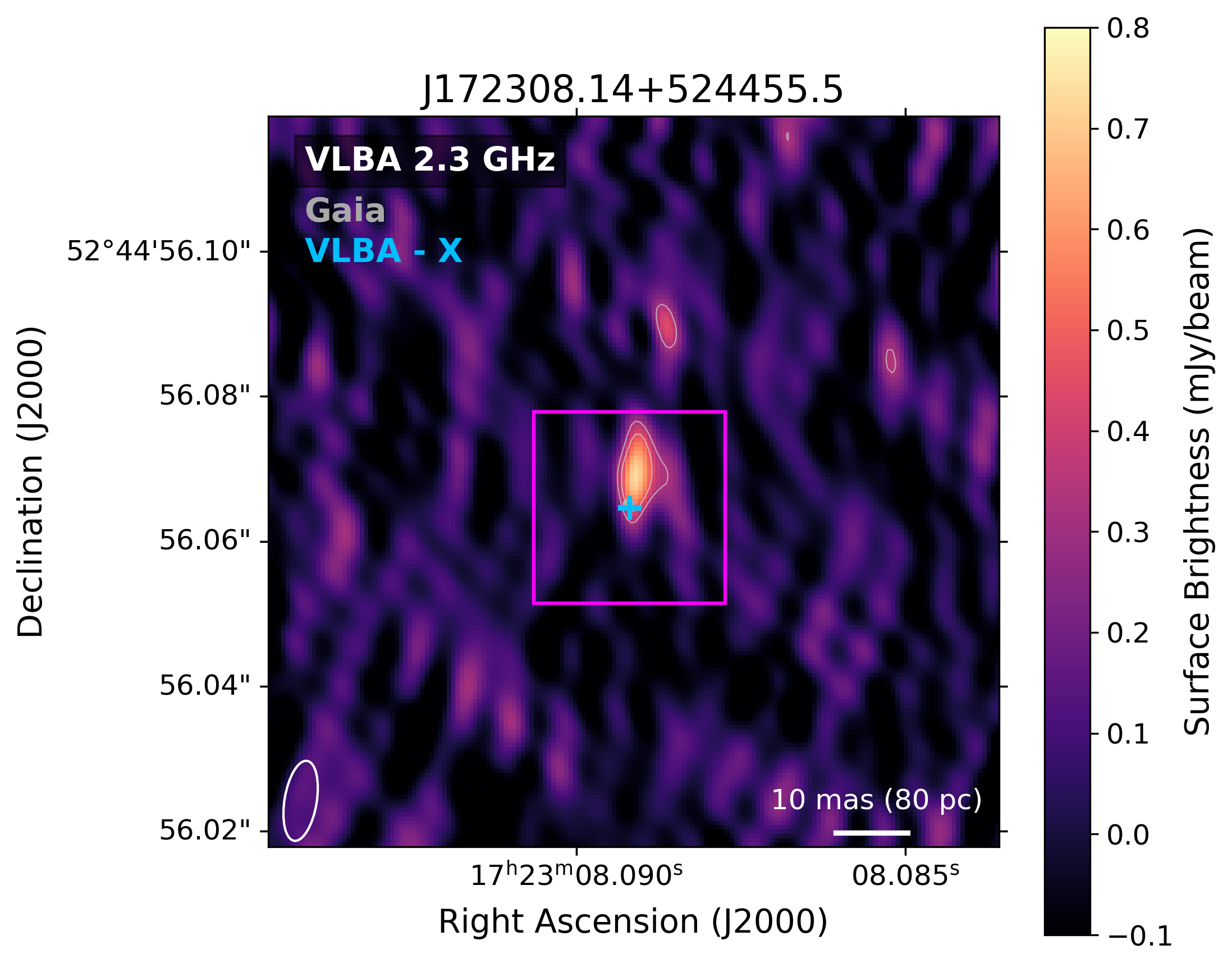}{0.5\textwidth}{(a) Observations at 2.3 GHz with a 0.011$\arcsec{}$ $\times$ 0.004$\arcsec{}$ beam. Image is 0.1'' square. Briggs weighting with a robust factor of 0.5 was used.}
          \fig{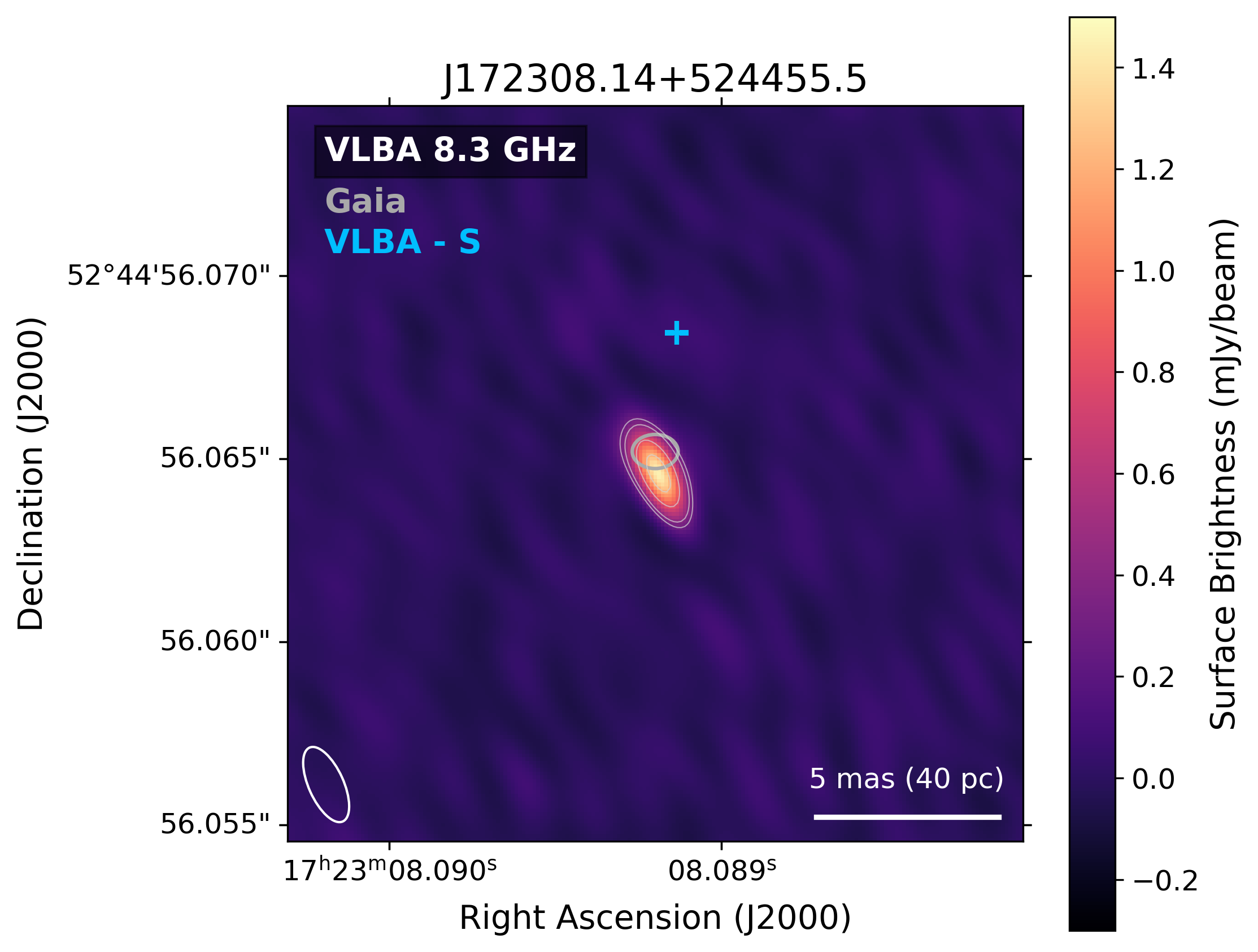}{0.5\textwidth}{(b) Observations at 8.3 GHz with a 0.002$\arcsec{}$ $\times$ 0.001$\arcsec{}$ beam. Image is 0.02'' square. Briggs weighting with a robust factor of 0.5 was used.}}
\caption{VLBA observations of \textbf{J172308.14+524455.5}. This source is unresolved at milliarcsecond scales. The target displays significant radio-optical offset. However, it has also been identified as a star+quasar superposition, and that remains the most likely driver of the excess astrometric variability. This target displays a spectral index of $\alpha_{2.3 GHz}^{8.3 GHz}$ = 0.50.} 
\label{fig:VLBA_172308}
\end{figure*}

\begin{figure*}[!htb]
\gridline{\fig{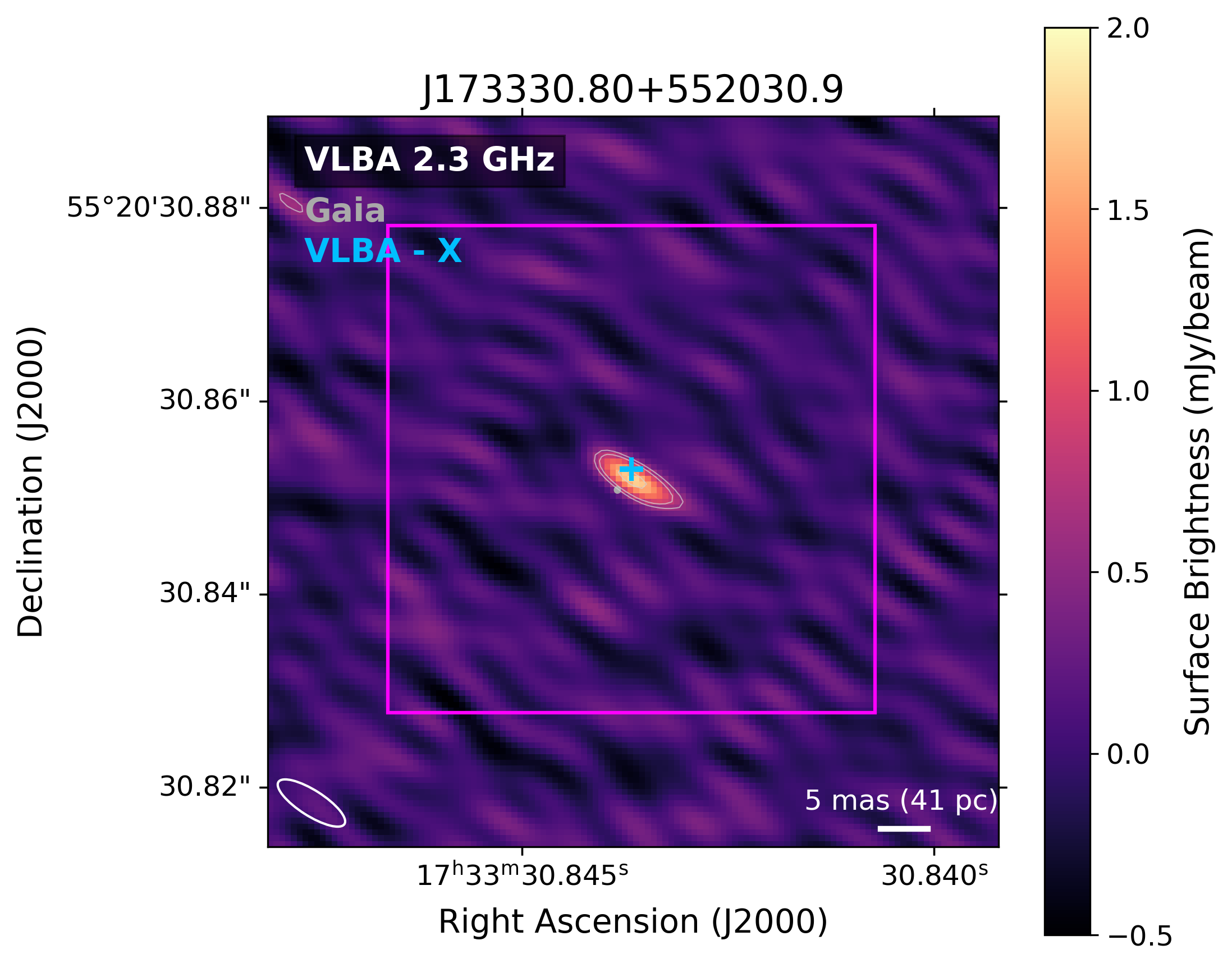}{0.5\textwidth}{(a) Observations at 2.3 GHz with a 0.008$\arcsec{}$ $\times$ 0.003$\arcsec{}$ beam. Image is 0.075'' square. Briggs weighting with a robust factor of 0.5 was used.}
          \fig{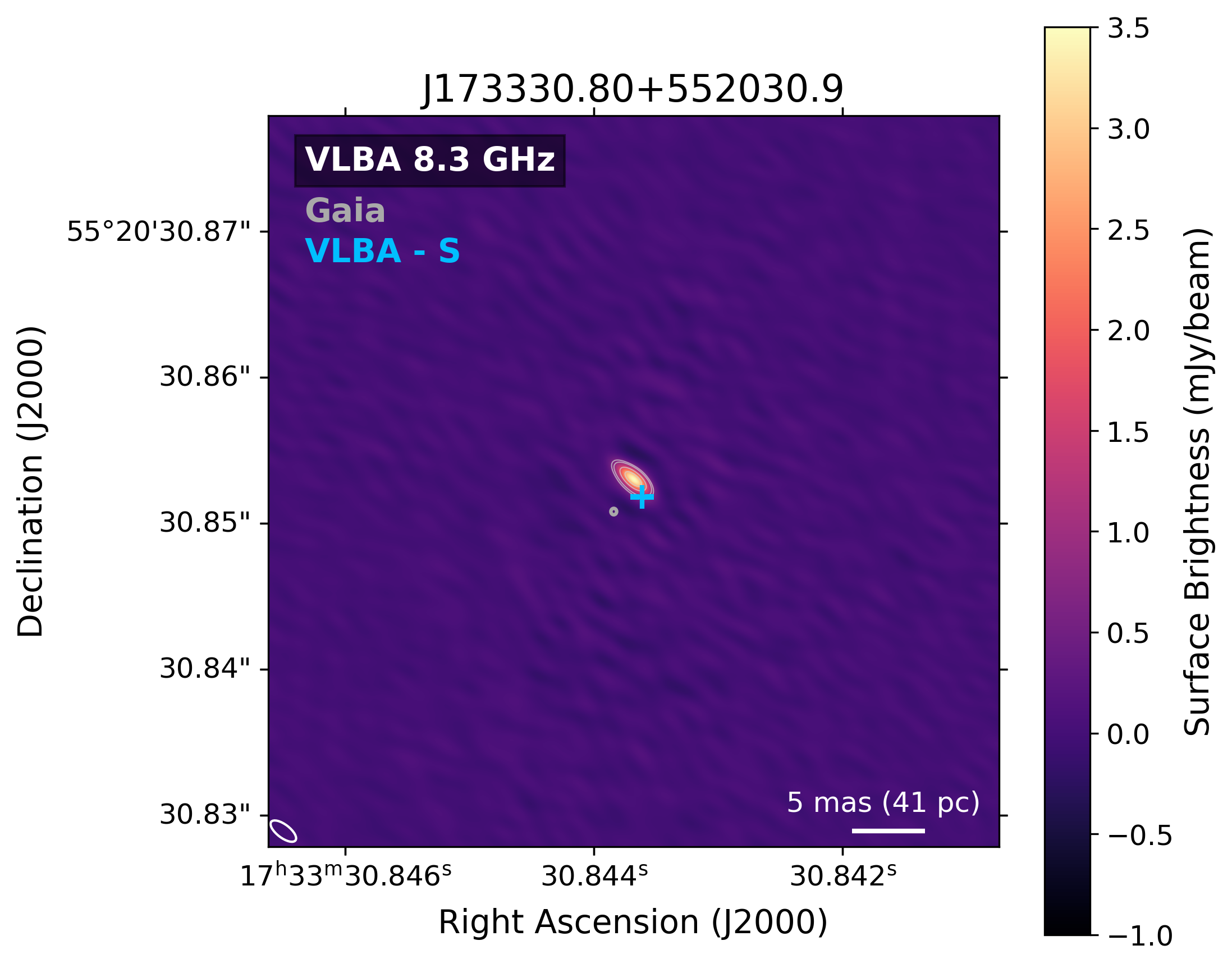}{0.5\textwidth}{(b) Observations at 8.3 GHz with a 0.002$\arcsec{}$ $\times$ 0.001$\arcsec{}$ beam. Image is 0.02'' square. Briggs weighting with a robust factor of 0.5 was used.}}
\caption{VLBA observations of \textbf{J173330.80+552030.9}. This source is unresolved at milliarcsecond scales. The target displays significant radio-optical offset, and is thus identified as a candidate multi-AGN. This is further supported by the target's steeply positive spectral index of $\alpha_{2.3 GHz}^{8.3 GHz}$ = 0.54.} 
\label{fig:VLBA_173330}
\end{figure*}

\clearpage

\section{Appendix B: Phase Calibrators} \label{sec:phasecalimages}

\begin{figure*}[!htb]
\gridline{
          \fig{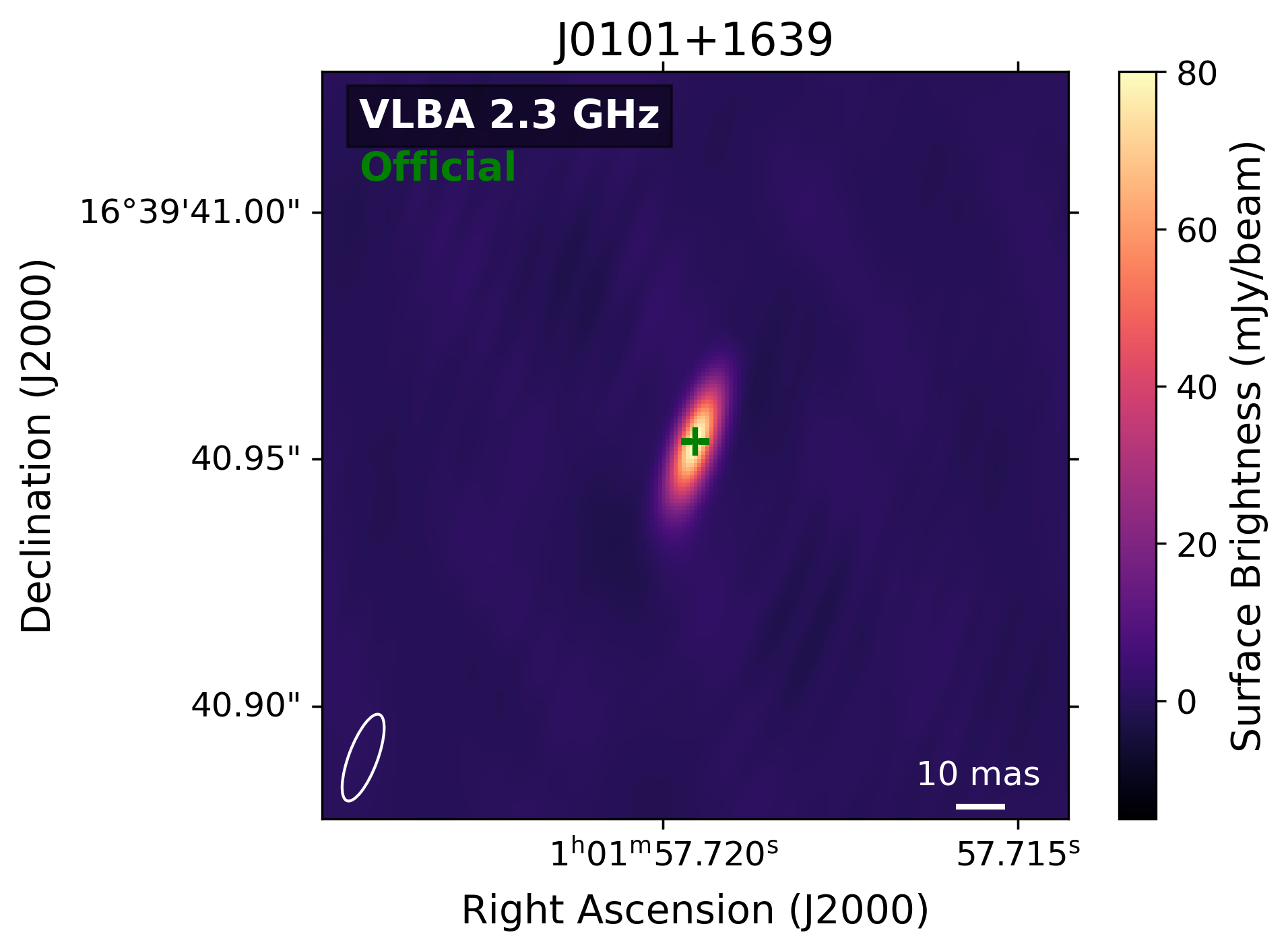}{0.485\textwidth}{(a) Observations at 2.3 GHz with a 0.019$\arcsec{}$ $\times$ 0.006$\arcsec{}$ beam. Image is 0.15'' square.}
          \fig{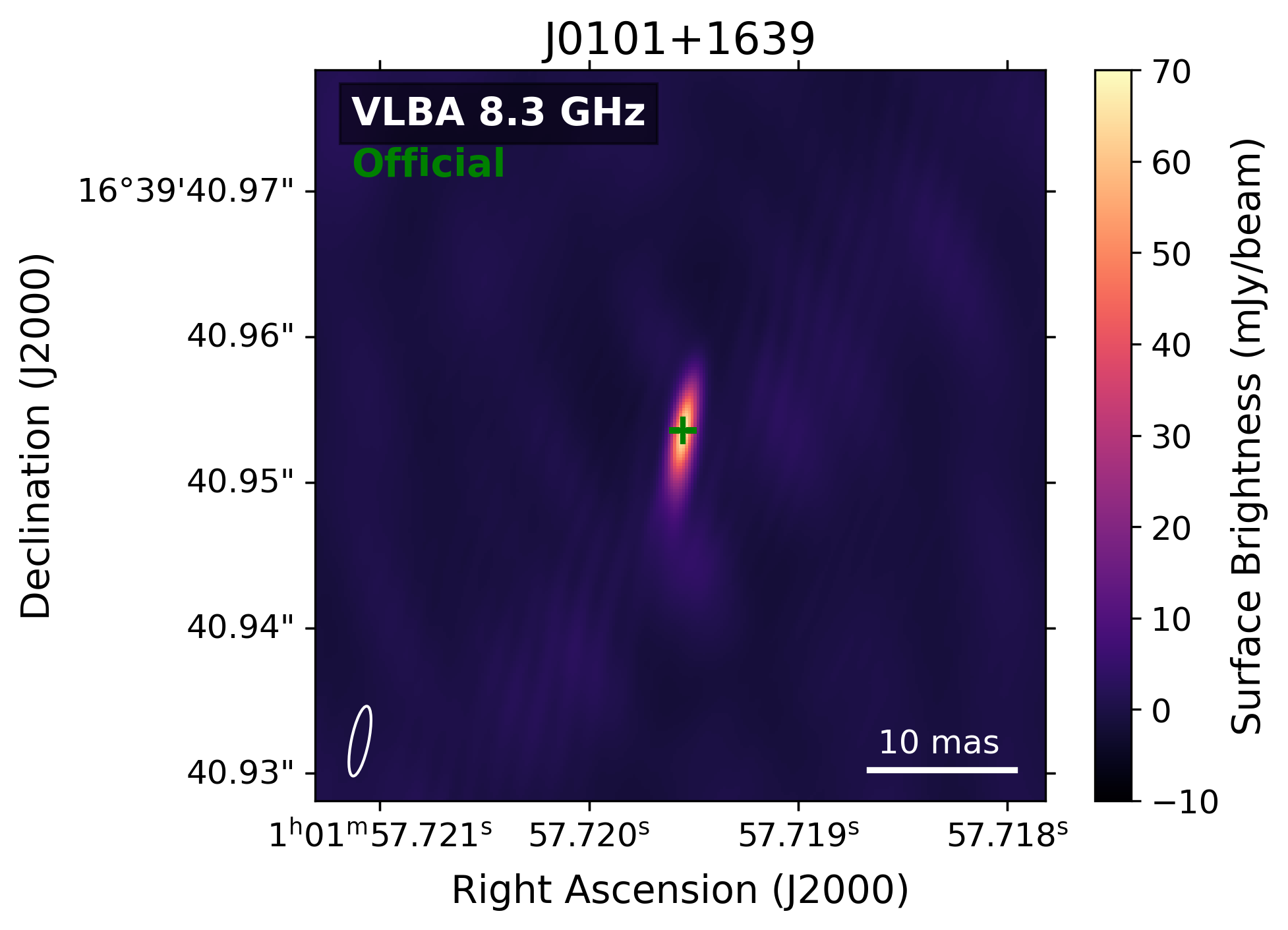}{0.5\textwidth}{(b) Observations at 8.3 GHz with a 0.005$\arcsec{}$ $\times$ 0.001$\arcsec{}$ beam. Image is 0.06'' square.}}
\caption{VLBA observations of \textbf{J0101+1639}, complex gain calibrator for J011114.41+171328.5.} 
\label{fig:011114_phase}
\end{figure*}

\begin{figure*}[!htb]
\gridline{\fig{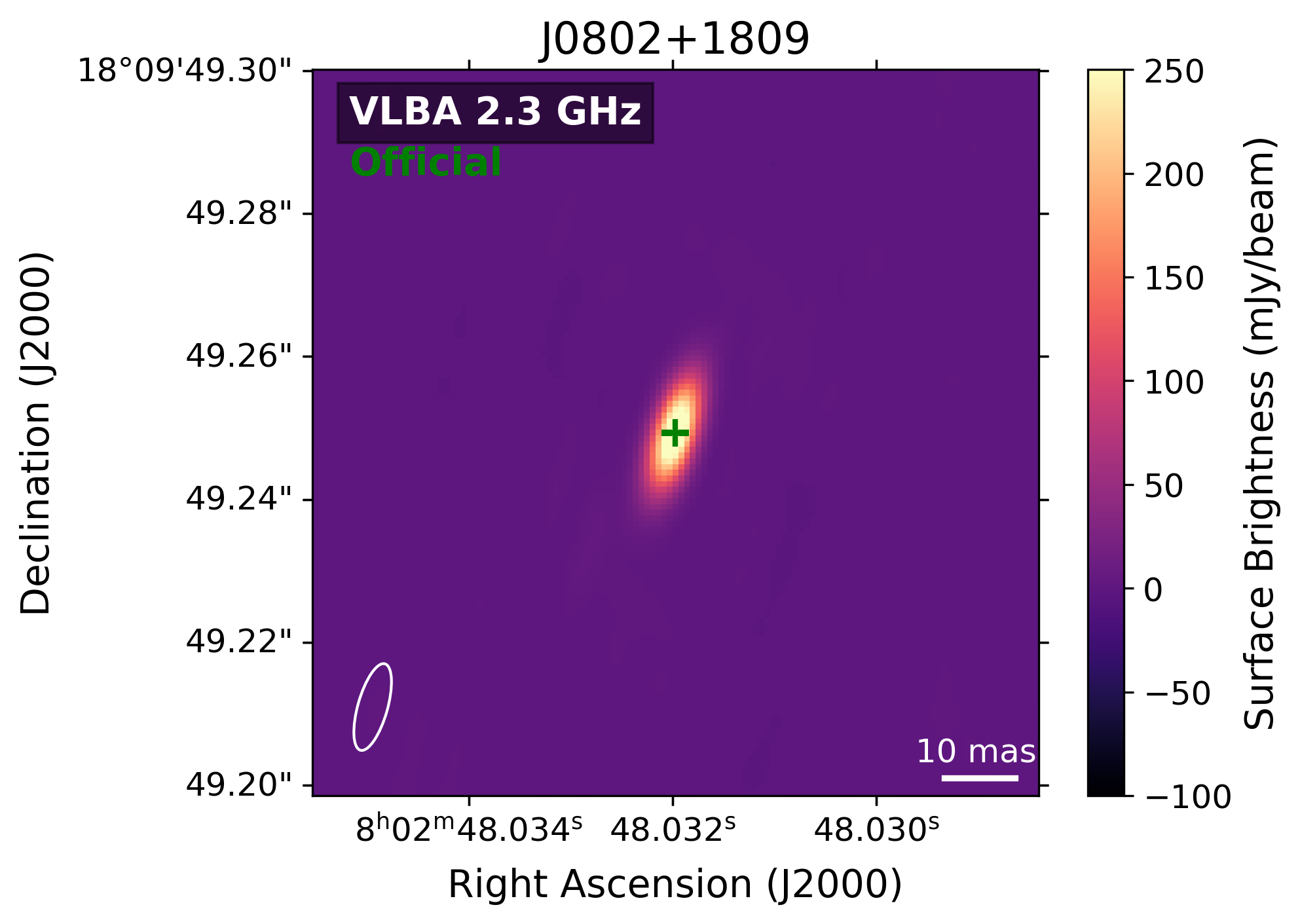}{0.5\textwidth}{(a) Observations at 2.3 GHz with a 0.013$\arcsec{}$ $\times$ 0.004$\arcsec{}$ beam. Image is 0.15'' square.}
          \fig{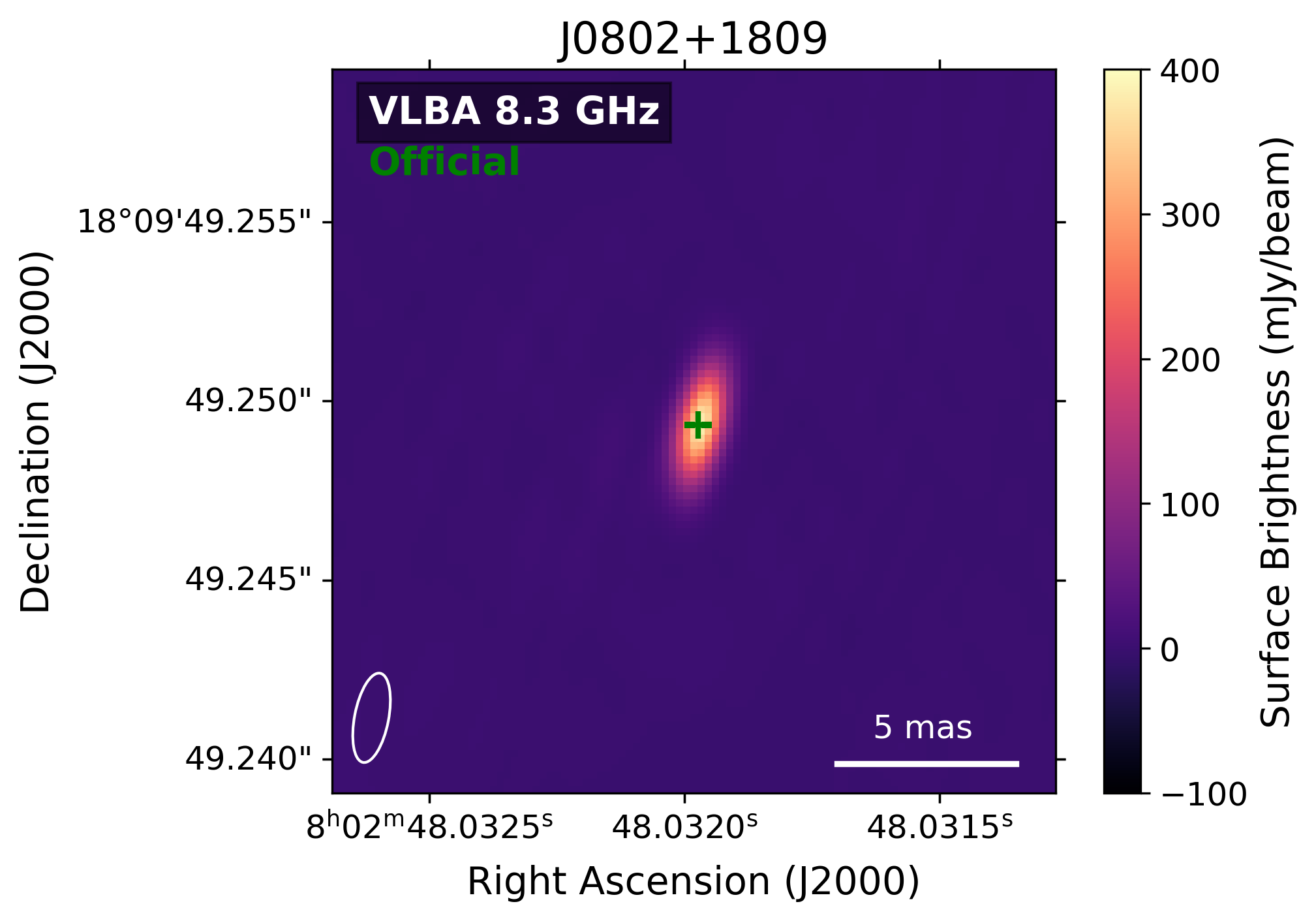}{0.5\textwidth}{(b) Observations at 8.3 GHz with a 0.003$\arcsec{}$ $\times$ 0.001$\arcsec{}$ beam. Image is 0.055'' square.}}
\caption{VLBA observations of \textbf{J0802+1809}, complex gain calibrator for J080009.98+165509.4.} 
\label{fig:080009_phase}
\end{figure*}

\begin{figure*}[!htb]
\gridline{\fig{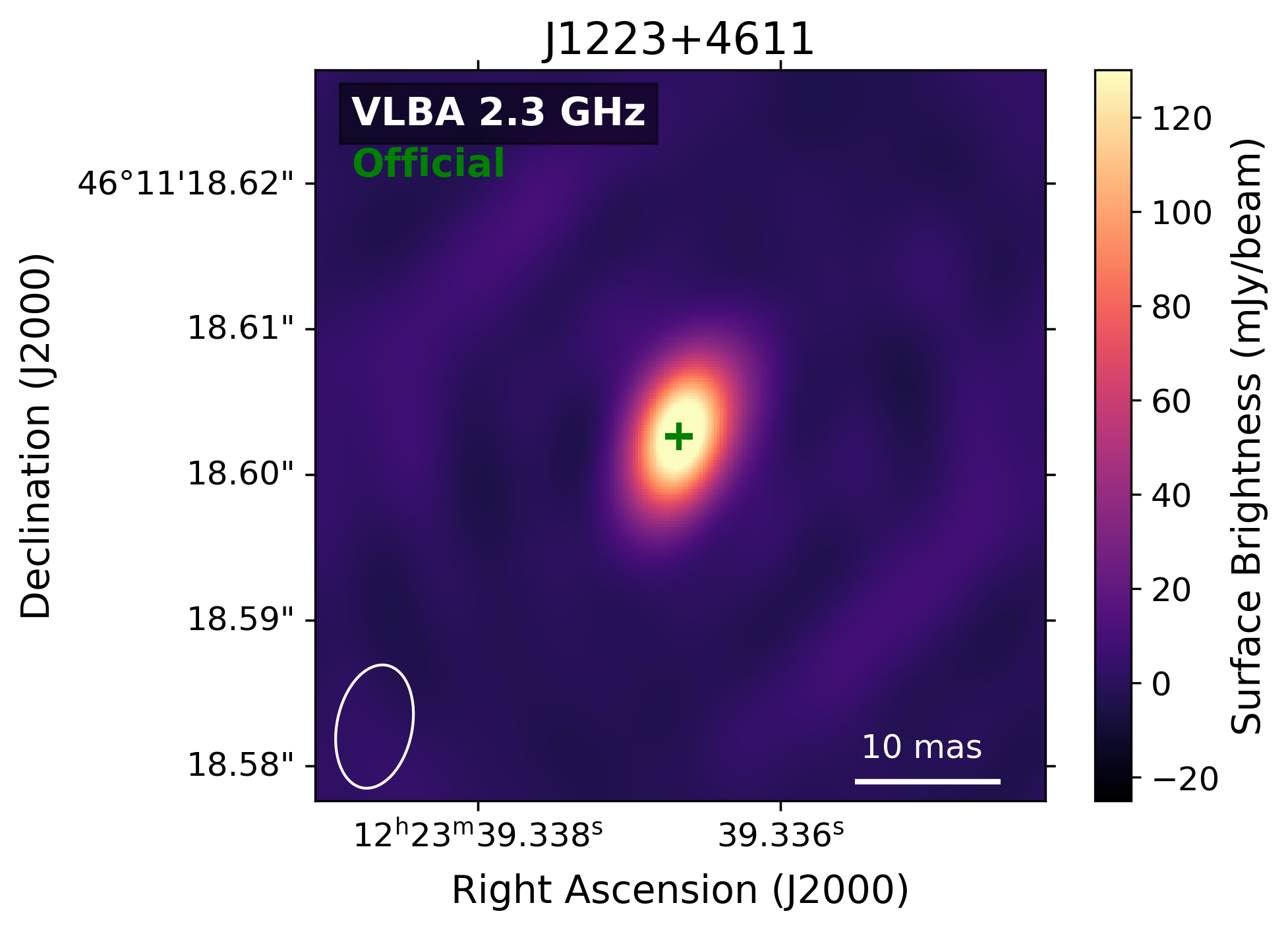}{0.5\textwidth}{(a) Observations at 2.3 GHz with a 0.008$\arcsec{}$ $\times$ 0.005$\arcsec{}$ beam. Image is 0.15'' square.}
          \fig{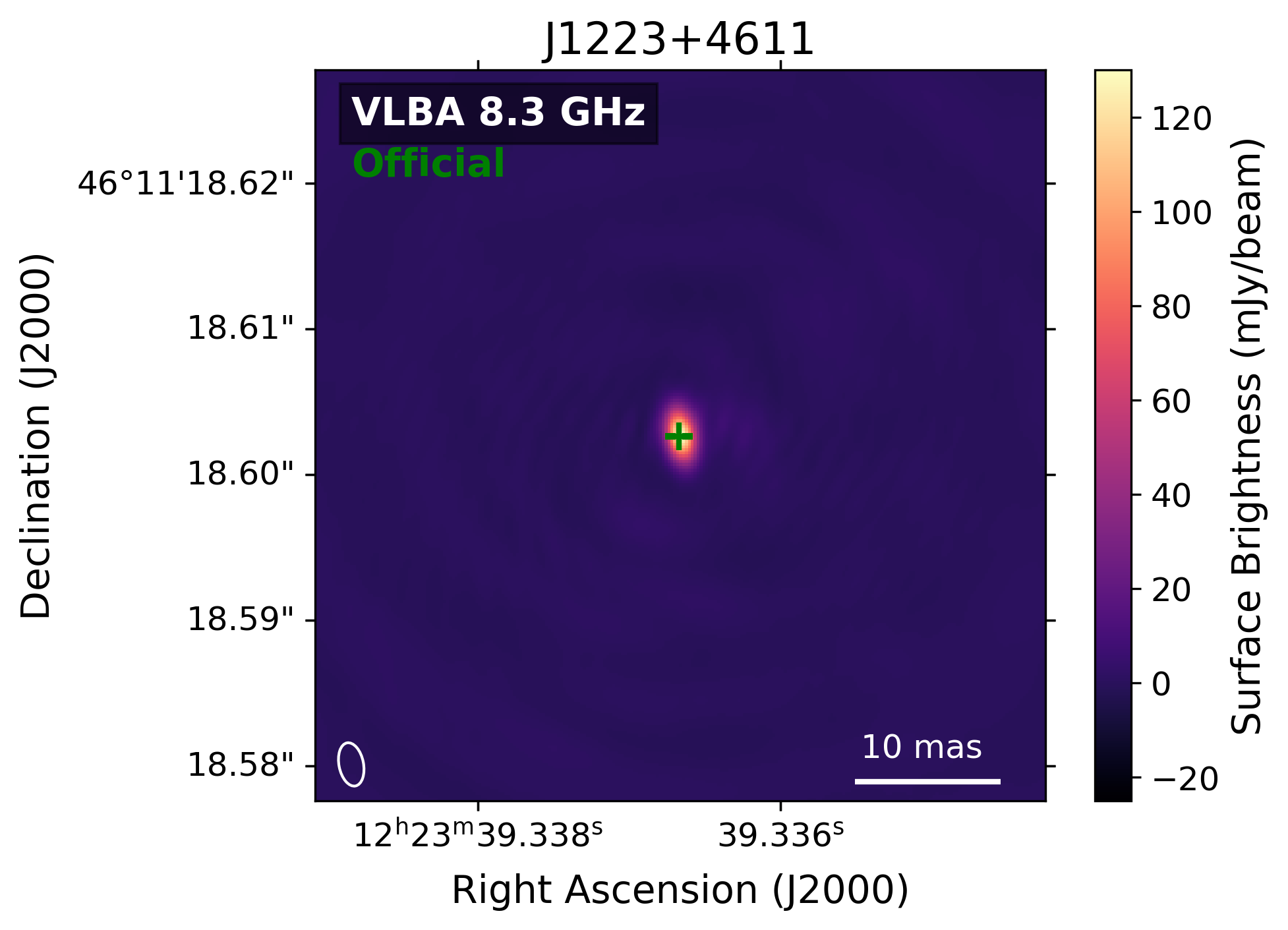}{0.5\textwidth}{(b) Observations at 8.3 GHz with a 0.003$\arcsec{}$ $\times$ 0.002$\arcsec{}$ beam. Image is 0.055'' square.}}
\caption{VLBA observations of \textbf{J1223+4611}, complex gain calibrator for J121544.36+452912.7.} 
\label{fig:121544_phase}
\end{figure*}

\begin{figure*}[!htb]
\gridline{\fig{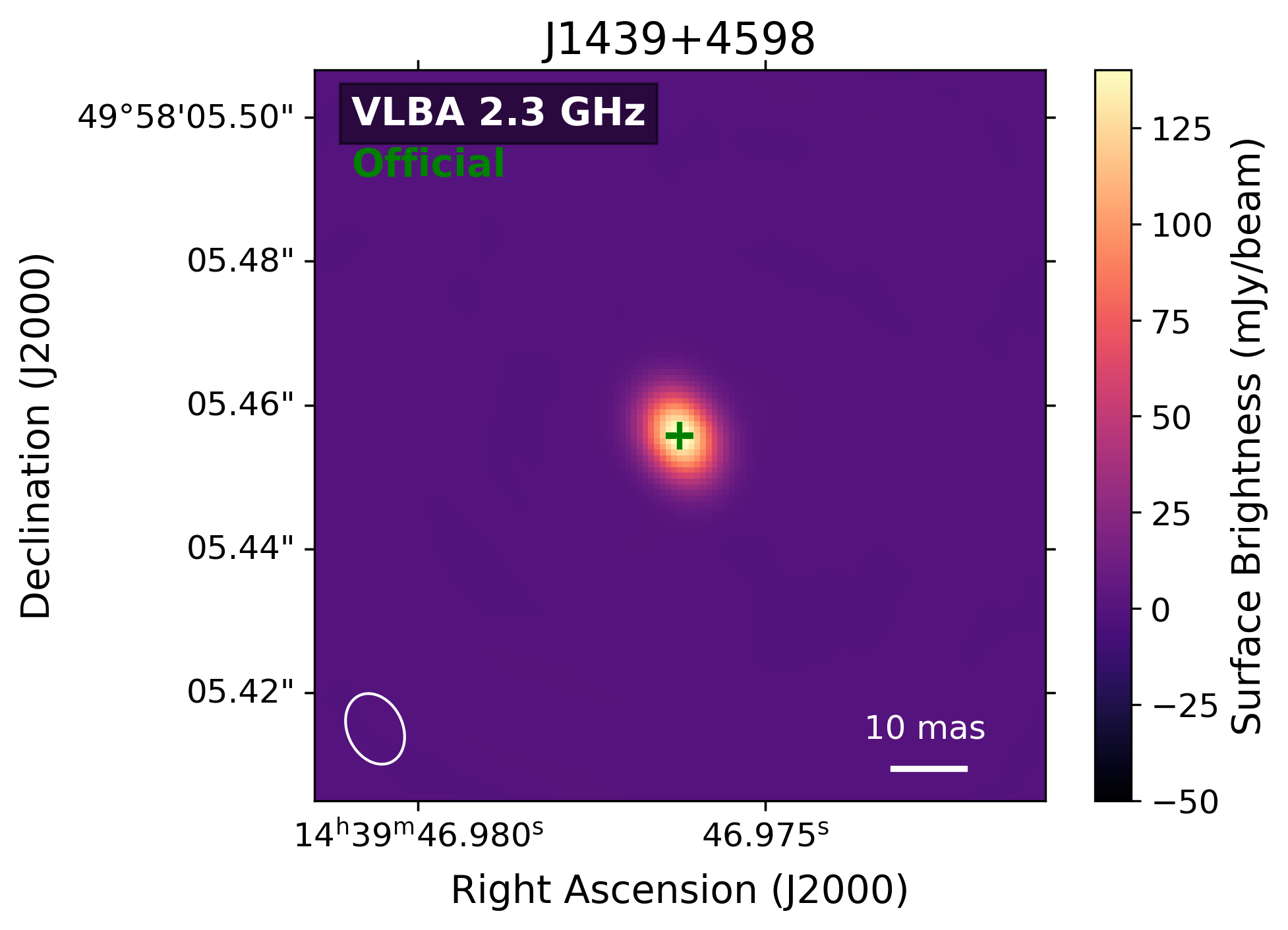}{0.5\textwidth}{(a) Observations at 2.3 GHz with a 0.010$\arcsec{}$ $\times$ 0.008$\arcsec{}$ beam. Image is 0.2'' square.}
          \fig{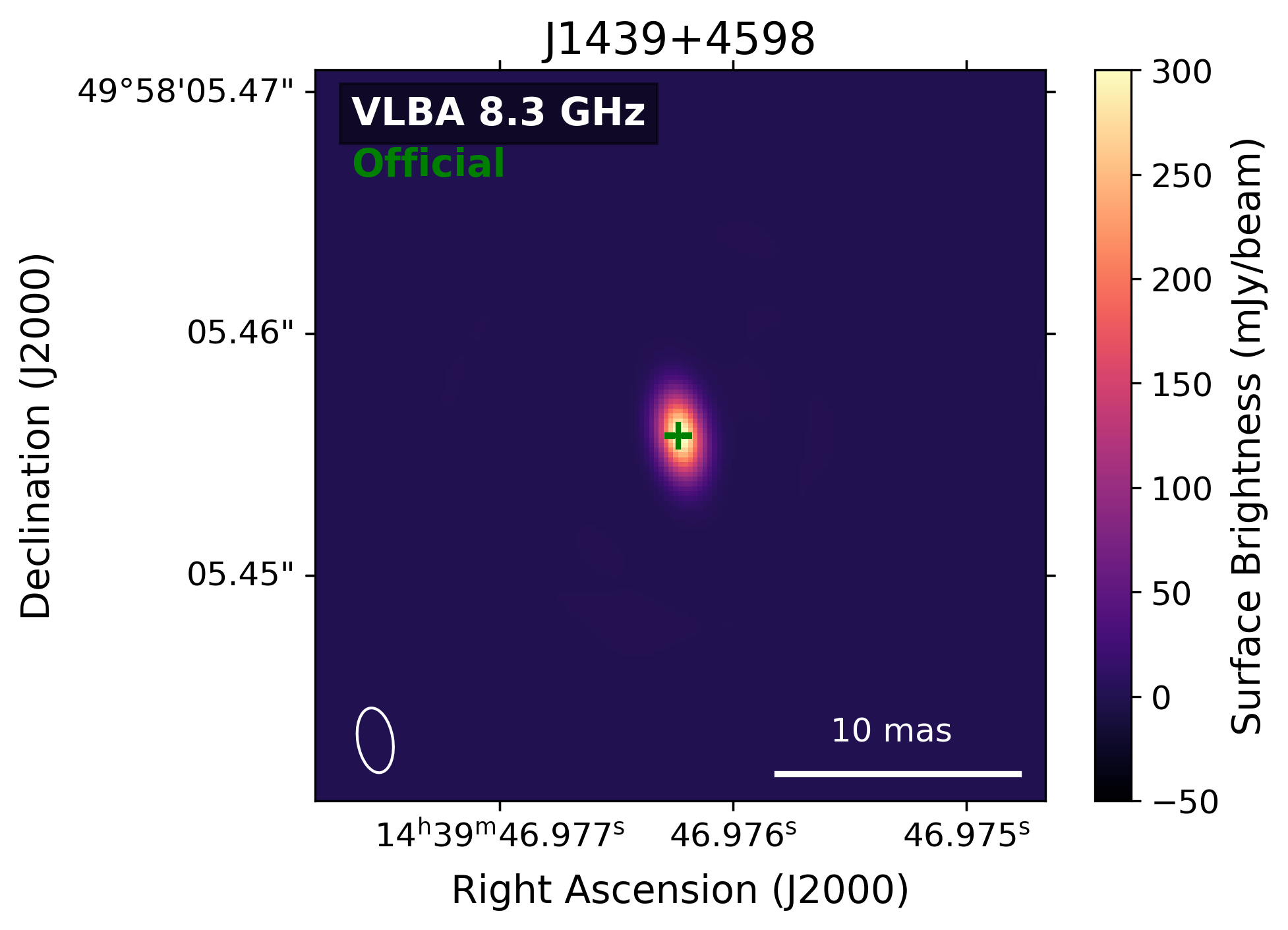}{0.5\textwidth}{(b) Observations at 8.3 GHz with a 0.003$\arcsec{}$ $\times$ 0.001$\arcsec{}$ beam. Image is 0.03'' square.}}
\caption{VLBA observations of \textbf{J1439+4958}, complex gain calibrator for J143333.02+484227.7.} 
\label{fig:143333_phase}
\end{figure*}

\begin{figure*}[!htb]
\gridline{\fig{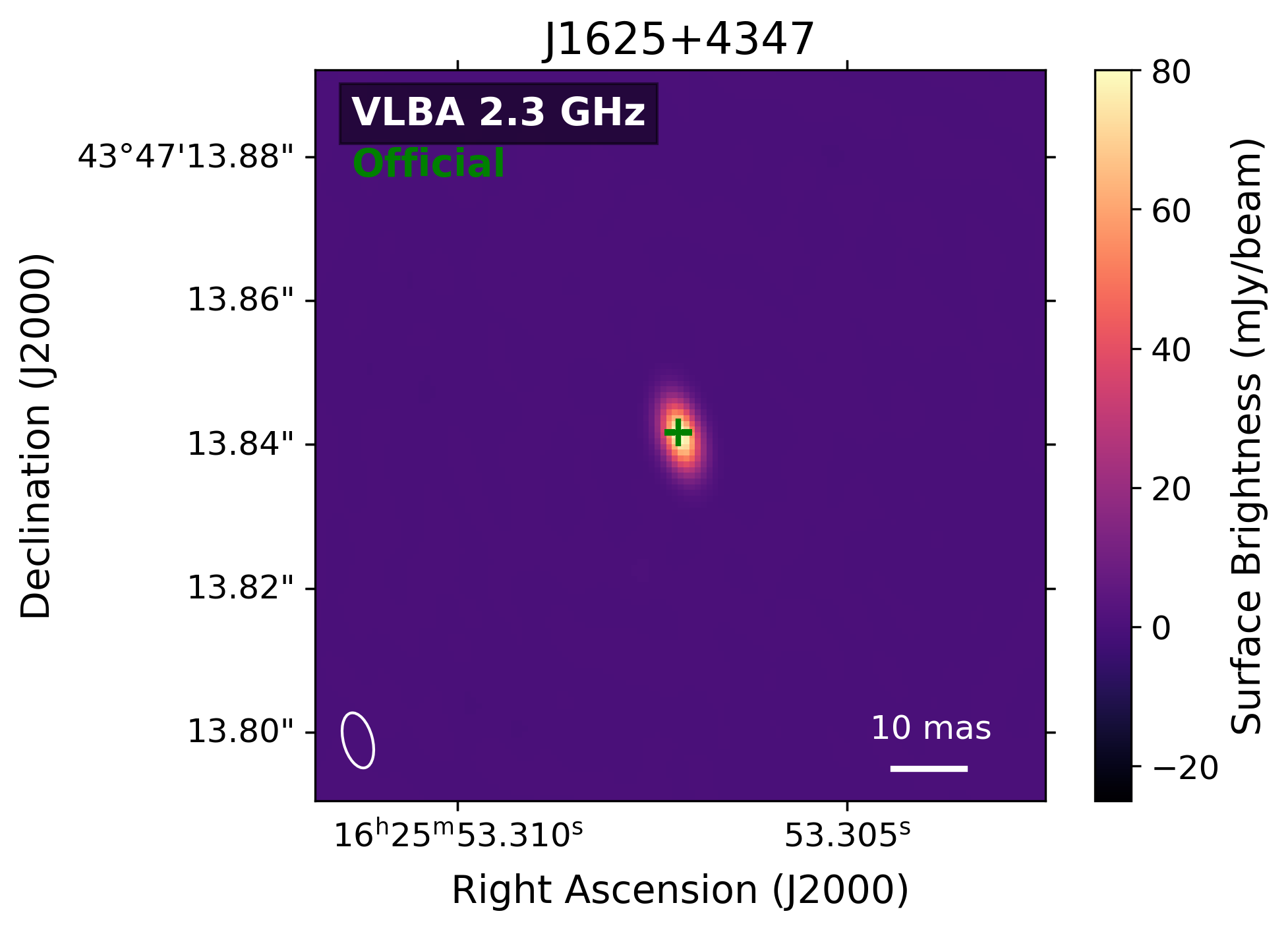}{0.5\textwidth}{(a) Observations at 2.3 GHz with a 0.008$\arcsec{}$ $\times$ 0.004$\arcsec{}$ beam. Image is 0.1'' square.}
          \fig{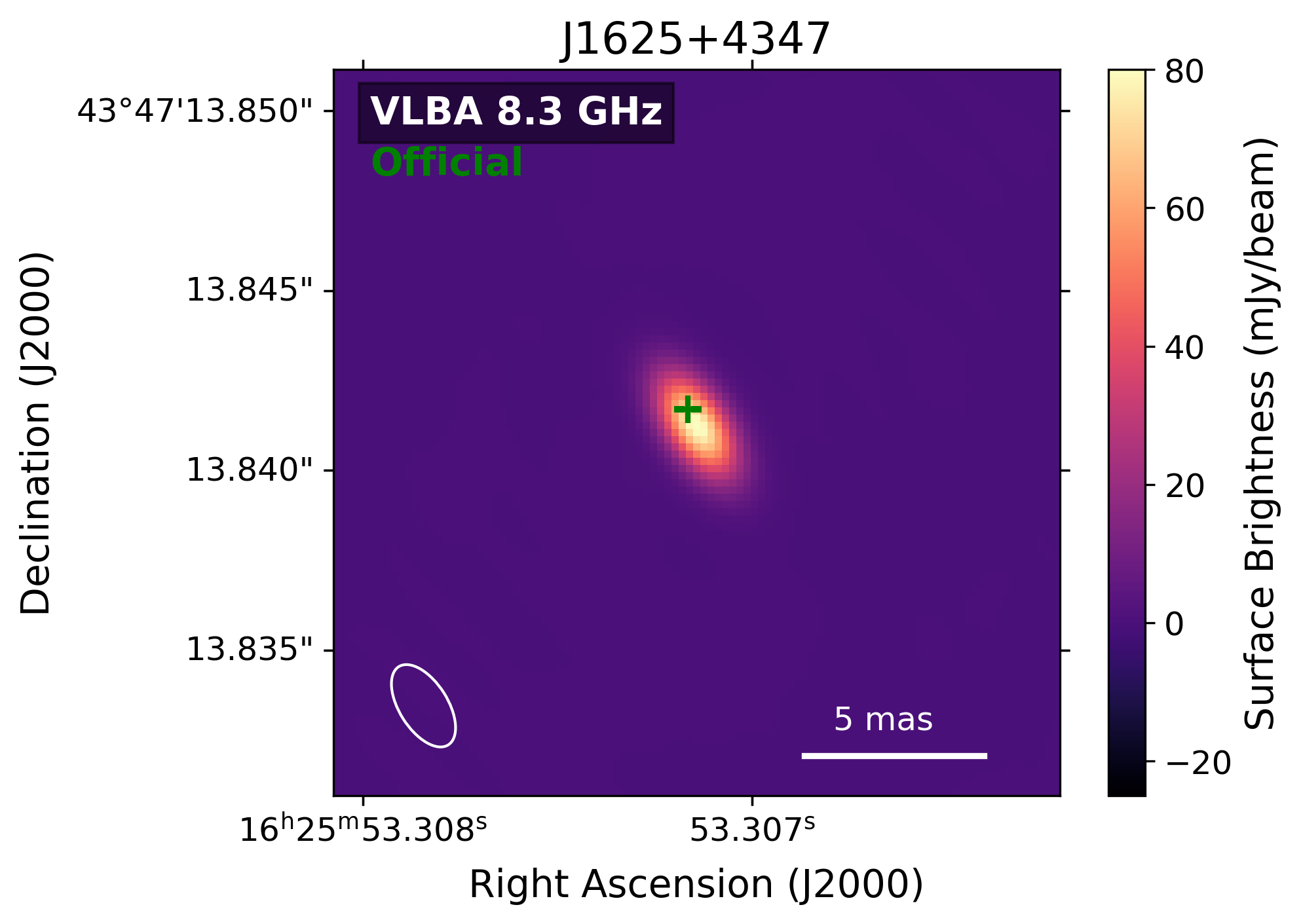}{0.5\textwidth}{(b) Observations at 8.3 GHz with a 0.003$\arcsec{}$ $\times$ 0.001$\arcsec{}$ beam. Image is 0.05'' square.}}
\caption{VLBA observations of \textbf{J1625+4347}, complex gain calibrator for J162501.98+430931.6.} 
\label{fig:162501_phase}
\end{figure*}

\begin{figure*}[!htb]
\gridline{\fig{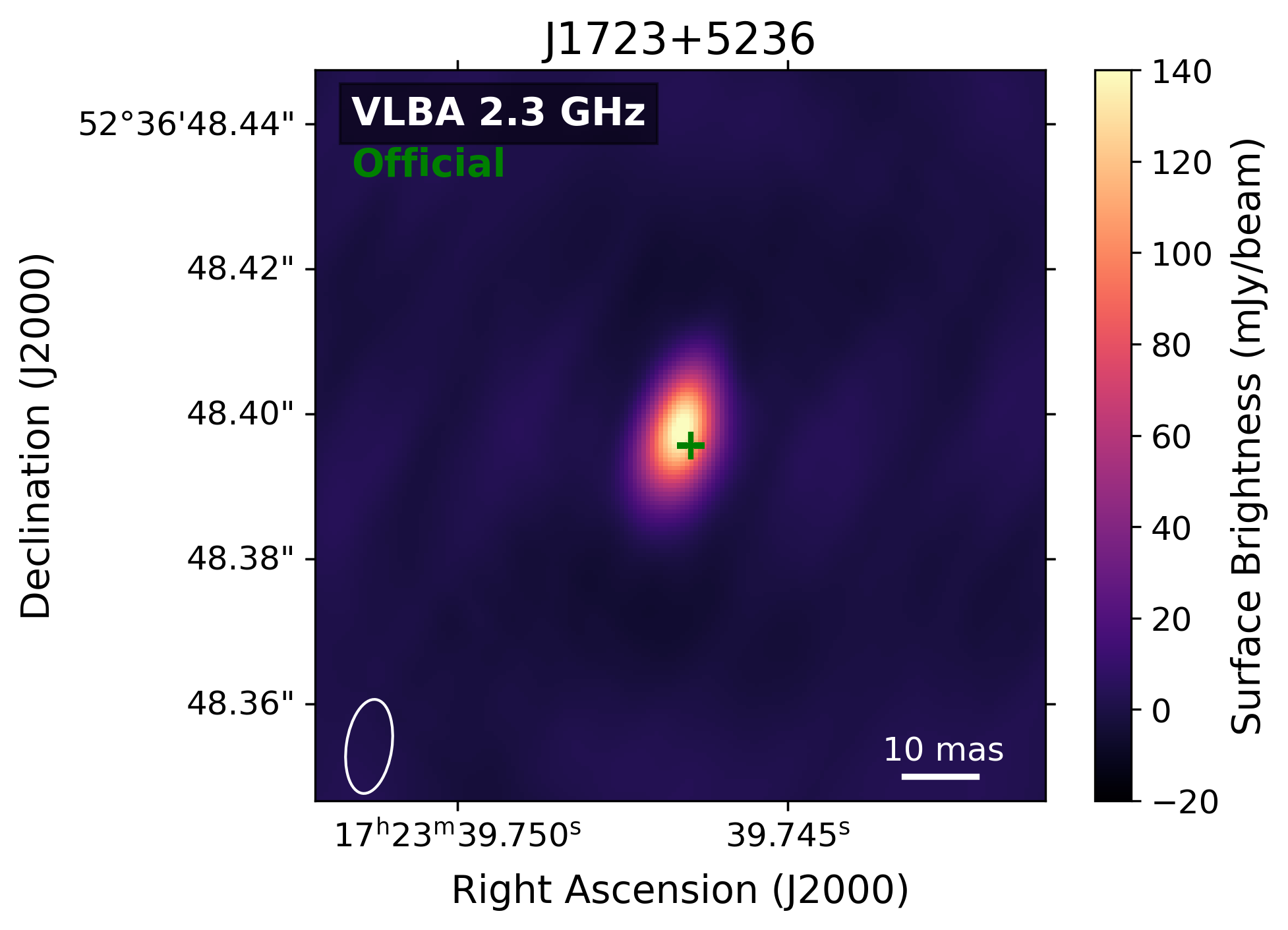}{0.5\textwidth}{(a) Observations at 2.3 GHz with a 0.013$\arcsec{}$ $\times$ 0.006$\arcsec{}$ beam. Image is 0.1'' square.}
          \fig{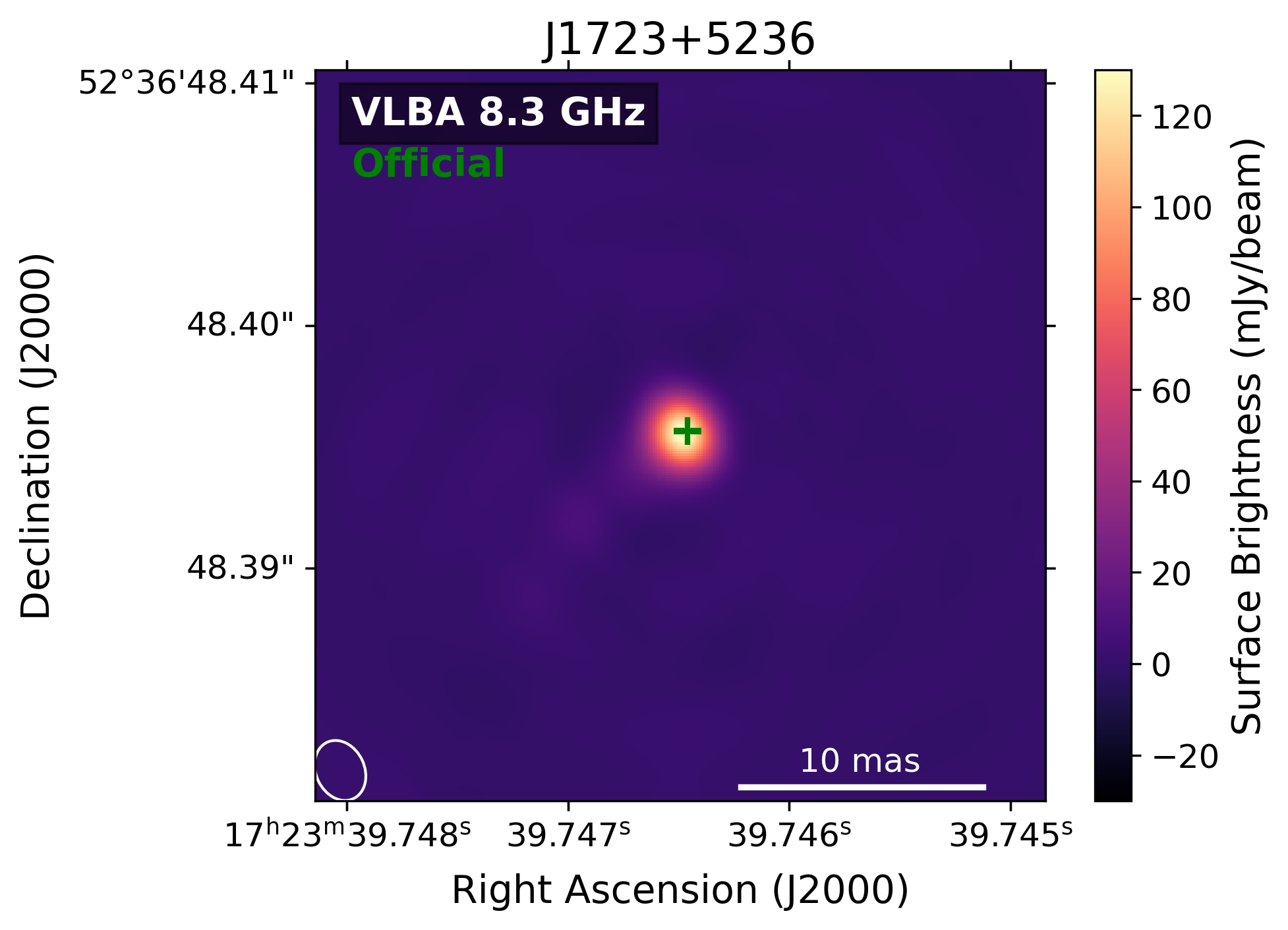}{0.5\textwidth}{(b) Observations at 8.3 GHz with a 0.003$\arcsec{}$ $\times$ 0.002$\arcsec{}$ beam. Image is 0.03'' square.}}
\caption{VLBA observations of \textbf{J1723+5236}, complex gain calibrator for J172308.14+524455.5.} 
\label{fig:172308_phase}
\end{figure*}

\begin{figure*}[!htb]
\gridline{\fig{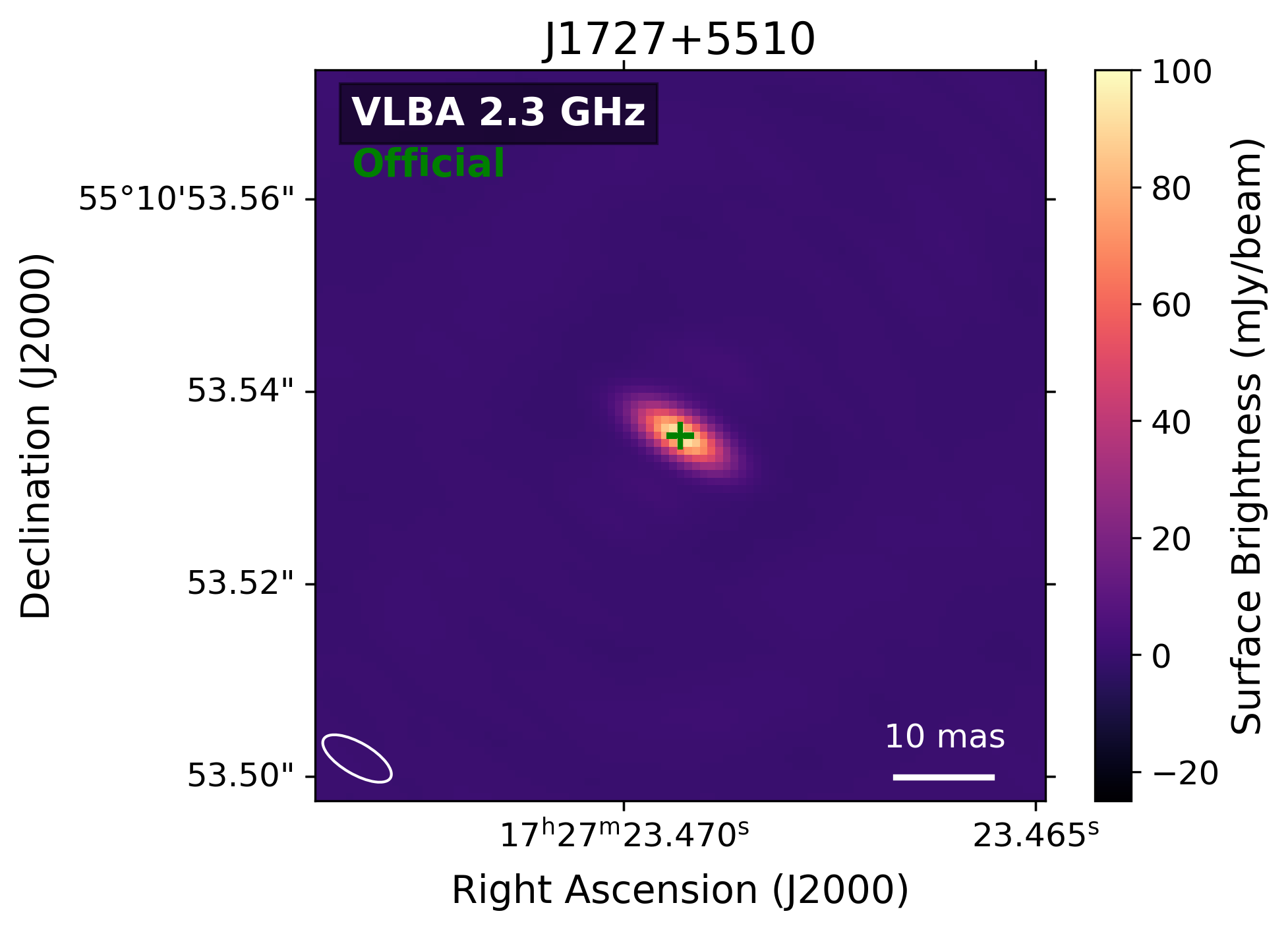}{0.5\textwidth}{(a) Observations at 2.3 GHz with a 0.008$\arcsec{}$ $\times$ 0.003$\arcsec{}$ beam. Image is 0.075'' square.}
          \fig{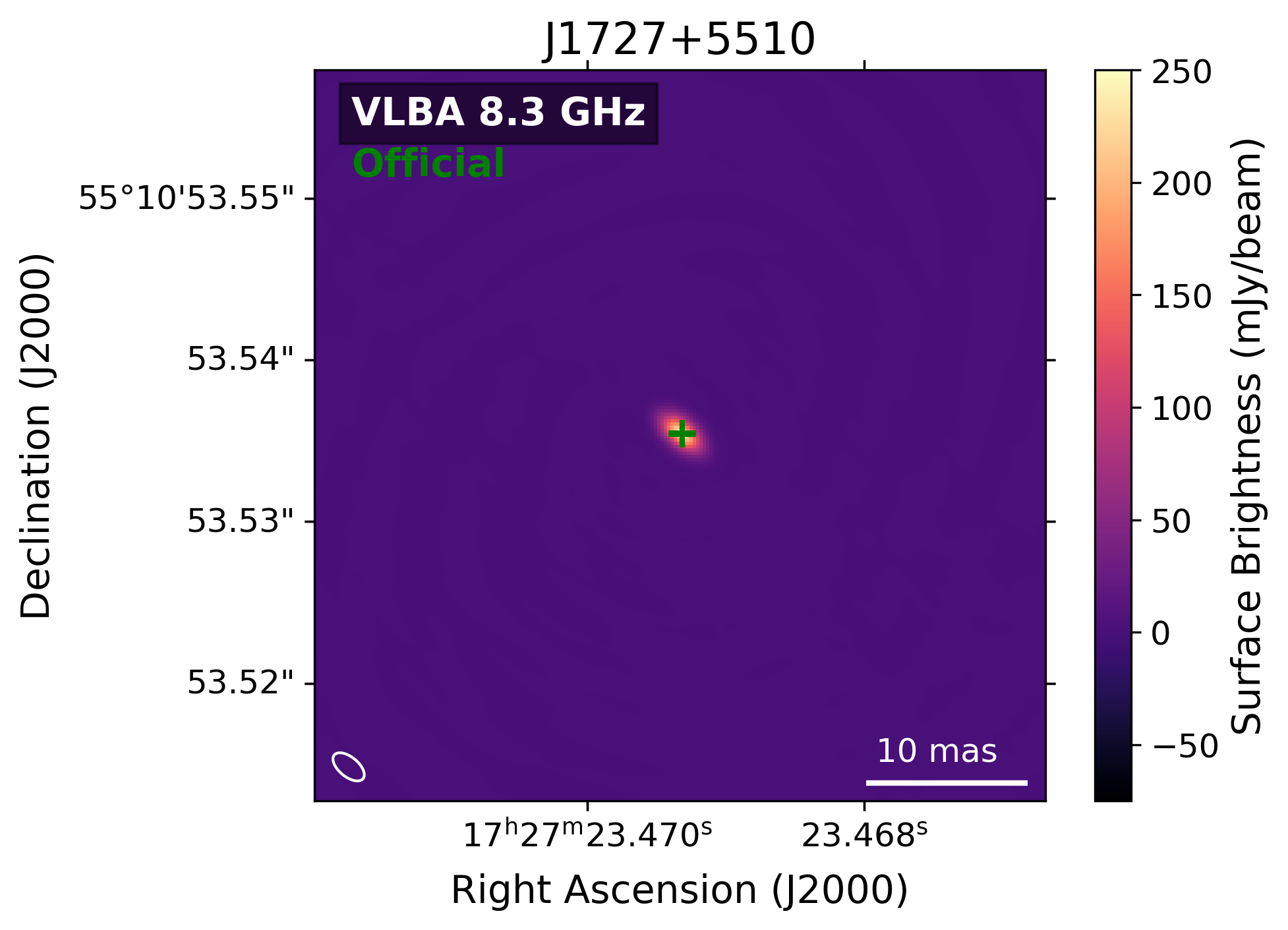}{0.5\textwidth}{(b) Observations at 8.3 GHz with a 0.002$\arcsec{}$ $\times$ 0.001$\arcsec{}$ beam. Image is 0.045'' square.}}
\caption{VLBA observations of \textbf{J1727+5510}, complex gain calibrator for J173330.80+552030.9.} 
\label{fig:173330_phase}
\end{figure*}

Figures \ref{fig:011114_phase} - \ref{fig:173330_phase} images of the phase calibrators associated with each source, for reference. All images are labeled with the appropriate frequency, a scale bar labeled with the appropriate number of milliarcseconds, and a beam in the lower left-hand corner. The green cross marks the location of the official position from the Radio Fundamental Catalog \citep[][]{petrov2024}.

\clearpage

\section{Appendix C: Optical Images} \label{sec:decals}

\begin{figure*}
\centering
\begin{minipage}{0.32\textwidth}
    \includegraphics[width=\linewidth]{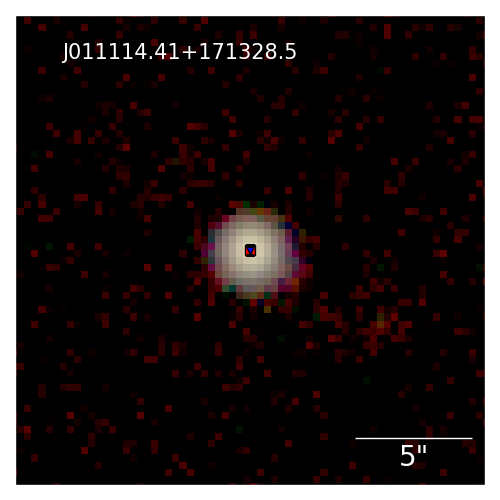}
\end{minipage}
\begin{minipage}{0.32\textwidth}
    \includegraphics[width=\linewidth]{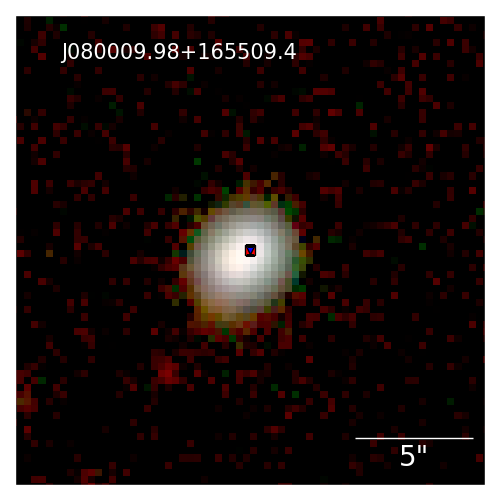}
\end{minipage}

\begin{minipage}{0.32\textwidth}
    \includegraphics[width=\linewidth]{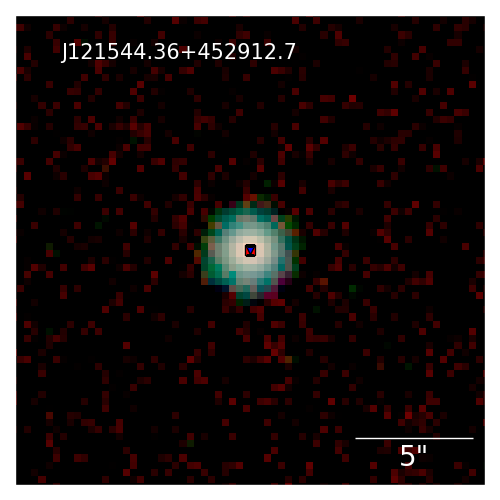}
\end{minipage}
\begin{minipage}{0.32\textwidth}
    \includegraphics[width=\linewidth]{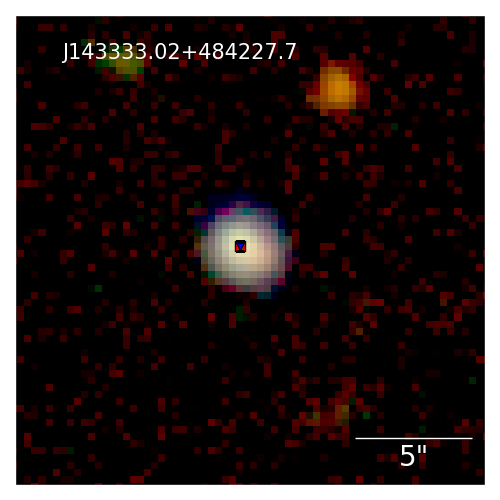}
\end{minipage}
\begin{minipage}{0.32\textwidth}
    \includegraphics[width=\linewidth]{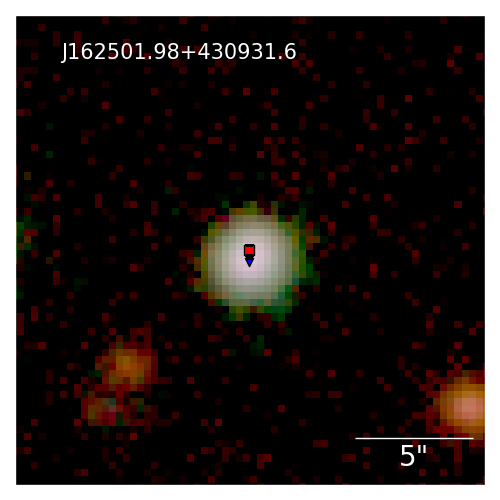}
\end{minipage}


\begin{minipage}{0.32\textwidth}
    \includegraphics[width=\linewidth]{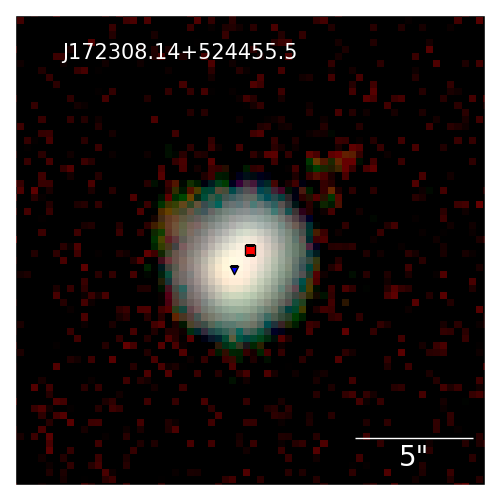}
\end{minipage}
\begin{minipage}{0.32\textwidth}
    \includegraphics[width=\linewidth]{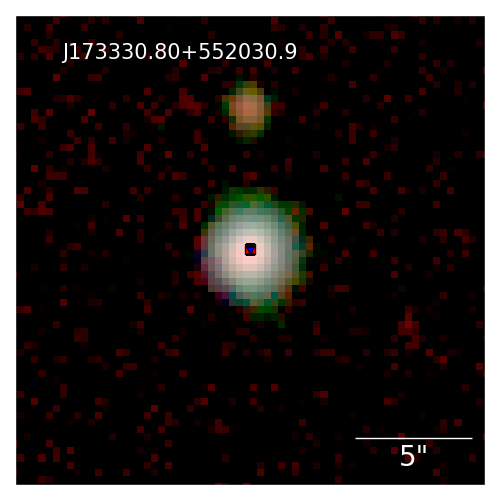}
\end{minipage}

\caption{Optical tricolor ugz images drawn from the DeCaLs Legacy Viewer for the full sample. The SDSS designation for each object is listed in the top left corner of each panel, while the scale bar in the bottom right corner indicates 5''. Notice in several cases that the sources appear elongated rather than pointlike, which may potentially point to underlying complex host morphologies and/or a multiplicity of sources, neither of which are resolvable in these images. The VLBA 8.3 GHz positions are marked with red squares, while the \textit{Gaia} optical positions are marked with blue triangles. Note that J172308.14+524455.5 has been identified as star+quasar superposition.}
\label{fig:decals}
\end{figure*}
\clearpage
\section{Appendix D: CASA Calibration}
\label{sec:vlba_calib_appendix}

\begin{figure*}
    \centering
     \includegraphics[width=14cm]{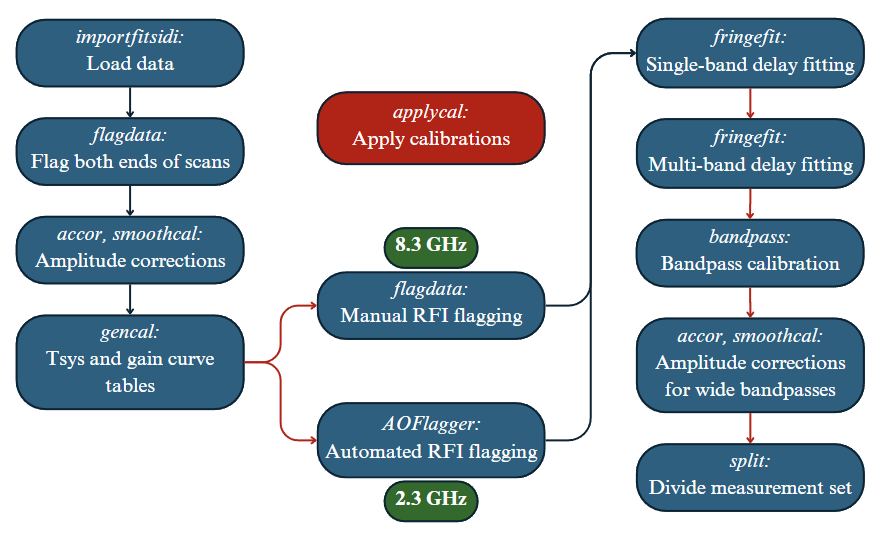}
        \vspace*{1mm}
          \caption{Flowchart depicting the overall data calibration process in CASA, following the method described in \cite[][]{Hunt_2021}. Titles of each step in italics represent CASA tasks, with the exception of AOFlagger. Red arrows represent points where calibration solutions were applied between steps. Different RFI flagging measures were applied depending on the frequency of the observations.}
    \label{fig:flowchart_casa}
\end{figure*}

The new VLBA observations presented in this paper were manually calibrated using CASA, following standard VLBA procedures for phase-referenced observations \citep[][]{VLBA_SciMemo38,vanbemmel2022}. This section describes the calibration process in detail, as well as presents a flowchart of the calibration process in Figure \ref{fig:flowchart_casa}.

The visibilities were loaded into CASA and inspected. Reference antennas were chosen from the most central antennas in the array (Kitt Peak, Fort Davis, Los Alamos, and Pie Town) and selected to have as little contribution from radio frequency interference (RFI) as possible. The timerange for instrumental delay calibration was chosen from the amplitude calibrator (see Table \ref{tab:obsdetails_vlba}) scans to be $\sim$60 seconds with as little RFI as possible.

The first and last four seconds of each scan were flagged. Amplitude corrections were determined from autocorrelations with a 30 second solution interval, and were smoothed with a median filter for 1800 seconds. The \textit{a priori} calibration tables and other initial solutions were applied with parallactic angle corrections. At this point, the considerable contributions from RFI were flagged. For all 8.3 GHz visibilities, \textsc{AOFlagger} \citep[][]{AOFlagger} was used for automated flagging. In the case of the 2.3 GHz visibilities, RFI was too severe for automated flagging, and so in-depth manual flagging was performed. The final flagged percentages of each visibility are recorded in Table \ref{tab:imgdetails_vlba}. The significant amount of flagging is not unexpected for the VLBA at these frequencies, particularly at S-band \citep[][]{ericVLBA_SciMemo41}.

Instrumental, single-band delay corrections were applied to the amplitude and complex gain calibrators following the fringe-fitting procedures. Corrections were determined using an infinite solution interval, zero delay rates, parallactic angle corrections, and a minimum signal-to-noise ratio (SNR) of ten. Similarly, multi-band delay corrections were applied, having been determined using a 30-second solution interval, parallactic angle corrections, and a minimum SNR of between 3 and 5 (depending on the severity of the RFI impacting the amplitude calibrator). Finally, bandpass calibration was performed for the amplitude calibrator. The amplitude calibrations were determined from the antenna information, as calibrators of constant, known brightness were not available at these resolutions. Thus, the system equivalent flux density (SEFD) was used for each antenna to calibrate amplitudes \citep[][]{hunt2021,VLBA_SciMemo38}. Corrections were determined using an infinite solution interval, normalized amplitudes and complex gains for each spectral window, and parallactic angle correction. Given the relatively wide bandwidth of the observations, amplitude corrections were made for a second time after the bandpass calibration was complete in order to correctly account for the wide bandpasses. The second round of amplitude corrections were derived with a two minute solution interval, smoothed with a median filter for 1800 seconds, and with parallactic angle corrections. 

\end{document}